\documentclass[aps,prl,nofootinbib,twocolumn,superscriptaddress,preprintnumbers,balancelastpage,longbibliography]{revtex4-1}

\usepackage{placeins}
\usepackage{amsmath,amssymb,mathtools,bm}
\usepackage{comment}
\usepackage[normalem]{ulem}
\usepackage{gensymb}
\usepackage{cancel}
\usepackage{caption}
\usepackage{graphicx, color, hepunits}
\usepackage[dvipsnames]{xcolor}
\usepackage{float}
\usepackage{multirow}
\usepackage{slashed}
\usepackage{pifont}

\usepackage[colorlinks=true,urlcolor=purple,anchorcolor=blue,citecolor=blue,
filecolor=magenta,linkcolor=blue,menucolor=blue,linktocpage=true,
pdfproducer=medialab,pdfa=true]{hyperref}
\usepackage[utf8]{inputenc}
\usepackage[english]{babel}
\usepackage{soul}
\usepackage{tabularx}
\usepackage{array}
\usepackage{booktabs}

\makeatletter
\def\hlinewd#1{
\noalign{\ifnum0=`}\fi\hrule \@height #1 \futurelet
\reserved@a\@xhline}
\makeatother

\newcommand{\mpl}{M_{\rm pl}}

\makeatletter
\newcounter{savesection}
\newcounter{apdxsection}
\renewcommand\appendix{\par
  \setcounter{savesection}{\value{section}}
  \setcounter{section}{\value{apdxsection}}
  \setcounter{subsection}{0}
  \gdef\thesection{\@Alph\c@section}}
\newcommand\unappendix{\par
  \setcounter{apdxsection}{\value{section}}
  \setcounter{section}{\value{savesection}}
  \setcounter{subsection}{0}
  \gdef\thesection{\@arabic\c@section}}
\makeatother

\restylefloat{table}

\begin{document}

\preprint{CERN-TH-2026-098}

\title{
Heterotic String Theory Suggests a QCD Axion Near 0.5 neV
}

\author{Joshua N. Benabou}
\affiliation{Theoretical Physics Group, Lawrence Berkeley National
  Laboratory, Berkeley, CA 94720, U.S.A.}
\affiliation{Berkeley Center for Theoretical Physics, University of
  California, Berkeley, CA 94720, U.S.A.}

\author{Giulio Alvise Dainelli}
\affiliation{Theoretical Physics Group, Lawrence Berkeley National
  Laboratory, Berkeley, CA 94720, U.S.A.}
\affiliation{Berkeley Center for Theoretical Physics, University of
  California, Berkeley, CA 94720, U.S.A.}

\author{Mario Reig}
\affiliation{Theoretical Physics Department, CERN, 1211 Geneva 23,
  Switzerland}

\author{Benjamin R. Safdi}
\affiliation{Theoretical Physics Group, Lawrence Berkeley National
  Laboratory, Berkeley, CA 94720, U.S.A.}
\affiliation{Berkeley Center for Theoretical Physics, University of
  California, Berkeley, CA 94720, U.S.A.}

\date{\today}

\begin{abstract}
We show that in heterotic string theory --- and dual corners of the landscape including Type I string theory --- the QCD axion mass is bounded from below by $m_a \gtrsim 0.5$~neV, a direct consequence of the model-independent axion whose decay constant is fixed by the grand unified theory (GUT) gauge coupling.
We explicitly compute the mass of the QCD axion in an ensemble of heterotic compactifications on Calabi-Yau
hypersurfaces of toric varieties sampled from the Kreuzer-Skarke (KS) 
ensemble, as well as on complete intersection Calabi-Yau manifolds.
We then perform an
extensive search over the K\"ahler moduli space of KS
compactifications with up to $11$ axions --- the maximum we identify as consistent with unification in our sample. We
establish that for all but a handful of manifolds the QCD axion mass is precisely
the model-independent value, lying in $[0.5, 0.8]\,\mathrm{neV}$, depending on the GUT gauge coupling. 
This window
should be a high-priority target for future lumped-element detectors such as
DMRadio-GUT. We show that the heavy axion population in our
heterotic ensemble generically decays before big bang nucleosynthesis and can naturally
accommodate leptogenesis, unlike in Type IIB axiverse constructions.
\end{abstract}

\maketitle

\textbf{\textit{Introduction.}}---The quantum chromodynamics (QCD) axion has emerged as the leading contender to solve the Strong $CP$ problem of the neutron electric dipole moment (EDM), while also explaining the observed dark matter (DM) of the Universe~\cite{Peccei:1977hh,Weinberg:1977ma,Wilczek:1977pj,Preskill:1982cy,Abbott:1982af,Dine:1982ah}.  Laboratory searches for QCD axion DM are underway worldwide (see~\cite{Irastorza:2018dyq,Adams:2022pbo,Berlin:2024pzi}
for reviews), but they are stymied by the currently unknown mass of the axion, $m_a$.
Recent work, however, suggests that if the QCD axion emerges from string theory~\cite{Svrcek:2006yi,Arvanitaki:2009fg}, which is the most natural setting for generating high-quality QCD axions that can solve the Strong $CP$ problem (see, {\it e.g.},~\cite{Kamionkowski:1992mf}), and if the compactification allows for grand unification, then $10^{-11} \, {\rm eV} \lesssim m_a \lesssim 10^{-8} \, {\rm eV}$~\cite{Benabou:2025kgx}.  In this work, we argue 
that 
in
heterotic string theory the QCD axion mass is even more sharply bounded, with the lower bound $m_a \gtrsim \left(5.2 \times 10^{-10} \, \mathrm{eV} \right) (\alpha_\mathrm{GUT}^{-1} /25)$, depending on the grand unified theory (GUT)~\cite{Georgi:1974sy,Fritzsch:1974nn} gauge coupling $\alpha_{{\rm GUT}}$, with upward deviations from this value possible but difficult to achieve. This result strongly motivates experiments such as DMRadio-GUT~\cite{DMRadio:2022pkf}, based on the successful ABRACADABRA program~\cite{Kahn:2016aff,Ouellet:2018beu,Salemi:2021gck,Benabou:2022qpv}, whose projected sensitivity window contains this mass (see also~\cite{Graham:2013gfa,Budker:2013hfa,Berlin:2019ahk,Giaccone:2022pke}). 
\begin{figure}[t]
    \centering
\includegraphics[width=1.0\linewidth]{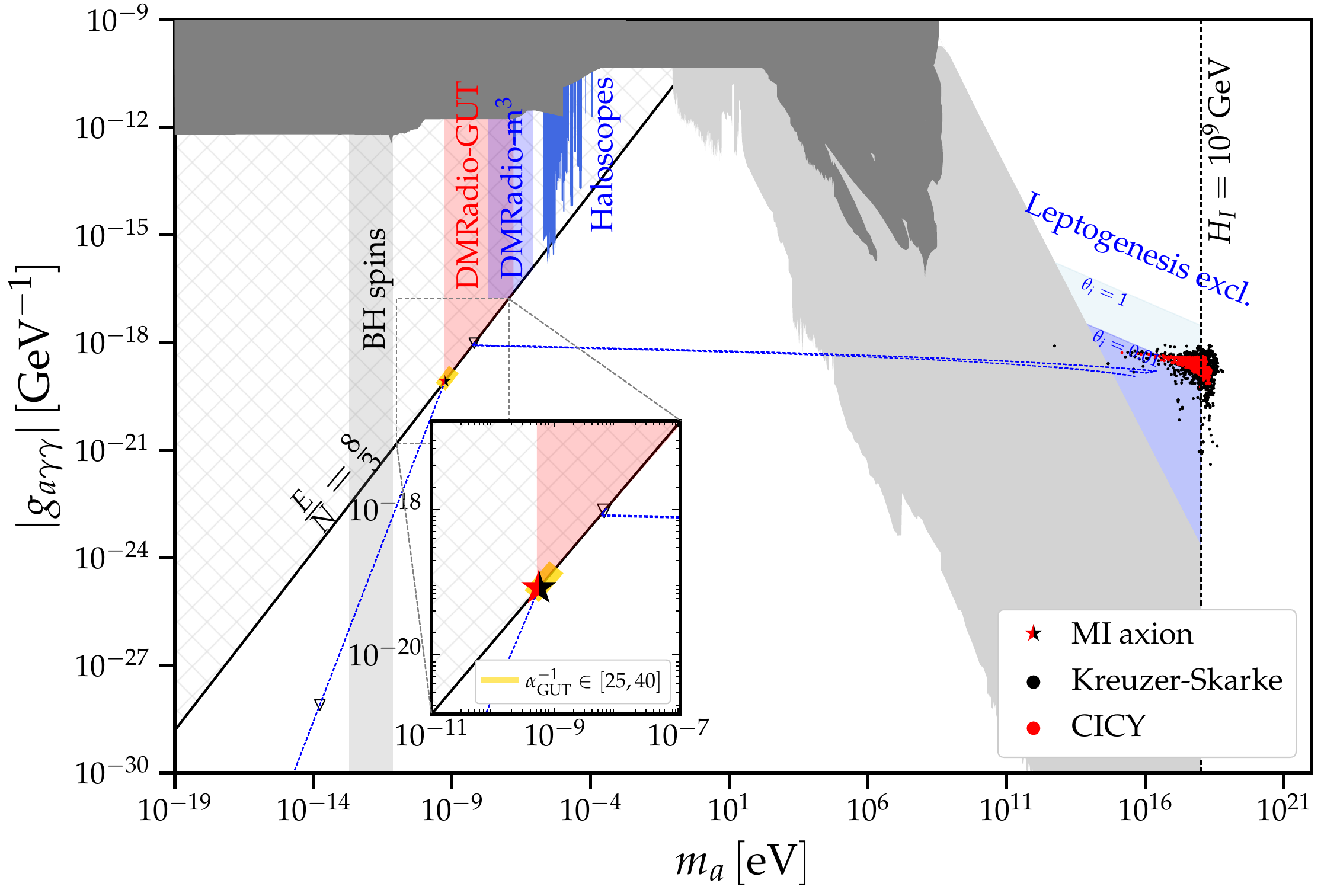}
    \caption{
    Axion-photon couplings for all axions in our GUT-compatible ensemble of 2027 
    (375) 
    KS (CICY) heterotic compactifications (points), fixing $\alpha_\mathrm{GUT}^{-1}=27$ and $g_s=1$, and the location in K\"ahler moduli space to be along the ray connecting the origin to the tip of the SKC. The MI QCD axion mass range is shaded in gold. Dotted curves show the two lightest eigenstates, varying within moduli space, for the three $h^{1,1}=2$ compactifications where the QCD axion mass may deviate from the MI value. We shade constraints from astrophysical and cosmological probes without (dark gray) and with (light gray) cosmological assumptions, along with leptogenesis-disfavored regions (see text).
    }
    \label{fig:gayy_heterotic}
    \vspace{-0.2in}
\end{figure}

Heterotic string theory~\cite{Gross:1984dd}
compactified on Calabi-Yau (CY) 3-folds is a compelling pathway to obtaining the Standard Model low-energy effective field theory (EFT) in part because it provides a natural framework
for GUTs~\cite{Green:1987mn}. The $E_8 \times E_8$ gauge group
contains GUT subgroups including $SO(10)$,
whose spinor representation $\mathbf{16}$ automatically accommodates a
right-handed neutrino, which may enable the seesaw
mechanism and thermal leptogenesis~\cite{Fukugita:1986hr,Davidson:2008bu,Buchmuller:2004nz}.
Alongside this, the Kalb-Ramond $B_2$ field and its 10D dual $B_6$ generate axions upon
dimensional reduction on the CY 3-fold~\cite{Witten:1984Dg,Choi:1985bz}.
The \textit{model-independent} (MI) axion arises from $B_6$, and its decay constant is fixed
entirely by the GUT fine-structure constant $\alpha_{\rm GUT}$~\cite{Svrcek:2006yi}:
\begin{equation}
    f_\mathrm{MI} = \frac{\alpha_\mathrm{GUT}}{2\pi}\frac{\mpl}{\sqrt{2}} \,.
    \label{eq:MI_fa_main}
\end{equation}
(The MI decay constant is independent of the warp factor in warped heterotic compactifications~\cite{Kim:2006aq}.) For $\alpha_\mathrm{GUT}^{-1} \in [25, 30]$, which is the range consistent
with supersymmetric (SUSY) grand unification as we discuss, this gives a QCD axion mass $m_a \in
[5.2, 6.3]\times 10^{-10}$~eV.
In this Letter we show that the {\it model-dependent} (MD) axions, which are those that arise from the dimensional reduction of $B_2$ on holomorphic 2-cycles of the CY, may mix with the MI axion but that this can only increase $m_a$ relative to the MI value.  This is a straightforward consequence of the
quadratic sum rule for axion decay constants. Moreover, in all cases we have tested, it is difficult to make the MD axions light enough to efficiently mix with the MI axion, such that for almost all the scenarios that we construct the QCD axion mass is simply given by the MI value.  Additionally, in almost all scenarios we find no ultralight axions lighter than the QCD axion, 
as needed for {\it e.g.} fuzzy DM~\cite{Hui:2016ltb,Sheridan:2024vtt,Leedom:2025mlr}. 

The reason that mixing is difficult is that it
requires a worldsheet instanton action $S_\mathrm{ws} = 2\pi \mathrm{Vol}(C_2)$ with a large effective curve volume $\mathrm{Vol}(C_2) \gtrsim 20$ 
,
so that the mass contribution to the MD axions from worldsheet instantons is subdominant relative to that from QCD. However, grand unification
fixes the total CY volume to $\mathcal{V}_6 \approx \alpha_{\rm GUT}^{-1} \sim 25$ in string
units --- assuming $g_s = 1$, which is conservative as we show later ---  making such large individual curve volumes impossible for
almost all CY 3-folds that we construct in this work. In particular, we verify this explicitly
across 2027 heterotic compactifications from the Kreuzer-Skarke (KS) toric ensemble~\cite{Kreuzer:2000xy} that are consistent with grand unification, as well as across
a sample of complete intersection CY manifolds (CICYs), finding only
six exceptions (all with $h^{1,1} =2$ or $h^{1,1} =3$ MD axions). All other
compactifications have a QCD axion mass equal to the MI value across the full moduli space that is under perturbative control.

The MI axion shift
symmetry is explicitly broken by exponentially-suppressed Euclidean NS5-brane instantons with
action $S_\mathrm{NS5} = 2\pi/\alpha_\mathrm{GUT}$. For the axion to
solve the Strong $CP$ problem, these must be subdominant to QCD
instantons.
Intriguingly, we find that while predictions for $\alpha_\mathrm{GUT}$ in scenarios such as Split SUSY~\cite{Giudice:2004tc} and Mini-Split SUSY~\cite{Arvanitaki:2012ps} satisfy current Peccei-Quinn (PQ) quality constraints, a non-zero neutron EDM could be detectable by near-term experiments in these models~\cite{nEDM:2019qgk}.

\begin{figure}[!t]
    \centering
\includegraphics[width=1.0\linewidth]{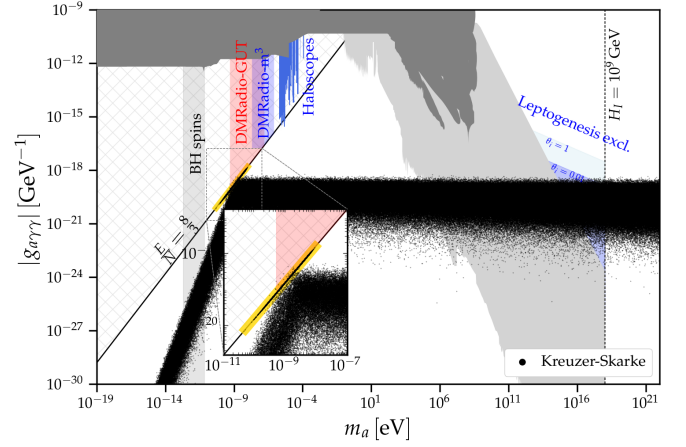}
    \caption{
    As in Fig.~\ref{fig:gayy_heterotic}, but for 79,014 O3/O7
    GUT-compatible orientifold compactifications of Type IIB string theory on CY 3-folds
    from the KS ensemble, fixing the point in moduli space
    to the tip of the SKC (points).
    The range of the QCD axion
    mass distribution, $[3 \times 10^{-11}, 10^{-8}]$~eV, is shaded in
    gold.
    }
    \label{fig:gayy_typeIIB}
\end{figure}

A further virtue of weakly coupled heterotic compactifications is
their cosmological cleanliness in terms of the spectrum of heavy axions (see~\cite{Baryakhtar:2026oun} for recent work in this direction). We find that worldsheet instanton actions are generically of order
unity, making MD axions typically very massive (i.e., above $\sim 10^8$ GeV). These heavy axions
decay before big bang nucleosynthesis (BBN) and cause little to no entropy
dilution. We show that this allows leptogenesis to proceed
unimpeded. In contrast, we show that in generic Type IIB constructions (those based on~\cite{Demirtas:2018akl,Demirtas:2022hqf}), long-lived
heavy axions generically threaten the baryon asymmetry, overproduce the DM abundance, and spoil the precision predictions of BBN. The
qualitative difference in the $(m_a,g_{a\gamma\gamma})$ plane between
heterotic and Type IIB ensembles consistent with grand unification is
shown in Figs.~\ref{fig:gayy_heterotic} and~\ref{fig:gayy_typeIIB}.

\textbf{\textit{Axions in weakly coupled heterotic string theory.}}---We
consider heterotic $E_8 \times E_8$ string theory compactified on a
CY 3-fold $X_6$. 
The MI axion is
obtained by integrating the dual of the NS-NS field over the entire CY, $a = \int_{X_6} B_6$. MD axions arise
from integrating $B_2$ over a basis of effective 2-cycles,
$b_i = \int_{C_i} B_2$. The total number of axions is $1 + h^{1,1}$, with $h^{1,1}$ the Hodge number.

The MI axion couples universally to all unbroken gauge groups in the
4D EFT via the Green-Schwarz counterterm~\cite{Choi:1985bz,Witten:1984Dg},
with decay constant given by~\eqref{eq:MI_fa_main}. 
Independently of $h^{1,1}$, only two linear combinations of MI and MD axions couple
to gauge bosons~\cite{Agrawal:2024ejr,Reig:2025dqb}.
Since the Standard Model gauge group is embedded in the first $E_8$, the QCD axion is the linear combination
\begin{align}
\theta_1 = a + \sum_i n_i\, b_i \,,
\label{eq:QCD_combo}
\end{align}
where $n_i$ are anomaly coefficients fixed by the vector bundle data.
For a standard embedding one has $n_i = \frac{1}{2}\int_{X_6} \beta^{(i)}
\wedge c_2(TX_6)$, with $\beta^{(i)}$ the basis of harmonic $(1,1)$-forms
on $X_6$ dual to $C_i$, $c_2(TX_6)$ the second Chern class of the tangent bundle, and $n_i$ values typically $\mathcal{O}(10\text{--}30)$ in our
ensemble (see Supplementary Material (SM) Fig.~\ref{fig:distribution_anomaly_coeffs}).

Kinetic mixing between MD axions is governed by the K\"ahler metric, 
which is a function of the K\"ahler parameters $t^i$ and the triple intersection numbers $\kappa_{ijk}$ (see SM).
The total CY volume in string units is
related to the unified gauge coupling by
\begin{equation}
\mathcal{V}_6 = \frac{1}{6}\kappa_{ijk}\,t^i t^j t^k
= \frac{g_s^2}{\alpha_\mathrm{GUT}} \approx 25\,,
\label{eq:vol_GUT}
\end{equation}
for $g_s \sim 1$ and $\alpha_\mathrm{GUT}^{-1} \sim 25$.

MD axions acquire masses from worldsheet instantons: Euclidean strings
wrapping holomorphic 2-cycles~\cite{Wen:1985jz}. These generate a potential
\begin{equation}
V = \sum_\alpha \Lambda_\mathrm{UV}^{(\alpha)4}
\,e^{-2\pi\, t_i Q_{i\alpha}} \cos\!\left(Q_{i\alpha} b_i\right)\,,
\label{eq:axion_pot_main}
\end{equation}
where the sum runs over effective curve
classes $C_\alpha$ in the Mori cone of $X_6$, labeled by $\alpha$,
with instanton action $S_\mathrm{ws}^{(\alpha)} = 2\pi\, t_i Q_{i\alpha}$. Here
$Q_{i\alpha} = \int_{C_\alpha} \beta_i$ are the Mori charge matrix
entries. The instanton scales are given by $\Lambda_\mathrm{UV}^{(\alpha)4} \approx A_\alpha\, m_{3/2}
M_s^3$, with $m_{3/2}$ the gravitino mass, $M_s$ the string scale,
and $A_\alpha$ a one-loop prefactor.
For a given compactification, the typical worldsheet instanton
action $S_\mathrm{ws} = \mathcal{O}(1)$ when $\mathcal{V}_6 \sim 25$,
making most MD axions generically heavy; we give explicit computations illustrating  this effect below.

\textbf{\textit{Lower bound on the QCD axion mass.}}---For the QCD
axion mass to deviate from the MI value, a sufficiently light MD axion
must exist, with instanton scale satisfying, for a given $C_\alpha$, 
\begin{equation}
\Lambda_\mathrm{MD}^4 \sim m_{3/2} M_s^3\, e^{-2\pi\, t_i Q_{i\alpha}}
\lesssim \chi_{\mathrm{top}} \,,
\label{eq:LambdaMD}
\end{equation}
with 
$\chi_{\mathrm{top}} \approx (75.4 \, \mathrm{MeV})^4$ the QCD topological susceptibility~\cite{Borsanyi:2016ksw},
so that this MD axion can mix with the MI axion.
For $m_{3/2} \sim 10$~TeV, this requires an effective curve volume
$\mathbf{Q}\cdot\mathbf{t} \gtrsim 20$. The constraint~\eqref{eq:vol_GUT}
fixes $\mathcal{V}_6 \approx 25$, so large individual curve volumes
require special cancellations in the volume form.
In our exhaustive scan of the KS ensemble (see below), we
find no manifolds with $h^{1,1} > 3$ for which~\eqref{eq:LambdaMD}
can be satisfied.
To illustrate this point, let us consider a simple example.  Suppose that $h^{1,1} = 1$, so that there is a single K\"ahler parameter $t$. Then, ${\mathcal V}_6 = {1 \over 6} \kappa_{111} t^3 \geq {1 \over 6} t^3$, since the $\kappa_{ijk}$ are integer intersection numbers and the total volume must be positive. For ${\mathcal V}_6 = 25$, this then bounds $t \lesssim 5.3$, which is not nearly large enough to suppress the MD axion for mixing with the MI QCD axion.  
By contrast, in Type IIB the gauge coupling is set by a local 4-cycle volume while the total volume can be much larger (${\mathcal V}_6 \lesssim 10^3$~\cite{Benabou:2025kgx}), naturally yielding exponentially suppressed instanton potentials and lighter axions~\cite{Demirtas:2018akl,Demirtas:2022hqf,Gendler:2023kjt}.

When mixing does occur, the direction of the mass shift is unambiguous.
The effective QCD axion decay constant satisfies
\begin{equation}
\frac{1}{f_\mathrm{QCD}^2}
= \frac{1}{f_\mathrm{MI}^2} + \sum_{i\,\mathrm{light}} \frac{n_i^2}{f_i^2}
> \frac{1}{f_\mathrm{MI}^2} \,,
\label{eq:fa_bound}
\end{equation}
 giving
$m_\mathrm{QCD} > \Lambda_\mathrm{QCD}^2/f_\mathrm{MI}
= m_\mathrm{MI}$. The lower bound on the QCD axion mass is thus a
direct consequence of the structure of the anomaly coupling: mixing
with MD axions can only \textit{increase} the mass above the MI value.
Note that the linear combination orthogonal to the QCD axion, when
a light MD axion exists, is an axion-like particle, lighter than the QCD axion, with suppressed couplings to gauge bosons~\cite{Agrawal:2022lsp}.

\begin{figure}[!t]
    \centering
\includegraphics[width=1.0\linewidth]{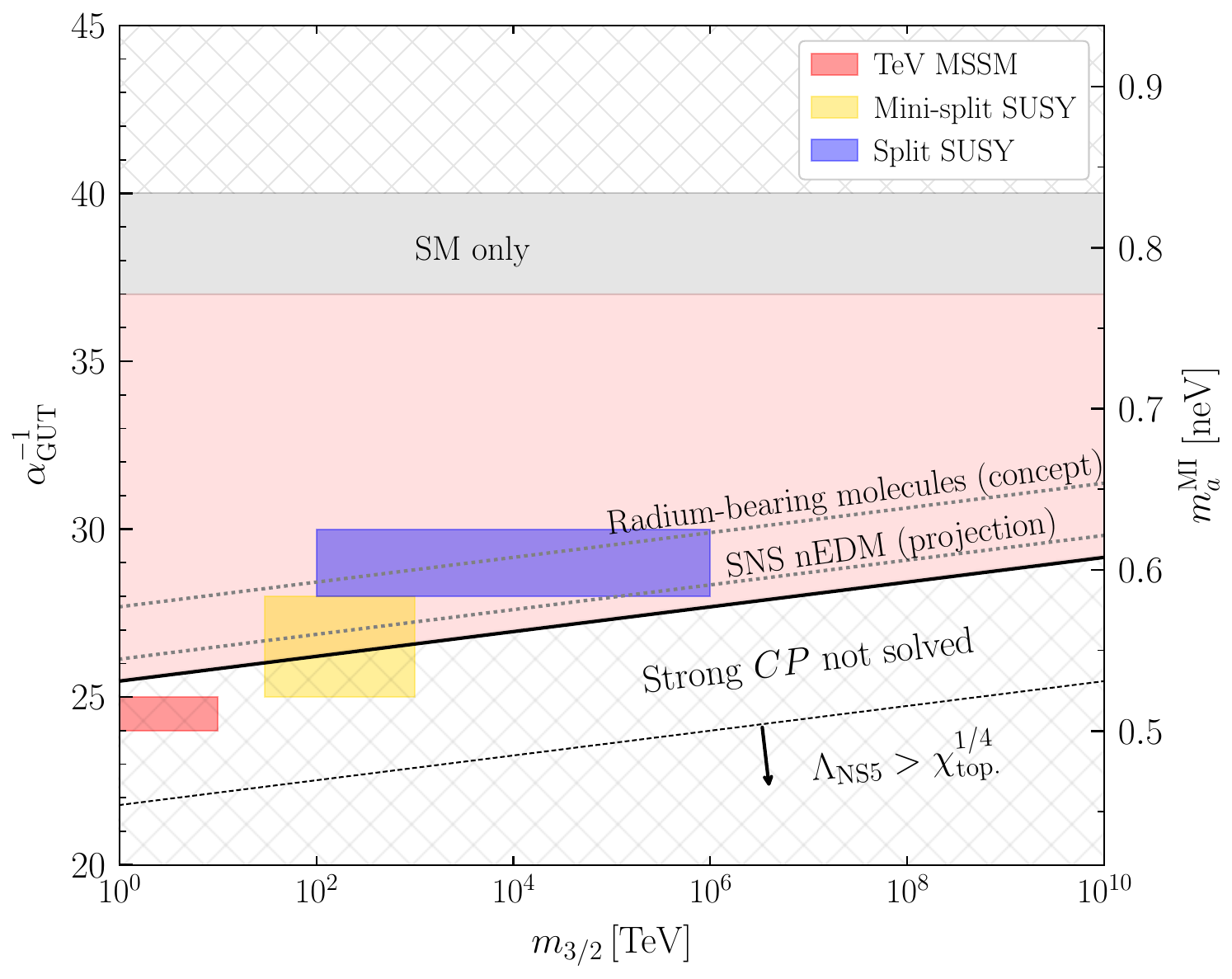}
    \caption{
    Relationship between $m_{3/2}$, $\alpha_\mathrm{GUT}$ and the neutron EDM due to Euclidean NS5-brane contributions to the MI QCD axion potential.  
    We show the mass of the MI axion on the right axis. We indicate approximate parameter space for three
    benchmark scenarios: Split SUSY,
    Mini-Split SUSY, and the TeV
    MSSM. Below the solid line, the QCD
    axion does not solve the Strong $CP$ problem (hatched). 
    We indicate 
    the projected sensitivity of future experiments (see text). Note that the contours apply more broadly than the heterotic MI axion: $S_{\rm NS5} = 2\pi/\alpha_{\rm GUT}$ coincides with the small-QCD-instanton action at $M_{\rm GUT}$, so analogous predictions hold for any QCD axion in a SUSY GUT with generic UV $CP$ violation. Note that the apparent exclusion of the TeV MSSM assumes $\kappa, \delta_{\rm NS5} \sim \mathcal{O}(1)$; smaller values of $\kappa$ would relax this conclusion.
    }
    \label{fig:PQ_quality}
\end{figure}

Without mixing, the QCD axion mass is completely determined by $\alpha_{\rm GUT}$ through~\eqref{eq:MI_fa_main}.  We may bound $\alpha_{\rm GUT}$ to determine the allowable range for the MI QCD axion mass.  Without SUSY, precision unification does not occur, but we estimate $\alpha_{\rm GUT}^{-1} \sim 37 - 40$ in this case (see Fig.~\ref{fig:PQ_quality}) from the value at which  $\alpha_1^{-1}$ and $\alpha_3^{-1}$ meet.
Note that threshold corrections from integrating out heavy, charged scalars and fermions can only {\it decrease} $\alpha_{\rm GUT}^{-1}$, so that~\eqref{eq:MI_fa_main} with $\alpha_{\rm GUT}^{-1} \approx 40$ gives an upper bound on the MI QCD axion mass of $8.3 \times 10^{-10}$ eV.

In SUSY extensions of the Standard Model, $\alpha_{\rm GUT}^{-1}$ is lower than $40$.  For example, in the Minimal Supersymmetric Standard Model (MSSM) with all superpartners at the TeV scale, $\alpha_{\rm GUT}^{-1} \sim 25$, as illustrated in Fig.~\ref{fig:PQ_quality}.  
Split SUSY and related constructs decouple the scalar superpartners while keeping the fermionic superpartners near the TeV scale~\cite{Wells:2003tf,Giudice:2004tc,Arkani-Hamed:2004ymt,Arvanitaki:2012ps,Arkani-Hamed:2012fhg}.
Because split-spectrum SUSY models decouple the scalars to higher mass scales, the value of $\alpha_{\rm GUT}^{-1}$ {\it increases} slightly versus the MSSM and thus lands between the MSSM and Standard Model predictions (see Fig.~\ref{fig:PQ_quality}). 
In these models, the gaugino masses are fixed near the TeV scale and the scalar masses are decoupled with $m_{3/2}$; Mini-Split has a more constrained scale separation than Split SUSY.

The values of $\alpha_{\rm GUT}$ and $m_{3/2}$ also determine the leading non-perturbative contribution to the axion potential from NS5-brane instantons wrapping the entire CY:
\begin{equation}
V(a) \sim -m_{3/2} M_\mathrm{GUT}^3\, e^{-2\pi/\alpha_\mathrm{GUT}}
\cos(a + \delta_\mathrm{NS5}) \,,
\label{eq:NS5_pot}
\end{equation}
with $\delta_\mathrm{NS5}$ an arbitrary phase. (Note that we assume the coefficient in front of~\eqref{eq:NS5_pot} is order unity, though it could be parametrically suppressed~\cite{Csaki:2023ziz}, as would be needed for the MSSM to be consistent with QCD axions.)    The potential should be sufficiently suppressed to not spoil the axion solution to the Strong $CP$ problem. 
Assuming $\delta_{\rm NS5}\sim \mathcal{O}(1)$ and no chiral suppression beyond low-scale SUSY (see SM), for $m_{3/2} > 1 \, {\rm TeV}$, this restricts $\alpha_{\rm GUT}^{-1} \gtrsim 26$. We note that for sufficiently low $\alpha_{\rm GUT}^{-1}$ the NS5-brane contribution to the axion's potential would dominate over that of QCD, as indicated by the dashed line in Fig.~\ref{fig:PQ_quality}.  Interestingly, we find that future measurements of the $\bar \theta$ parameter through the neutron EDM by the SNS nEDM experiment~\cite{nEDM:2019qgk} ($|\bar \theta| \sim 10^{-11}$) and by radium-bearing molecule experiments~\cite{Arrowsmith-Kron:2023hcr,Wilkins:2023hua} ($|\bar{\theta}| \sim 10^{-16}$) may detect a signal for nearly the entire SUSY-motivated parameter space.
We emphasize, however, that the relationship between the SUSY spectrum and $\bar\theta$ shown in Fig.~\ref{fig:PQ_quality} is generic: the NS5-brane action $2\pi/\alpha_{\rm GUT}$ coincides with the action of a small QCD instanton at the GUT scale, and the resulting contribution to $\bar\theta$ is the same as that arising from ultraviolet (UV) QCD instantons in SUSY GUTs with $\mathcal{O}(1)$ $CP$-violating Wilson coefficients in the Standard Model EFT~\cite{Demirtas:2021gsq}.

\textbf{\textit{Heterotic compactification ensembles.}}---We sample $\sim 3\times 10^7$ heterotic compactifications on CY 3-fold
hypersurfaces of toric varieties constructed from the KS
database~\cite{Kreuzer:2000xy}, using the \texttt{CYTools} software
package~\cite{Demirtas:2022hqf}; we find that 2027 are compatible with  weakly coupled heterotic string theory with $\alpha_{\rm GUT}^{-1} = 27$. For each $h^{1,1} \le 16$, we scan
over favorable polytopes and sample fine, regular, star
triangulations (FRSTs) \cite{Demirtas:2022hqf}. Our sample is topologically exhaustive for
$h^{1,1} \le 8$ (covering all 6,084,574 FRSTs). For $9 \le h^{1,1} \le
16$ we sample a subset of FRSTs. 

An FRST generates an acceptable heterotic compactification
if there exists at least one point satisfying \eqref{eq:vol_GUT} within the \textit{stretched K\"ahler cone}
(SKC)~\cite{Demirtas:2018akl}, defined as the set of K\"ahler
parameters for which all effective curve volumes are at least unity in
string units.
(Note that we use $\alpha_{\rm GUT}^{-1} = 27$, $g_s = 1$ as benchmarks but discuss variations in the SM. Decreasing $g_s$ simply decreases ${\mathcal V}_6$, which makes it only more difficult to find manifolds with low $\Lambda_{\rm MD}$ scales.) We find {2027}
such compactifications and identify no acceptable manifolds with
$h^{1,1} > 10$.
The
counts as a function of $h^{1,1}$ are given in SM Table ~\ref{tab:number_polytops_triangulations}.

For each compactification, we perform a numerical scan, using stochastic global optimization, over the SKC
to find the point realizing the largest effective curve volume subject
to $\mathcal{V}_6 = 27$. We find that the
condition~\eqref{eq:LambdaMD} is satisfied (with $m_{3/2} = 10$~TeV)
only for three manifolds with $h^{1,1}=2$ and three with $h^{1,1}=3$. For all remaining
compactifications, 
the worldsheet instanton mass scale is sufficiently large
that all MD axions are heavy and the QCD axion mass equals the MI value
to high accuracy. For the six exceptional manifolds, the QCD axion
mass can be heavier than the MI value in some tuned regions of moduli
space near the boundaries of the SKC.

We also analyze 375 ``favorable'' CICYs~\cite{Bull:2019cij} as an independent ensemble. The largest
effective curve volume attained in this ensemble is $13.33$ 
(for
$\alpha_\mathrm{GUT}^{-1}=27$, $g_s=1$), insufficient to satisfy
\eqref{eq:LambdaMD}. We therefore find no CICY compactification for
which the QCD axion mass deviates from the MI value. (See the SM for further details.)

\textbf{\textit{Heavy axions, cosmology, and leptogenesis.}}---Our heterotic ensembles contain populations
of heavy MD axions, with masses typically well above a TeV and axion-photon couplings distributed about $g_{a\gamma\gamma} \sim 10^{-19}$ GeV$^{-1}$; see Fig.~\ref{fig:gayy_heterotic}.
It is worth contrasting the heterotic axiverse with the Type IIB axiverse, which has been extensively studied previously~\cite{Cicoli:2012sz,Demirtas:2018akl,Demirtas:2021gsq,Gendler:2023kjt,Fallon:2025lvn,Loladze:2025uvf,Petrossian-Byrne:2025mto,Sheridan:2024vtt,Mehta:2021pwf,Halverson:2019cmy,Gendler:2024adn,Broeckel:2021dpz,Jain:2025vfh} and which is illustrated in Fig.~\ref{fig:gayy_typeIIB}.  Note that in that figure we show the distribution of axion masses and photon couplings summed over an ensemble of 79,014 O3/O7 Type IIB orientifold compactifications constructed from the KS ensemble, with the point in moduli space fixed to be at the tip of the SKC, and with the requirement $M_s > M_{\rm GUT}$.  (This is the same ensemble constructed in~\cite{Benabou:2025kgx}, which found that $h^{1,1} \leq 47$ is required to satisfy the GUT requirement on $M_s$.) Most importantly, and in contrast to the heterotic examples, the Type IIB constructions show a roughly log-uniformly distributed range of axion masses, extending well below the TeV scale and indeed below the QCD axion mass scale.

In Figs.~\ref{fig:gayy_heterotic} and~\ref{fig:gayy_typeIIB} we illustrate, in dark (light) gray, constraints on axion-like particles that arise from astrophysical and cosmological probes that do not (do) require cosmological assumptions.  The dark gray constraints at high axion masses arise primarily from the irreducible axion background~\cite{Langhoff:2022bij} and gamma-ray signals from supernovae~\cite{Jaeckel:2017tud,Hoof:2022xbe,Muller:2023vjm,Lella:2024dmx,Manzari:2024jns,Benabou:2024jlj} (see~\cite{Caputo:2024oqc} for a summary of the constraints at lower $m_a$).  The light gray constraints arise from heavy axion decay; recall that the decay rate satisfies $\Gamma \propto m_a^3/f_a^2$.  Heavy axions acquire significant relic abundances through the misalignment mechanism assuming $H_I \gtrsim m_a$, with $H_I$ the Hubble scale during inflation~\cite{Preskill:1982cy,Abbott:1982af,Dine:1982ah,Marsh:2015xka}.  
Under the assumption that, through {\it e.g.} fine tuning, the heavy axion relic abundance is reduced to saturate the observed DM abundance, strong constraints still arise from DM decays, which give observable signatures in the CMB~\cite{Poulin:2016anj,Liu:2023nct}
 and high-energy photon data~\cite{Calore:2022pks,Foster:2021ngm,Foster:2022nva,Cohen:2016uyg,Blanco:2018esa,Roach:2022lgo,Kalashev:2020hqc,Das:2023wtk}.  Even shorter-lived axions may inject energy around the epoch of BBN; the BBN constraint sets the high-$g_{a\gamma\gamma}$ boundary of the gray band in Fig.~\ref{fig:gayy_heterotic} and comes from requiring that the reheat temperature from axion-induced early matter domination be above roughly 5 MeV~\cite{Hasegawa:2019jsa}.
Heavy axions that decay prior to BBN are not directly observable through current probes but can lead to entropy dilution that may disfavor leptogenesis, which is otherwise naturally accommodated in {\it e.g.} heterotic string theory through $SO(10)$ unification.  The leptogenesis constraints are discussed further in the End Matter and in the SM.  For $H_I \sim 10^9$ GeV ($H_I \sim 10^8$ GeV) and assuming $m_{3/2} = 10$~TeV, roughly 9\% (92\%) of the KS compactifications we consider are compatible with leptogenesis for $\mathcal{O}(1)$ initial misalignment angles.

\textbf{\textit{Discussion.}}---The heterotic lower bound on $m_a$ appears to be one instance of a broader pattern.
The corners of the string landscape connected by dualities (see
SM Fig.~\ref{fig:string_dualities}) divide into two classes. In the first ---
heterotic $E_8\times E_8$, heterotic $SO(32)$, Type~I (related to
 heterotic $SO(32)$ by S-duality~\cite{Witten:1995ex}), and strongly
coupled heterotic $E_8 \times E_8$ as described by Ho\v{r}ava-Witten
M-theory~\cite{Horava:1995qa} (where the gauge coupling is set by the
CY volume at the boundary) --- the gauge coupling is set by the total compactification volume rather than by a local cycle.
In
each of these theories, a 6-form field ($B_6$ in heterotic, $C_6$ in
Type~I and M-theory) can be integrated over the full compact space to
produce a MI axion with universal Green-Schwarz
couplings and decay constant as in~\eqref{eq:MI_fa_main}. Together,
the volume constraint and the decay constant sum rule enforce $m_a \gtrsim
m_{\mathrm{MI}} \approx 0.5$~neV (see the SM for the explicit
verification in each dual frame). Moreover, because the total
compactification volume is fixed~\eqref{eq:vol_GUT}, the volume
budget available to individual sub-cycles is tightly constrained. This
suggests that worldsheet (or $D1$-brane) instanton actions are
generically $\mathcal{O}(1)$, making MD axions
generically heavy and short-lived.

In the second class --- Type~IIA/IIB with $D$-branes, F-theory with seven-branes wrapping divisors of a
3-fold base, and M-theory on
$G_2$-holonomy manifolds --- gauge fields are localized on
submanifolds and the gauge coupling is set by a local cycle volume.
No 6-form integrated over the full compact space couples universally
to all gauge groups; axions instead arise from $C_4$ on 4-cycles
(Type~IIB/F-theory) or $C_3$ on 3-cycles ($G_2$).
Conversely, the total
volume can be much larger than $\alpha_{\mathrm{GUT}}^{-1}$, allowing
for exponentially suppressed instanton potentials and a broad distribution of
axion masses that generically includes light and long-lived species.

The heterotic compactifications have the additional advantage that the moduli fields paired with the MD axions are also generically massive, since they receive supersymmetric mass contributions equal to the MD axion masses.  This greatly simplifies the cosmological moduli problem relative to {\it e.g.} the situation in Type IIB constructions. On the other hand, the dilaton, corresponding to the modulus field paired with the MI axion, should receive its dominant mass contributions from SUSY breaking and may parametrically have a mass of order $m_{3/2}$; verifying that this modulus field does not cause cosmological problems deserves further study. 

On the other hand, our analysis applies directly to compactifications on simply-connected CY 3-folds, including non-standard embeddings---line bundle sums~\cite{Anderson:2011ns,Anderson:2012yf}, monad bundles, and spectral cover models---that realize the Standard Model without discrete Wilson lines. In scenarios where Wilson-line breaking on a quotient $X = \tilde X/\Gamma$ is required to obtain three generations, the volume condition applies on $X$, so $\mathcal{V}_{\tilde X} = |\Gamma|\,\alpha_{\rm GUT}^{-1}$ on the cover. The MI axion is unaffected: $f_{\rm MI}$ and the NS5-brane action $2\pi/\alpha_{\rm GUT}$ are intrinsic to $\mathcal{V}_X$. MD axions may in principle become lighter on $X$ if $\Gamma$-invariant curves attain sufficiently large $X$-volumes to trigger mixing; an explicit determination requires equivariant cohomology data which we leave for future work. 
Independently, mixing scenarios introduce lighter MD axions, whose abundances we show are constrained by a variety of cosmological constraints (BBN, CMB, decaying DM, leptogenesis, etc.); restricting to scenarios without mixing reduces cosmological tension and sharpens the MI axion prediction.

It may be challenging to achieve $H_I \lesssim 10^9$ GeV, as required by isocurvature constraints when the MI axion saturates the DM abundance. The interplay of axions with inflation in the context of heterotic compactifications is an interesting direction for future work, including {\it e.g.} the possibility that the compactification volume is smaller during inflation, which would enhance the NS5-brane instanton contribution and make the MI axion heavy enough to suppress isocurvature perturbations.

\begin{acknowledgments}
\textit{Acknowledgments.}---We thank Sebastian Vander Ploeg Fallon, Naomi Gendler, Thomas Harvey, Nick Hutzler, Andrew Jayich, Soubhik Kumar, Liam McAllister, Jakob Moritz, Fernando Quevedo, Matthew Reece, Elijah Sheridan, and Timo
Weigand for useful discussions. 
J.B.\ and B.R.S.\ are supported in
part by the DOE award
DESC0025293. This research used
resources of NERSC, a U.S.\ DOE Office of Science User Facility at
LBNL, under Contract No.\ DE-AC02-05CH11231 using NERSC award
HEP-ERCAP0023978. Additional computations used the Lawrencium cluster
at LBNL (supported by the Director, Office of Science, Office of Basic
Energy Sciences, of the U.S.\ DOE under Contract No.\ DE-AC02-05CH11231).
\end{acknowledgments}

\bibliography{Bibliography}

@article{Svrcek:2006yi,
    author = "Svrcek, Peter and Witten, Edward",
    title = "{Axions In String Theory}",
    eprint = "hep-th/0605206",
    archivePrefix = "arXiv",
    reportNumber = "SLAC-PUB-11894",
    doi = "10.1088/1126-6708/2006/06/051",
    journal = "JHEP",
    volume = "06",
    pages = "051",
    year = "2006"
}

@article{Nee:2026,
    author ="Nee, Michael and Reig, Mario and Weigand, Timo" ,
    title = "{In preparation}"
}

@article{Lee:2026umy,
    author = "Lee, Sung Mook and Ramos, Maria and Vilches, Fuensanta",
    title = "{How well can the QCD axion hide?}",
    eprint = "2604.08657",
    archivePrefix = "arXiv",
    primaryClass = "hep-ph",
    reportNumber = "CERN-TH-2026-087",
    month = "4",
    year = "2026"
}

@article{Carralot:2026kps,
    author = "Carralot, Florie and Diego-Palazuelos, Patricia and Duivenvoorden, Adriaan J. and Komatsu, Eiichiro and Krachmalnicoff, Nicoletta and Baccigalupi, Carlo",
    title = "{Is cosmic birefringence due to dark energy or dark matter? Simulation-based inference}",
    eprint = "2602.12019",
    archivePrefix = "arXiv",
    primaryClass = "astro-ph.CO",
    month = "2",
    year = "2026"
}

@article{Gavela:2023tzu,
    author = "Gavela, Bel{\'e}n and Qu{\'\i}lez, Pablo and Ramos, Maria",
    title = "{The QCD axion sum rule}",
    eprint = "2305.15465",
    archivePrefix = "arXiv",
    primaryClass = "hep-ph",
    reportNumber = "IFT-UAM/CSIC-23-58",
    doi = "10.1007/JHEP04(2024)056",
    journal = "JHEP",
    volume = "04",
    pages = "056",
    year = "2024"
}

@article{Horava:1996ma,
    author = "Horava, Petr and Witten, Edward",
    title = "{Eleven-dimensional supergravity on a manifold with boundary}",
    eprint = "hep-th/9603142",
    archivePrefix = "arXiv",
    reportNumber = "IASSNS-HEP-96-17, PUPT-1597",
    doi = "10.1016/0550-3213(96)00308-2",
    journal = "Nucl. Phys. B",
    volume = "475",
    pages = "94--114",
    year = "1996"
}

@book{Green:1987mn,
    author = "Green, Michael B. and Schwarz, J. H. and Witten, Edward",
    title = "{SUPERSTRING THEORY. VOL. 2: LOOP AMPLITUDES, ANOMALIES AND PHENOMENOLOGY}",
    isbn = "978-0-521-35753-1",
    month = "7",
    year = "1988"
}

@article{Blumenhagen:2006ux,
    author = "Blumenhagen, Ralph and Moster, Sebastian and Weigand, Timo",
    title = "{Heterotic GUT and standard model vacua from simply connected Calabi-Yau manifolds}",
    eprint = "hep-th/0603015",
    archivePrefix = "arXiv",
    reportNumber = "MPP-2006-16, LMU-ASC-11-06",
    doi = "10.1016/j.nuclphysb.2006.06.005",
    journal = "Nucl. Phys. B",
    volume = "751",
    pages = "186--221",
    year = "2006"
}

@article{Berlin:2024pzi,
    author = "Berlin, Asher and Kahn, Yonatan",
    title = "{New Technologies for Axion and Dark Photon Searches}",
    eprint = "2412.08704",
    archivePrefix = "arXiv",
    primaryClass = "hep-ph",
    reportNumber = "FERMILAB-PUB-24-0933-T",
    doi = "10.1146/annurev-nucl-121423-101015",
    journal = "Ann. Rev. Nucl. Part. Sci.",
    volume = "75",
    number = "1",
    pages = "83--108",
    year = "2025"
}

@article{Irastorza:2018dyq,
    author = "Irastorza, Igor G. and Redondo, Javier",
    title = "{New experimental approaches in the search for axion-like particles}",
    eprint = "1801.08127",
    archivePrefix = "arXiv",
    primaryClass = "hep-ph",
    doi = "10.1016/j.ppnp.2018.05.003",
    journal = "Prog. Part. Nucl. Phys.",
    volume = "102",
    pages = "89--159",
    year = "2018"
}

@article{Reig:2025dpz,
    author = "Reig, Mario and Ruiz, Ignacio",
    title = "{The dark dimension, proton decay, and the length of the M-theory interval}",
    eprint = "2510.25832",
    archivePrefix = "arXiv",
    primaryClass = "hep-th",
    reportNumber = "CERN-TH-2025-207",
    month = "10",
    year = "2025"
}

@article{Sheridan:2024vtt,
    author = "Sheridan, Elijah and Carta, Federico and Gendler, Naomi and Jain, Mudit and Marsh, David J. E. and McAllister, Liam and Righi, Nicole and Rogers, Keir K. and Schachner, Andreas",
    title = "{Fuzzy axions and associated relics}",
    eprint = "2412.12012",
    archivePrefix = "arXiv",
    primaryClass = "hep-th",
    reportNumber = "KCL-PH-TH/2024-75, KCL-PH-TH/2024-75",
    doi = "10.1007/JHEP09(2025)016",
    journal = "JHEP",
    volume = "09",
    pages = "016",
    year = "2025"
}

@article{Hui:2016ltb,
    author = "Hui, Lam and Ostriker, Jeremiah P. and Tremaine, Scott and Witten, Edward",
    title = "{Ultralight scalars as cosmological dark matter}",
    eprint = "1610.08297",
    archivePrefix = "arXiv",
    primaryClass = "astro-ph.CO",
    doi = "10.1103/PhysRevD.95.043541",
    journal = "Phys. Rev. D",
    volume = "95",
    number = "4",
    pages = "043541",
    year = "2017"
}

@article{Leedom:2025mlr,
    author = "Leedom, Jacob M. and Putti, Margherita and Westphal, Alexander",
    title = "{Towards a Heterotic Axiverse}",
    eprint = "2509.03578",
    archivePrefix = "arXiv",
    primaryClass = "hep-th",
    reportNumber = "DESY 25-113",
    month = "9",
    year = "2025"
}

@article{Kamionkowski:1992mf,
    author = "Kamionkowski, Marc and March-Russell, John",
    title = "{Planck scale physics and the Peccei-Quinn mechanism}",
    eprint = "hep-th/9202003",
    archivePrefix = "arXiv",
    reportNumber = "IASSNS-HEP-92-9, PUPT-92-1309",
    doi = "10.1016/0370-2693(92)90492-M",
    journal = "Phys. Lett. B",
    volume = "282",
    pages = "137--141",
    year = "1992"
}

@article{Baryakhtar:2026oun,
    author = "Baryakhtar, Masha and Cyncynates, David and Henry, Ella",
    title = "{Axiverse Lampposts}",
    eprint = "2602.23424",
    archivePrefix = "arXiv",
    primaryClass = "hep-ph",
    month = "2",
    year = "2026"
}

@article{Choi:1985bz,
    author = "Choi, Kiwoon and Kim, Jihn E.",
    title = "{Compactification and Axions in E(8) x E(8)-prime Superstring Models}",
    reportNumber = "SNUHE 85/10",
    doi = "10.1016/0370-2693(85)90693-8",
    journal = "Phys. Lett. B",
    volume = "165",
    pages = "71--75",
    year = "1985"
}

@article{Conlon:2006tq,
    author = "Conlon, Joseph P.",
    title = "{The QCD axion and moduli stabilisation}",
    eprint = "hep-th/0602233",
    archivePrefix = "arXiv",
    reportNumber = "DAMTP-2006-17",
    doi = "10.1088/1126-6708/2006/05/078",
    journal = "JHEP",
    volume = "05",
    pages = "078",
    year = "2006"
}

@article{Dunsky:2025sgz,
    author = "Dunsky, David I. and Manzari, Claudio Andrea and Qu{\'\i}lez, Pablo and Ramos, Maria and S{\o}rensen, Philip",
    title = "{Resonant Landau-Zener conversion in multi-axion systems}",
    eprint = "2507.06287",
    archivePrefix = "arXiv",
    primaryClass = "hep-ph",
    reportNumber = "CERN-TH-2025-131",
    doi = "10.1007/JHEP01(2026)077",
    journal = "JHEP",
    volume = "01",
    pages = "077",
    year = "2026"
}

@article{Cyncynates:2021xzw,
    author = "Cyncynates, David and Giurgica-Tiron, Tudor and Simon, Olivier and Thompson, Jedidiah O.",
    title = "{Resonant nonlinear pairs in the axiverse and their late-time direct and astrophysical signatures}",
    eprint = "2109.09755",
    archivePrefix = "arXiv",
    primaryClass = "hep-ph",
    doi = "10.1103/PhysRevD.105.055005",
    journal = "Phys. Rev. D",
    volume = "105",
    number = "5",
    pages = "055005",
    year = "2022"
}

@article{Anderson:2017aux,
    author = "Anderson, Lara B. and Gao, Xin and Gray, James and Lee, Seung-Joo",
    title = "{Fibrations in CICY Threefolds}",
    eprint = "1708.07907",
    archivePrefix = "arXiv",
    primaryClass = "hep-th",
    doi = "10.1007/JHEP10(2017)077",
    journal = "JHEP",
    volume = "10",
    pages = "077",
    year = "2017"
}

@article{Benabou:2024jlj,
    author = "Benabou, Joshua N. and Manzari, Claudio Andrea and Park, Yujin and Prabhakar, Garima and Safdi, Benjamin R. and Savoray, Inbar",
    title = "{Time-delayed gamma-ray signatures of heavy axions from core-collapse supernovae}",
    eprint = "2412.13247",
    archivePrefix = "arXiv",
    primaryClass = "hep-ph",
    month = "12",
    year = "2024"
}

@article{Bull:2019cij,
    author = "Bull, Kieran and He, Yang-Hui and Jejjala, Vishnu and Mishra, Challenger",
    title = "{Getting CICY High}",
    eprint = "1903.03113",
    archivePrefix = "arXiv",
    primaryClass = "hep-th",
    doi = "10.1016/j.physletb.2019.06.067",
    journal = "Phys. Lett. B",
    volume = "795",
    pages = "700--706",
    year = "2019"
}

@article{Planck2018Parameters,
  author = {Aghanim, N. and others},
  collaboration = {Planck},
  title = {Planck 2018 results. VI. Cosmological parameters},
  journal = {Astronomy and Astrophysics},
  volume = {641},
  pages = {A6},
  year = {2020},
  doi = {10.1051/0004-6361/201833910},
  eprint = {1807.06209},
  archivePrefix = {arXiv},
  primaryClass = {astro-ph.CO}
}

@article{Robles-Llana:2007bbv,
    author = "Robles-Llana, Daniel and Saueressig, Frank and Theis, Ulrich and Vandoren, Stefan",
    title = "{Membrane instantons from mirror symmetry}",
    eprint = "0707.0838",
    archivePrefix = "arXiv",
    primaryClass = "hep-th",
    reportNumber = "ITP-UU-07-32, SPIN-07-22",
    doi = "10.4310/CNTP.2007.v1.n4.a3",
    journal = "Commun. Num. Theor. Phys.",
    volume = "1",
    pages = "681--711",
    year = "2007"
}

@article{Hashimoto:2015zqm,
    author = "Hashimoto, Kenji and Kanazawa, Atsushi",
    title = "{Calabi-Yau threefolds of type K (II): mirror symmetry}",
    eprint = "1511.08778",
    archivePrefix = "arXiv",
    primaryClass = "math.AG",
    doi = "10.4310/CNTP.2016.v10.n2.a1",
    journal = "Commun. Num. Theor. Phys.",
    volume = "10",
    pages = "157--192",
    year = "2016"
}

@article{Fischer:2012qj,
    author = "Fischer, Maximilian and Ratz, Michael and Torrado, Jesus and Vaudrevange, Patrick K. S.",
    title = "{Classification of symmetric toroidal orbifolds}",
    eprint = "1209.3906",
    archivePrefix = "arXiv",
    primaryClass = "hep-th",
    reportNumber = "DESY-12-147, TUM-HEP-855-12, FLAVOUR(267104)-ERC-29, CETUP*-12-012",
    doi = "10.1007/JHEP01(2013)084",
    journal = "JHEP",
    volume = "01",
    pages = "084",
    year = "2013"
}

@article{Gendler:2026uux,
    author = "Gendler, Naomi and Sheridan, Elijah and Stillman, Michael and Wu, David H.",
    title = "{Holes in Calabi-Yau Effective Cones}",
    eprint = "2603.11173",
    archivePrefix = "arXiv",
    primaryClass = "hep-th",
    month = "3",
    year = "2026"
}

@article{nEDM:2019qgk,
    author = "Ahmed, M. W. and others",
    collaboration = "nEDM",
    title = "{A New Cryogenic Apparatus to Search for the Neutron Electric Dipole Moment}",
    eprint = "1908.09937",
    archivePrefix = "arXiv",
    primaryClass = "physics.ins-det",
    doi = "10.1088/1748-0221/14/11/P11017",
    journal = "JINST",
    volume = "14",
    number = "11",
    pages = "P11017",
    year = "2019"
}

@article{Arvanitaki:2012ps,
    author = "Arvanitaki, Asimina and Craig, Nathaniel and Dimopoulos, Savas and Villadoro, Giovanni",
    title = "{Mini-Split}",
    eprint = "1210.0555",
    archivePrefix = "arXiv",
    primaryClass = "hep-ph",
    doi = "10.1007/JHEP02(2013)126",
    journal = "JHEP",
    volume = "02",
    pages = "126",
    year = "2013"
}

@article{Arkani-Hamed:2004ymt,
    author = "Arkani-Hamed, Nima and Dimopoulos, Savas",
    title = "{Supersymmetric unification without low energy supersymmetry and signatures for fine-tuning at the LHC}",
    eprint = "hep-th/0405159",
    archivePrefix = "arXiv",
    doi = "10.1088/1126-6708/2005/06/073",
    journal = "JHEP",
    volume = "06",
    pages = "073",
    year = "2005"
}

@article{Giudice:2004tc,
    author = "Giudice, G. F. and Romanino, A.",
    title = "{Split supersymmetry}",
    eprint = "hep-ph/0406088",
    archivePrefix = "arXiv",
    reportNumber = "CERN-PH-TH-2004-100",
    doi = "10.1016/j.nuclphysb.2004.08.001",
    journal = "Nucl. Phys. B",
    volume = "699",
    pages = "65--89",
    year = "2004",
    note = "[Erratum: Nucl.Phys.B 706, 487--487 (2005)]"
}

@article{Blinov:2019rhb,
    author = "Blinov, Nikita and Dolan, Matthew J and Draper, Patrick and Kozaczuk, Jonathan",
    title = "{Dark matter targets for axionlike particle searches}",
    eprint = "1905.06952",
    archivePrefix = "arXiv",
    primaryClass = "hep-ph",
    reportNumber = "FERMILAB-PUB-19-197-A-T",
    doi = "10.1103/PhysRevD.100.015049",
    journal = "Phys. Rev. D",
    volume = "100",
    number = "1",
    pages = "015049",
    year = "2019"
}

@article{Graf:2010tv,
    author = "Graf, Peter and Steffen, Frank Daniel",
    title = "{Thermal axion production in the primordial quark-gluon plasma}",
    eprint = "1008.4528",
    archivePrefix = "arXiv",
    primaryClass = "hep-ph",
    reportNumber = "MPP-2010-20",
    doi = "10.1103/PhysRevD.83.075011",
    journal = "Phys. Rev. D",
    volume = "83",
    pages = "075011",
    year = "2011"
}

@article{DINE198555,
title = {Gluino condensation in superstring models},
journal = {Physics Letters B},
volume = {156},
number = {1},
pages = {55-60},
year = {1985},
issn = {0370-2693},
doi = {https://doi.org/10.1016/0370-2693(85)91354-1},
url = {https://www.sciencedirect.com/science/article/pii/0370269385913541},
author = {M. Dine and R. Rohm and N. Seiberg and E. Witten}
}

@article{Planck:2018jri,
    author = "Akrami, Y. and others",
    collaboration = "Planck",
    title = "{Planck 2018 results. X. Constraints on inflation}",
    eprint = "1807.06211",
    archivePrefix = "arXiv",
    primaryClass = "astro-ph.CO",
    doi = "10.1051/0004-6361/201833887",
    journal = "Astron. Astrophys.",
    volume = "641",
    pages = "A10",
    year = "2020"
}

@article{Hasegawa:2019jsa,
    author = "Hasegawa, Takuya and Hiroshima, Nagisa and Kohri, Kazunori and Hansen, Rasmus S. L. and Tram, Thomas and Hannestad, Steen",
    title = "{MeV-scale reheating temperature and thermalization of oscillating neutrinos by radiative and hadronic decays of massive particles}",
    eprint = "1908.10189",
    archivePrefix = "arXiv",
    primaryClass = "hep-ph",
    reportNumber = "KEK-TH-2149, KEK-Cosmo-242, RIKEN-iTHEMS-Report-19, IPMU19-0120",
    doi = "10.1088/1475-7516/2019/12/012",
    journal = "JCAP",
    volume = "12",
    pages = "012",
    year = "2019"
}

@article{Kim:2006aq,
    author = "Kim, Ian-Woo and Kim, Jihn E.",
    title = "{Modification of decay constants of superstring axions: Effects of flux compactification and axion mixing}",
    eprint = "hep-th/0605256",
    archivePrefix = "arXiv",
    doi = "10.1016/j.physletb.2006.06.033",
    journal = "Phys. Lett. B",
    volume = "639",
    pages = "342--347",
    year = "2006"
}

@article{Davidson:2002qv,
    author = "Davidson, Sacha and Ibarra, Alejandro",
    title = "{A Lower bound on the right-handed neutrino mass from leptogenesis}",
    eprint = "hep-ph/0202239",
    archivePrefix = "arXiv",
    reportNumber = "OUTP-02-10P, IPPP-02-16, DCPT-02-32",
    doi = "10.1016/S0370-2693(02)01735-5",
    journal = "Phys. Lett. B",
    volume = "535",
    pages = "25--32",
    year = "2002"
}

@article{Csaki:2023ziz,
    author = "Cs{\'a}ki, Csaba and D'Agnolo, Raffaele Tito and Kuflik, Eric and Ruhdorfer, Maximilian",
    title = "{Instanton NDA and applications to axion models}",
    eprint = "2311.09285",
    archivePrefix = "arXiv",
    primaryClass = "hep-ph",
    doi = "10.1007/JHEP04(2024)074",
    journal = "JHEP",
    volume = "04",
    pages = "074",
    year = "2024"
}

@article{Anderson:2011ns,
    author = "Anderson, Lara B. and Gray, James and Lukas, Andre and Palti, Eran",
    title = "{Two Hundred Heterotic Standard Models on Smooth Calabi-Yau Threefolds}",
    eprint = "1106.4804",
    archivePrefix = "arXiv",
    primaryClass = "hep-th",
    doi = "10.1103/PhysRevD.84.106005",
    journal = "Phys. Rev. D",
    volume = "84",
    pages = "106005",
    year = "2011"
}

@article{Anderson:2012yf,
    author = "Anderson, Lara B. and Gray, James and Lukas, Andre and Palti, Eran",
    title = "{Heterotic Line Bundle Standard Models}",
    eprint = "1202.1757",
    archivePrefix = "arXiv",
    primaryClass = "hep-th",
    doi = "10.1007/JHEP06(2012)113",
    journal = "JHEP",
    volume = "06",
    pages = "113",
    year = "2012"
}

@article{He:2010uj,
    author = "He, Yang-Hui.",
    title = "{An Algorithmic Approach to Heterotic String Phenomenology}",
    eprint = "1001.2419",
    archivePrefix = "arXiv",
    primaryClass = "hep-th",
    doi = "10.1142/S0217732310032731",
    journal = "Mod. Phys. Lett. A",
    volume = "25",
    pages = "79--90",
    year = "2010"
}

@article{Benabou:2025kgx,
    author = "Benabou, Joshua N. and Fraser, Katherine and Reig, Mario and Safdi, Benjamin R.",
    title = "{String Theory and Grand Unification Suggest a Sub-Microelectronvolt QCD Axion}",
    eprint = "2505.15884",
    archivePrefix = "arXiv",
    primaryClass = "hep-ph",
    month = "5",
    year = "2025"
}

@article{Benabou:2025viy,
    author = "Benabou, Joshua N. and Hook, Anson and Manzari, Claudio Andrea and Murayama, Hitoshi and Safdi, Benjamin R.",
    title = "{Clearing up the Strong $CP$ problem}",
    eprint = "2510.18951",
    archivePrefix = "arXiv",
    primaryClass = "hep-ph",
    month = "10",
    year = "2025"
}

@article{ReeceRudeliusTudball:toappear,
  author = "Reece, Matt and Rudelius, Tom and Tudball, Christopher",
  title  = "{To appear}",
  note   = "to appear"
}

@article{Reece:2025zva,
    author = "Reece, Matthew and Rudelius, Tom and Tudball, Christopher",
    title = "{Co-scaling and alignment of electric and magnetic towers}",
    eprint = "2505.22713",
    archivePrefix = "arXiv",
    primaryClass = "hep-th",
    doi = "10.1007/JHEP09(2025)146",
    journal = "JHEP",
    volume = "09",
    pages = "146",
    year = "2025"
}

@article{Demirtas:2023als,
    author = "Demirtas, Mehmet and Kim, Manki and McAllister, Liam and Moritz, Jakob and Rios-Tascon, Andres",
    title = "{Computational Mirror Symmetry}",
    eprint = "2303.00757",
    archivePrefix = "arXiv",
    primaryClass = "hep-th",
    reportNumber = "MIT-CTP/5528",
    doi = "10.1007/JHEP01(2024)184",
    journal = "JHEP",
    volume = "01",
    pages = "184",
    year = "2024"
}

@article{Beasley:2003fx,
    author = "Beasley, Chris and Witten, Edward",
    title = "{Residues and world sheet instantons}",
    eprint = "hep-th/0304115",
    archivePrefix = "arXiv",
    reportNumber = "PUPT-2081",
    doi = "10.1088/1126-6708/2003/10/065",
    journal = "JHEP",
    volume = "10",
    pages = "065",
    year = "2003"
}

@article{Gopakumar:1998ii,
    author = "Gopakumar, Rajesh and Vafa, Cumrun",
    title = "{M theory and topological strings. 1.}",
    eprint = "hep-th/9809187",
    archivePrefix = "arXiv",
    reportNumber = "HUTP-98-A069",
    month = "9",
    year = "1998"
}

@article{Gopakumar:1998jq,
    author = "Gopakumar, Rajesh and Vafa, Cumrun",
    title = "{M theory and topological strings. 2.}",
    eprint = "hep-th/9812127",
    archivePrefix = "arXiv",
    reportNumber = "HUTP-98-A070",
    month = "12",
    year = "1998"
}

@article{Storn:1997uea,
    author = "Storn, Rainer and Price, Kenneth",
    title = "{Differential Evolution {\textendash} A Simple and Efficient Heuristic for global Optimization over Continuous Spaces}",
    doi = "10.1023/A:1008202821328",
    journal = "J. Global Optim.",
    volume = "11",
    number = "4",
    pages = "341--359",
    year = "1997"
}

@article{Alonso:2017avz,
    author = "Alonso, Rodrigo and Urbano, Alfredo",
    title = "{Wormholes and masses for Goldstone bosons}",
    eprint = "1706.07415",
    archivePrefix = "arXiv",
    primaryClass = "hep-ph",
    reportNumber = "CERN-TH-2017-135",
    doi = "10.1007/JHEP02(2019)136",
    journal = "JHEP",
    volume = "02",
    pages = "136",
    year = "2019"
}

@article{Alim:2021vhs,
    author = "Alim, Murad and Heidenreich, Ben and Rudelius, Tom",
    title = "{The Weak Gravity Conjecture and BPS Particles}",
    eprint = "2108.08309",
    archivePrefix = "arXiv",
    primaryClass = "hep-th",
    reportNumber = "ACFI-T21-09",
    doi = "10.1002/prop.202100125",
    journal = "Fortsch. Phys.",
    volume = "69",
    number = "11-12",
    pages = "2100125",
    year = "2021"
}

@article{Benabou:2023npn,
    author = "Benabou, Joshua N. and Bonnefoy, Quentin and Buschmann, Malte and Kumar, Soubhik and Safdi, Benjamin R.",
    title = "{Cosmological dynamics of string theory axion strings}",
    eprint = "2312.08425",
    archivePrefix = "arXiv",
    primaryClass = "hep-ph",
    doi = "10.1103/PhysRevD.110.035021",
    journal = "Phys. Rev. D",
    volume = "110",
    number = "3",
    pages = "035021",
    year = "2024"
}

@article{Benabou:2022qpv,
    author = "Benabou, Joshua N. and Foster, Joshua W. and Kahn, Yonatan and Safdi, Benjamin R. and Salemi, Chiara P.",
    title = "{Lumped-element axion dark matter detection beyond the magnetoquasistatic limit}",
    eprint = "2211.00008",
    archivePrefix = "arXiv",
    primaryClass = "hep-ph",
    reportNumber = "MIT-CTP/5490",
    doi = "10.1103/PhysRevD.108.035009",
    journal = "Phys. Rev. D",
    volume = "108",
    number = "3",
    pages = "035009",
    year = "2023"
}

@article{Fukugita:1986hr,
      author       = "Fukugita, Masataka and Yanagida, Tsutomu",
      title        = "{Baryogenesis Without Grand Unification}",
      journal      = "Phys. Lett. B",
      volume       = "174",
      pages        = "45--47",
      year         = "1986",
      doi          = "10.1016/0370-2693(86)91126-3",
}

@article{Davidson:2008bu,
      author       = "Davidson, Sacha and Nardi, Enrico and Nir, Yosef",
      title        = "{Leptogenesis}",
      journal      = "Phys. Rept.",
      volume       = "466",
      pages        = "105--177",
      year         = "2008",
      doi          = "10.1016/j.physrep.2008.06.002",
}

@article{Buchmuller:2004nz,
      author       = "Buchmuller, Wilfried and Di Bari, Pasquale and Pluemacher, Michael",
      title        = "{Leptogenesis for Pedestrians}",
      journal      = "Ann. Phys.",
      volume       = "315",
      pages        = "305--351",
      year         = "2005",
      doi          = "10.1016/j.aop.2004.02.003",
      eprint       = "hep-ph/0401240",
}

@article{Komatsu:2022nvu,
    author = "Komatsu, Eiichiro",
    title = "{New physics from the polarized light of the cosmic microwave background}",
    eprint = "2202.13919",
    archivePrefix = "arXiv",
    primaryClass = "astro-ph.CO",
    doi = "10.1038/s42254-022-00452-4",
    journal = "Nature Rev. Phys.",
    volume = "4",
    number = "7",
    pages = "452--469",
    year = "2022"
}

@article{Benabou:2025jcv,
    author = "Benabou, Joshua N. and Dessert, Christopher and Patra, Kishore C. and Brink, Thomas G. and Zheng, WeiKang and Filippenko, Alexei V. and Safdi, Benjamin R.",
    title = "{Search for Axions in Magnetic White Dwarf Polarization at Lick and Keck Observatories}",
    eprint = "2504.12377",
    archivePrefix = "arXiv",
    primaryClass = "hep-ph",
    month = "4",
    year = "2025"
}

@article{Hisano:2022qll,
    author = "Hisano, Junji",
    title = "{Proton decay in SUSY GUTs}",
    eprint = "2202.01404",
    archivePrefix = "arXiv",
    primaryClass = "hep-ph",
    doi = "10.1093/ptep/ptac017",
    journal = "PTEP",
    volume = "2022",
    number = "12",
    pages = "12B104",
    year = "2022"
}

@article{Georgi:1974sy,
    author = "Georgi, H. and Glashow, S. L.",
    title = "{Unity of All Elementary Particle Forces}",
    doi = "10.1103/PhysRevLett.32.438",
    journal = "Phys. Rev. Lett.",
    volume = "32",
    pages = "438--441",
    year = "1974"
}

@article{Fritzsch:1974nn,
    author = "Fritzsch, Harald and Minkowski, Peter",
    title = "{Unified Interactions of Leptons and Hadrons}",
    reportNumber = "CALT-68-467",
    doi = "10.1016/0003-4916(75)90211-0",
    journal = "Annals Phys.",
    volume = "93",
    pages = "193--266",
    year = "1975"
}

@article{Super-Kamiokande:2020wjk,
    author = "Takenaka, A. and others",
    collaboration = "Super-Kamiokande",
    title = "{Search for proton decay via $p\to e^+\pi^0$ and $p\to \mu^+\pi^0$ with an enlarged fiducial volume in Super-Kamiokande I-IV}",
    eprint = "2010.16098",
    archivePrefix = "arXiv",
    primaryClass = "hep-ex",
    doi = "10.1103/PhysRevD.102.112011",
    journal = "Phys. Rev. D",
    volume = "102",
    number = "11",
    pages = "112011",
    year = "2020"
}

@article{Loladze:2025uvf,
    author = "Loladze, Vazha and Platschorre, Arthur and Reig, Mario",
    title = "{Higher Axion Strings}",
    eprint = "2503.18707",
    archivePrefix = "arXiv",
    primaryClass = "hep-ph",
    month = "3",
    year = "2025"
}

@article{Petrossian-Byrne:2025mto,
    author = "Petrossian-Byrne, Rudin and Villadoro, Giovanni",
    title = "{Open String Axiverse}",
    eprint = "2503.16387",
    archivePrefix = "arXiv",
    primaryClass = "hep-ph",
    month = "3",
    year = "2025"
}

@article{Choi:2011xt,
    author = "Choi, Kiwoon and Jeong, Kwang Sik and Okumura, Ken-Ichi and Yamaguchi, Masahiro",
    title = "{Mixed Mediation of Supersymmetry Breaking with Anomalous U(1) Gauge Symmetry}",
    eprint = "1104.3274",
    archivePrefix = "arXiv",
    primaryClass = "hep-ph",
    reportNumber = "KYUSHU-HET-125, TU-882",
    doi = "10.1007/JHEP06(2011)049",
    journal = "JHEP",
    volume = "06",
    pages = "049",
    year = "2011"
}

@inproceedings{Adams:2022pbo,
    author = "Adams, C. B. and others",
    title = "{Axion Dark Matter}",
    booktitle = "{Snowmass 2021}",
    eprint = "2203.14923",
    archivePrefix = "arXiv",
    primaryClass = "hep-ex",
    reportNumber = "FERMILAB-CONF-22-996-PPD-T",
    month = "3",
    year = "2022"
}

@article{Budker:2013hfa,
    author = "Budker, Dmitry and Graham, Peter W. and Ledbetter, Micah and Rajendran, Surjeet and Sushkov, Alex",
    title = "{Proposal for a Cosmic Axion Spin Precession Experiment (CASPEr)}",
    eprint = "1306.6089",
    archivePrefix = "arXiv",
    primaryClass = "hep-ph",
    doi = "10.1103/PhysRevX.4.021030",
    journal = "Phys. Rev. X",
    volume = "4",
    number = "2",
    pages = "021030",
    year = "2014"
}

@article{Graham:2013gfa,
    author = "Graham, Peter W. and Rajendran, Surjeet",
    title = "{New Observables for Direct Detection of Axion Dark Matter}",
    eprint = "1306.6088",
    archivePrefix = "arXiv",
    primaryClass = "hep-ph",
    doi = "10.1103/PhysRevD.88.035023",
    journal = "Phys. Rev. D",
    volume = "88",
    pages = "035023",
    year = "2013"
}

@article{DMRadio:2022pkf,
    author = "Brouwer, L. and others",
    collaboration = "DMRadio",
    title = "{Projected sensitivity of DMRadio-m3: A search for the QCD axion below 1\,\,\ensuremath{\mu}eV}",
    eprint = "2204.13781",
    archivePrefix = "arXiv",
    primaryClass = "hep-ex",
    doi = "10.1103/PhysRevD.106.103008",
    journal = "Phys. Rev. D",
    volume = "106",
    number = "10",
    pages = "103008",
    year = "2022"
}

@article{Kahn:2016aff,
    author = "Kahn, Yonatan and Safdi, Benjamin R. and Thaler, Jesse",
    title = "{Broadband and Resonant Approaches to Axion Dark Matter Detection}",
    eprint = "1602.01086",
    archivePrefix = "arXiv",
    primaryClass = "hep-ph",
    reportNumber = "MIT-CTP-4763, PUPT-2497",
    doi = "10.1103/PhysRevLett.117.141801",
    journal = "Phys. Rev. Lett.",
    volume = "117",
    number = "14",
    pages = "141801",
    year = "2016"
}

@article{Aulakh:2004hm,
    author = "Aulakh, Charanjit S. and Girdhar, Aarti",
    title = "{SO(10) MSGUT: Spectra, couplings and threshold effects}",
    eprint = "hep-ph/0405074",
    archivePrefix = "arXiv",
    doi = "10.1016/j.nuclphysb.2005.01.008",
    journal = "Nucl. Phys. B",
    volume = "711",
    pages = "275--313",
    year = "2005"
}

@article{Hartmann:2014fya,
    author = "Hartmann, Florian and Kilian, Wolfgang and Schnitter, Karsten",
    title = "{Multiple Scales in Pati-Salam Unification Models}",
    eprint = "1401.7891",
    archivePrefix = "arXiv",
    primaryClass = "hep-ph",
    reportNumber = "SI-HEP-2013-17",
    doi = "10.1007/JHEP05(2014)064",
    journal = "JHEP",
    volume = "05",
    pages = "064",
    year = "2014"
}

@article{Bertolini:2009qj,
    author = "Bertolini, Stefano and Di Luzio, Luca and Malinsky, Michal",
    title = "{Intermediate mass scales in the non-supersymmetric SO(10) grand unification: A Reappraisal}",
    eprint = "0903.4049",
    archivePrefix = "arXiv",
    primaryClass = "hep-ph",
    doi = "10.1103/PhysRevD.80.015013",
    journal = "Phys. Rev. D",
    volume = "80",
    pages = "015013",
    year = "2009"
}

@article{Salemi:2021gck,
    author = "Salemi, Chiara P. and others",
    title = "{Search for Low-Mass Axion Dark Matter with ABRACADABRA-10~cm}",
    eprint = "2102.06722",
    archivePrefix = "arXiv",
    primaryClass = "hep-ex",
    doi = "10.1103/PhysRevLett.127.081801",
    journal = "Phys. Rev. Lett.",
    volume = "127",
    number = "8",
    pages = "081801",
    year = "2021"
}

@article{Ouellet:2018beu,
      author         = "Ouellet, Jonathan L. and others",
      title          = "{First Results from ABRACADABRA-10 cm: A Search for
                        Sub-$\mu$eV Axion Dark Matter}",
      collaboration  = "ABRACADABRA",
      journal        = "Phys. Rev. Lett.",
      volume         = "122",
      year           = "2019",
      number         = "12",
      pages          = "121802",
      doi            = "10.1103/PhysRevLett.122.121802",
      eprint         = "1810.12257",
      archivePrefix  = "arXiv",
      primaryClass   = "hep-ex",
      SLACcitation   = "%%CITATION = ARXIV:1810.12257;%%"
}

@article{Giaccone:2022pke,
    author = "Giaccone, B. and others",
    title = "{Design of axion and axion dark matter searches based on ultra high Q SRF cavities}",
    eprint = "2207.11346",
    archivePrefix = "arXiv",
    primaryClass = "hep-ex",
    reportNumber = "FERMILAB-PUB-22-592-SQMS-T-TD",
    month = "7",
    year = "2022"
}

@article{Berlin:2019ahk,
    author = "Berlin, Asher and D'Agnolo, Raffaele Tito and Ellis, Sebastian A. R. and Nantista, Christopher and Neilson, Jeffrey and Schuster, Philip and Tantawi, Sami and Toro, Natalia and Zhou, Kevin",
    title = "{Axion Dark Matter Detection by Superconducting Resonant Frequency Conversion}",
    eprint = "1912.11048",
    archivePrefix = "arXiv",
    primaryClass = "hep-ph",
    doi = "10.1007/JHEP07(2020)088",
    journal = "JHEP",
    volume = "07",
    number = "07",
    pages = "088",
    year = "2020"
}

@article{Buchbinder:2014qca,
    author = "Buchbinder, Evgeny I. and Constantin, Andrei and Lukas, Andre",
    title = "{Heterotic QCD axion}",
    eprint = "1412.8696",
    archivePrefix = "arXiv",
    primaryClass = "hep-th",
    doi = "10.1103/PhysRevD.91.046010",
    journal = "Phys. Rev. D",
    volume = "91",
    number = "4",
    pages = "046010",
    year = "2015"
}

@article{Wen:1985jz,
    author = "Wen, X. G. and Witten, Edward",
    title = "{World Sheet Instantons and the {Peccei-Quinn} Symmetry}",
    reportNumber = "PRINT-85-0906 (PRINCETON)",
    doi = "10.1016/0370-2693(86)91587-X",
    journal = "Phys. Lett. B",
    volume = "166",
    pages = "397--401",
    year = "1986"
}

@article{MacFadden:2023cyf,
    author = "MacFadden, Nate",
    title = "{Efficient Algorithm for Generating Homotopy Inequivalent Calabi-Yaus}",
    eprint = "2309.10855",
    archivePrefix = "arXiv",
    primaryClass = "hep-th",
    month = "9",
    year = "2023"
}

@article{Fallon:2025lvn,
    author = "Fallon, Sebastian Vander Ploeg and Halverson, James and McAllister, Liam and Zhu, Yunhao",
    title = "{F-theory Axiverse}",
    eprint = "2511.20458",
    archivePrefix = "arXiv",
    primaryClass = "hep-th",
    month = "11",
    year = "2025"
}

@article{Reig:2025dqb,
    author = "Reig, Mario and Weigand, Timo",
    title = "{Testing the Heterotic String with the Axion-Photon Coupling}",
    eprint = "2509.08042",
    archivePrefix = "arXiv",
    primaryClass = "hep-th",
    month = "9",
    year = "2025"
}

@article{Ho:2018qur,
    author = "Ho, Shu-Yu and Saikawa, Ken'ichi and Takahashi, Fuminobu",
    title = "{Enhanced photon coupling of ALP dark matter adiabatically converted from the QCD axion}",
    eprint = "1806.09551",
    archivePrefix = "arXiv",
    primaryClass = "hep-ph",
    reportNumber = "IPMU18-0110, MPP-2018-140, TU-1065, MIT-CTP/5026",
    doi = "10.1088/1475-7516/2018/10/042",
    journal = "JCAP",
    volume = "10",
    pages = "042",
    year = "2018"
}

@article{Gendler:2023kjt,
    author = "Gendler, Naomi and Marsh, David J. E. and McAllister, Liam and Moritz, Jakob",
    title = "{Glimmers from the Axiverse}",
    eprint = "2309.13145",
    archivePrefix = "arXiv",
    primaryClass = "hep-th",
    reportNumber = "KCL-PH-TH/2023-49",
    month = "9",
    year = "2023"
}

@article{pEDM:2022ytu,
    author = "Alexander, Jim and others",
    collaboration = "pEDM",
    title = "{The storage ring proton EDM experiment}",
    eprint = "2205.00830",
    archivePrefix = "arXiv",
    primaryClass = "hep-ph",
    reportNumber = "FERMILAB-PUB-22-611-PPD",
    month = "4",
    year = "2022"
}

@article{Wilkins:2023hua,
    author = "Wilkins, S. G. and others",
    title = "{Observation of the distribution of nuclear magnetization in a molecule}",
    eprint = "2311.04121",
    archivePrefix = "arXiv",
    primaryClass = "nucl-ex",
    doi = "10.1126/science.adm7717",
    journal = "Science",
    volume = "390",
    number = "6771",
    pages = "adm7717",
    year = "2025"
}

@article{Arrowsmith-Kron:2023hcr,
    author = "Arrowsmith-Kron, Gordon and others",
    title = "{Opportunities for fundamental physics research with radioactive molecules}",
    eprint = "2302.02165",
    archivePrefix = "arXiv",
    primaryClass = "nucl-ex",
    doi = "10.1088/1361-6633/ad1e39",
    journal = "Rept. Prog. Phys.",
    volume = "87",
    number = "8",
    pages = "084301",
    year = "2024"
}

@article{Giudice:2011cg,
    author = "Giudice, G. F. and Strumia, A.",
    title = "{Probing High-Scale and Split Supersymmetry with
             Higgs Mass Measurements}",
    eprint = "1108.6077",
    archivePrefix = "arXiv",
    primaryClass = "hep-ph",
    doi = "10.1016/j.nuclphysb.2012.01.001",
    journal = "Nucl. Phys. B",
    volume = "858",
    pages = "63--83",
    year = "2012"
}

@article{Aad:2012tfa,
    author = "Aad, Georges and others",
    collaboration = "ATLAS",
    title = "{Observation of a new particle in the search for the
             Standard Model Higgs boson with the ATLAS detector at the LHC}",
    eprint = "1207.7214",
    archivePrefix = "arXiv",
    primaryClass = "hep-ex",
    doi = "10.1016/j.physletb.2012.08.020",
    journal = "Phys. Lett. B",
    volume = "716",
    pages = "1--29",
    year = "2012"
}

@article{Chatrchyan:2012xdj,
    author = "Chatrchyan, Serguei and others",
    collaboration = "CMS",
    title = "{Observation of a New Boson at a Mass of 125 GeV with the
             CMS Experiment at the LHC}",
    eprint = "1207.7235",
    archivePrefix = "arXiv",
    primaryClass = "hep-ex",
    doi = "10.1016/j.physletb.2012.08.021",
    journal = "Phys. Lett. B",
    volume = "716",
    pages = "30--61",
    year = "2012"
}

@article{Bershadsky:1996nh, author = "Bershadsky, M. and Intriligator, K. A. and Kachru, S. and Morrison, D. R. and Sadov, V. and Vafa, C.", title = "{Geometric singularities and enhanced gauge symmetries}", eprint = "hep-th/9605200", archivePrefix = "arXiv", doi = "10.1016/S0550-3213(96)00491-2", journal = "Nucl. Phys. B", volume = "481", pages = "215--252", year = "1996" }

@article{Katz:2011qp, author = "Katz, Sheldon and Morrison, David R. and Schafer-Nameki, Sakura and Sully, James", title = "{Tate's algorithm and F-theory}", eprint = "1106.3854", archivePrefix = "arXiv", primaryClass = "hep-th", doi = "10.1007/JHEP08(2011)094", journal = "JHEP", volume = "08", pages = "094", year = "2011" }

@article{Beasley:2008dc, author = "Beasley, Chris and Heckman, Jonathan J. and Vafa, Cumrun", title = "{GUTs and Exceptional Branes in F-theory --- I}", eprint = "0802.3391", archivePrefix = "arXiv", primaryClass = "hep-th", doi = "10.1088/1126-6708/2009/01/058", journal = "JHEP", volume = "01", pages = "058", year = "2009" }

@article{Beasley:2008kw, author = "Beasley, Chris and Heckman, Jonathan J. and Vafa, Cumrun", title = "{GUTs and Exceptional Branes in F-theory --- II: Experimental Predictions}", eprint = "0806.0102", archivePrefix = "arXiv", primaryClass = "hep-th", doi = "10.1088/1126-6708/2009/01/059", journal = "JHEP", volume = "01", pages = "059", year = "2009" }

@article{Donagi:2008ca, author = "Donagi, Ron and Wijnholt, Martijn", title = "{Model Building with F-Theory}", eprint = "0802.2969", archivePrefix = "arXiv", primaryClass = "hep-th", doi = "10.4310/ATMP.2011.v15.n5.a2", journal = "Adv. Theor. Math. Phys.", volume = "15", pages = "1237--1317", year = "2011" }

@article{Weigand:2018rez, author = "Weigand, Timo", title = "{F-theory}", eprint = "1806.01854", archivePrefix = "arXiv", primaryClass = "hep-th", doi = "10.22323/1.305.0016", journal = "PoS", volume = "TASI2017", pages = "016", year = "2018" }

@article{Chen:2010ts, author = "Chen, Ching-Ming and Knapp, Johanna and Kreuzer, Maximilian and Mayrhofer, Christoph", title = "{Global SO(10) F-theory GUTs}", eprint = "1005.5735", archivePrefix = "arXiv", primaryClass = "hep-th", doi = "10.1007/JHEP10(2010)057", journal = "JHEP", volume = "10", pages = "057", year = "2010" }

@article{ParticleDataGroup:2024cfk,
    author         = "Navas, S. and others",
    collaboration  = "Particle Data Group",
    title          = "{Review of Particle Physics}",
    journal        = "Phys. Rev. D",
    volume         = "110",
    number         = "3",
    pages          = "030001",
    year           = "2024",
    note           = "and 2025 update",
    doi            = "10.1103/PhysRevD.110.030001"
}

@article{Arason:1991ic,
    author = "Arason, H. and Castano, D. J. and Keszthelyi, B. and Mikaelian, S. and Piard, E. J. and Ramond, Pierre and Wright, B. D.",
    title = "{Renormalization group study of the standard model and its extensions. 1. The Standard model}",
    reportNumber = "UFIFT-HEP-91-33-REV, UFIFT-HEP-91-33",
    doi = "10.1103/PhysRevD.46.3945",
    journal = "Phys. Rev. D",
    volume = "46",
    pages = "3945--3965",
    year = "1992"
}

@book{Polchinski:1998rr,
    author = "Polchinski, J.",
    title = "{String theory. Vol. 2: Superstring theory and beyond}",
    doi = "10.1017/CBO9780511618123",
    isbn = "978-0-511-25228-0, 978-0-521-63304-8, 978-0-521-67228-3",
    publisher = "Cambridge University Press",
    series = "Cambridge Monographs on Mathematical Physics",
    month = "12",
    year = "2007"
}

@article{Horava:1995qa,
    author = "Horava, Petr and Witten, Edward",
    title = "{Heterotic and Type I string dynamics from eleven dimensions}",
    eprint = "hep-th/9510209",
    archivePrefix = "arXiv",
    reportNumber = "IASSNS-HEP-95-86, PUPT-1571A",
    doi = "10.1016/0550-3213(95)00621-4",
    journal = "Nucl. Phys. B",
    volume = "460",
    pages = "506--524",
    year = "1996"
}

@article{Witten:1995ex,
    author = "Witten, Edward",
    title = "{String theory dynamics in various dimensions}",
    eprint = "hep-th/9503124",
    archivePrefix = "arXiv",
    reportNumber = "IASSNS-HEP-95-18",
    doi = "10.1016/0550-3213(95)00158-O",
    journal = "Nucl. Phys. B",
    volume = "443",
    pages = "85--126",
    year = "1995"
}

@article{Avram:1996pj,       author = "Avram, A. C. and Kreuzer, M. and Mandelberg, M. and Skarke, H.",       title = "{Searching for K3 fibrations}",       eprint = "hep-th/9610154",       journal = "Nucl. Phys. B",       volume = "494",       pages = "567--589",       year = "1997"   }

@article{Kreuzer:2000xy,
    author = "Kreuzer, Maximilian and Skarke, Harald",
    title = "{Complete classification of reflexive polyhedra in four-dimensions}",
    eprint = "hep-th/0002240",
    archivePrefix = "arXiv",
    reportNumber = "HUB-EP-00-13, TUW-00-07",
    doi = "10.4310/ATMP.2000.v4.n6.a2",
    journal = "Adv. Theor. Math. Phys.",
    volume = "4",
    pages = "1209--1230",
    year = "2000"
}

@article{Narain:1986am,
    author = "Narain, K. S. and Sarmadi, M. H. and Witten, Edward",
    title = "{A Note on Toroidal Compactification of Heterotic String Theory}",
    reportNumber = "Print-86-0870 (RUTHERFORD), RAL-86-022",
    doi = "10.1016/0550-3213(87)90001-0",
    journal = "Nucl. Phys. B",
    volume = "279",
    pages = "369--379",
    year = "1987"
}

@article{Gross:1984dd,
    author = "Gross, David J. and Harvey, Jeffrey A. and Martinec, Emil J. and Rohm, Ryan",
    title = "{The Heterotic String}",
    reportNumber = "PRINT-85-0029 (PRINCETON)",
    doi = "10.1103/PhysRevLett.54.502",
    journal = "Phys. Rev. Lett.",
    volume = "54",
    pages = "502--505",
    year = "1985"
}

@article{Caputo:2024oqc,
    author = "Caputo, Andrea and Raffelt, Georg",
    title = "{Astrophysical Axion Bounds: The 2024 Edition}",
    eprint = "2401.13728",
    archivePrefix = "arXiv",
    primaryClass = "hep-ph",
    reportNumber = "MPP-2024-13, CERN-TH-2024-013",
    doi = "10.22323/1.454.0041",
    journal = "PoS",
    volume = "COSMICWISPers",
    pages = "041",
    year = "2024"
}

@article{Dessert:2022yqq,
    author = "Dessert, Christopher and Dunsky, David and Safdi, Benjamin R.",
    title = "{Upper limit on the axion-photon coupling from magnetic white dwarf polarization}",
    eprint = "2203.04319",
    archivePrefix = "arXiv",
    primaryClass = "hep-ph",
    doi = "10.1103/PhysRevD.105.103034",
    journal = "Phys. Rev. D",
    volume = "105",
    number = "10",
    pages = "103034",
    year = "2022"
}

@article{Marsh:2015xka,
    author = "Marsh, David J. E.",
    title = "{Axion Cosmology}",
    eprint = "1510.07633",
    archivePrefix = "arXiv",
    primaryClass = "astro-ph.CO",
    reportNumber = "KCL-PH-TH-2015-50",
    doi = "10.1016/j.physrep.2016.06.005",
    journal = "Phys. Rept.",
    volume = "643",
    pages = "1--79",
    year = "2016"
}

@article{Halverson:2019cmy,
    author = "Halverson, James and Long, Cody and Nelson, Brent and Salinas, Gustavo",
    title = "{Towards string theory expectations for photon couplings to axionlike particles}",
    eprint = "1909.05257",
    archivePrefix = "arXiv",
    primaryClass = "hep-th",
    doi = "10.1103/PhysRevD.100.106010",
    journal = "Phys. Rev. D",
    volume = "100",
    number = "10",
    pages = "106010",
    year = "2019"
}

@article{Demirtas:2021gsq,
    author = "Demirtas, Mehmet and Gendler, Naomi and Long, Cody and McAllister, Liam and Moritz, Jakob",
    title = "{PQ axiverse}",
    eprint = "2112.04503",
    archivePrefix = "arXiv",
    primaryClass = "hep-th",
    doi = "10.1007/JHEP06(2023)092",
    journal = "JHEP",
    volume = "06",
    pages = "092",
    year = "2023"
}

@article{Demirtas:2022hqf,
    author = "Demirtas, Mehmet and Rios-Tascon, Andres and McAllister, Liam",
    title = "{CYTools: A Software Package for Analyzing Calabi-Yau Manifolds}",
    eprint = "2211.03823",
    archivePrefix = "arXiv",
    primaryClass = "hep-th",
    month = "11",
    year = "2022"
}

@article{Gendler:2024adn,
    author = "Gendler, Naomi and Marsh, David J. E.",
    title = "{QCD Axion Dark Matter in String Theory: Haloscopes and Helioscopes as Probes of the Landscape}",
    eprint = "2407.07143",
    archivePrefix = "arXiv",
    primaryClass = "hep-th",
    month = "7",
    year = "2024"
}

@article{Agrawal:2022lsp,
    author = "Agrawal, Prateek and Nee, Michael and Reig, Mario",
    title = "{Axion couplings in grand unified theories}",
    eprint = "2206.07053",
    archivePrefix = "arXiv",
    primaryClass = "hep-ph",
    doi = "10.1007/JHEP10(2022)141",
    journal = "JHEP",
    volume = "10",
    pages = "141",
    year = "2022"
}

@article{Langhoff:2022bij,
    author = "Langhoff, Kevin and Outmezguine, Nadav Joseph and Rodd, Nicholas L.",
    title = "{Irreducible Axion Background}",
    eprint = "2209.06216",
    archivePrefix = "arXiv",
    primaryClass = "hep-ph",
    reportNumber = "CERN-TH-2022-148",
    doi = "10.1103/PhysRevLett.129.241101",
    journal = "Phys. Rev. Lett.",
    volume = "129",
    number = "24",
    pages = "241101",
    year = "2022"
}

@article{Salvio:2013iaa,
    author = "Salvio, Alberto and Strumia, Alessandro and Xue, Wei",
    title = "{Thermal axion production}",
    eprint = "1310.6982",
    archivePrefix = "arXiv",
    primaryClass = "hep-ph",
    reportNumber = "FTUAM-13-29, IFT-UAM-CSIC-13-113",
    doi = "10.1088/1475-7516/2014/01/011",
    journal = "JCAP",
    volume = "01",
    pages = "011",
    year = "2014"
}

@article{Cicoli:2012sz,
    author = "Cicoli, Michele and Goodsell, Mark and Ringwald, Andreas",
    title = "{The type IIB string axiverse and its low-energy phenomenology}",
    eprint = "1206.0819",
    archivePrefix = "arXiv",
    primaryClass = "hep-th",
    reportNumber = "DESY-12-058, CERN-PH-TH-2012-153",
    doi = "10.1007/JHEP10(2012)146",
    journal = "JHEP",
    volume = "10",
    pages = "146",
    year = "2012"
}

@article{Demirtas:2018akl,
    author = "Demirtas, Mehmet and Long, Cody and McAllister, Liam and Stillman, Mike",
    title = "{The Kreuzer-Skarke Axiverse}",
    eprint = "1808.01282",
    archivePrefix = "arXiv",
    primaryClass = "hep-th",
    doi = "10.1007/JHEP04(2020)138",
    journal = "JHEP",
    volume = "04",
    pages = "138",
    year = "2020"
}

@article{Arvanitaki:2009fg,
    author = "Arvanitaki, Asimina and Dimopoulos, Savas and Dubovsky, Sergei and Kaloper, Nemanja and March-Russell, John",
    title = "{String Axiverse}",
    eprint = "0905.4720",
    archivePrefix = "arXiv",
    primaryClass = "hep-th",
    doi = "10.1103/PhysRevD.81.123530",
    journal = "Phys. Rev. D",
    volume = "81",
    pages = "123530",
    year = "2010"
}

@article{Broeckel:2021dpz,
    author = "Broeckel, Igor and Cicoli, Michele and Maharana, Anshuman and Singh, Kajal and Sinha, Kuver",
    title = "{Moduli stabilisation and the statistics of axion physics in the landscape}",
    eprint = "2105.02889",
    archivePrefix = "arXiv",
    primaryClass = "hep-th",
    doi = "10.1007/JHEP01(2022)191",
    journal = "JHEP",
    volume = "08",
    pages = "059",
    year = "2021",
    note = "[Addendum: JHEP 01, 191 (2022)]"
}

@article{Gendler:2023hwg,
    author = "Gendler, Naomi and Janssen, Oliver and Kleban, Matthew and La Madrid, Joan and Mehta, Viraf M.",
    title = "{Axion minima in string theory}",
    eprint = "2309.01831",
    archivePrefix = "arXiv",
    primaryClass = "hep-th",
    month = "9",
    year = "2023"
}

@article{Mehta:2021pwf,
    author = "Mehta, Viraf M. and Demirtas, Mehmet and Long, Cody and Marsh, David J. E. and McAllister, Liam and Stott, Matthew J.",
    title = "{Superradiance in string theory}",
    eprint = "2103.06812",
    archivePrefix = "arXiv",
    primaryClass = "hep-th",
    doi = "10.1088/1475-7516/2021/07/033",
    journal = "JCAP",
    volume = "07",
    pages = "033",
    year = "2021"
}

@article{Weinberg:1977ma,
	Author = {Weinberg, Steven},
	Doi = {10.1103/PhysRevLett.40.223},
	Journal = {Phys. Rev. Lett.},
	Pages = {223-226},
	Reportnumber = {HUTP-77/A074},
	Slaccitation = {%%CITATION = PRLTA,40,223;%%},
	Title = {{A New Light Boson?}},
	Volume = {40},
	Year = {1978}}

@article{Dine:1982ah,
	Author = {Dine, Michael and Fischler, Willy},
	Doi = {10.1016/0370-2693(83)90639-1},
	Journal = {Phys. Lett.},
	Pages = {137-141},
	Reportnumber = {UPR-0201T},
	Slaccitation = {%%CITATION = PHLTA,120B,137;%%},
	Title = {{The Not So Harmless Axion}},
	Volume = {120B},
	Year = {1983}}

@article{Abbott:1982af,
	Author = {Abbott, L. F. and Sikivie, P.},
	Doi = {10.1016/0370-2693(83)90638-X},
	Journal = {Phys. Lett.},
	Pages = {133-136},
	Reportnumber = {PRINT-82-0695 (BRANDEIS)},
	Slaccitation = {%%CITATION = PHLTA,120B,133;%%},
	Title = {{A Cosmological Bound on the Invisible Axion}},
	Volume = {120B},
	Year = {1983}}

@article{Preskill:1982cy,
	Author = {Preskill, John and Wise, Mark B. and Wilczek, Frank},
	Doi = {10.1016/0370-2693(83)90637-8},
	Journal = {Phys. Lett.},
	Pages = {127-132},
	Reportnumber = {HUTP-82-A048, NSF-ITP-82-103},
	Slaccitation = {%%CITATION = PHLTA,120B,127;%%},
	Title = {{Cosmology of the Invisible Axion}},
	Volume = {120B},
	Year = {1983}}

@article{Witten:1984dg,
    author = "Witten, Edward",
    title = "{Some Properties of O(32) Superstrings}",
    reportNumber = "Print-84-0838 (PRINCETON)",
    doi = "10.1016/0370-2693(84)90422-2",
    journal = "Phys. Lett. B",
    volume = "149",
    pages = "351--356",
    year = "1984"
}

@article{Wilczek:1977pj,
	Author = {Wilczek, Frank},
	Doi = {10.1103/PhysRevLett.40.279},
	Journal = {Phys. Rev. Lett.},
	Pages = {279-282},
	Reportnumber = {Print-77-0939 (COLUMBIA)},
	Slaccitation = {%%CITATION = PRLTA,40,279;%%},
	Title = {{Problem of Strong p and t Invariance in the Presence of Instantons}},
	Volume = {40},
	Year = {1978}}

@article{Peccei:1977hh,
	Author = {Peccei, R. D. and Quinn, Helen R.},
	Doi = {10.1103/PhysRevLett.38.1440},
	Journal = {Phys. Rev. Lett.},
	Pages = {1440-1443},
	Reportnumber = {ITP-568-STANFORD},
	Slaccitation = {%%CITATION = PRLTA,38,1440;%%},
	Title = {{CP Conservation in the Presence of Instantons}},
	Volume = {38},
	Year = {1977}}

@article{Jaeckel:2017tud,
    author = "Jaeckel, J. and Malta, P. C. and Redondo, J.",
    title = "{Decay photons from the axionlike particles burst of type II supernovae}",
    eprint = "1702.02964",
    archivePrefix = "arXiv",
    primaryClass = "hep-ph",
    doi = "10.1103/PhysRevD.98.055032",
    journal = "Phys. Rev. D",
    volume = "98",
    number = "5",
    pages = "055032",
    year = "2018"
}

@article{Arkani-Hamed:2012fhg,
    author = "Arkani-Hamed, Nima and Gupta, Arpit and Kaplan, David E. and Weiner, Neal and Zorawski, Tom",
    title = "{Simply Unnatural Supersymmetry}",
    eprint = "1212.6971",
    archivePrefix = "arXiv",
    primaryClass = "hep-ph",
    month = "12",
    year = "2012"
}

@inproceedings{Wells:2003tf,
    author = "Wells, James D.",
    title = "{Implications of supersymmetry breaking with a little hierarchy between gauginos and scalars}",
    booktitle = "{11th International Conference on Supersymmetry and the Unification of Fundamental Interactions}",
    eprint = "hep-ph/0306127",
    archivePrefix = "arXiv",
    reportNumber = "MCTP-03-30",
    month = "6",
    year = "2003"
}

@article{Hoof:2022xbe,
    author = "Hoof, Sebastian and Schulz, Lena",
    title = "{Updated constraints on axion-like particles from temporal information in supernova SN1987A gamma-ray data}",
    eprint = "2212.09764",
    archivePrefix = "arXiv",
    primaryClass = "hep-ph",
    reportNumber = "TTP22-072",
    doi = "10.1088/1475-7516/2023/03/054",
    journal = "JCAP",
    volume = "03",
    pages = "054",
    year = "2023"
}

@article{Muller:2023vjm,
    author = {M\"uller, Eike and Calore, Francesca and Carenza, Pierluca and Eckner, Christopher and Marsh, M. C. David},
    title = "{Investigating the gamma-ray burst from decaying MeV-scale axion-like particles produced in supernova explosions}",
    eprint = "2304.01060",
    archivePrefix = "arXiv",
    primaryClass = "astro-ph.HE",
    doi = "10.1088/1475-7516/2023/07/056",
    journal = "JCAP",
    volume = "07",
    pages = "056",
    year = "2023"
}

@article{Lella:2024dmx,
    author = "Lella, Alessandro and Ravensburg, Eike and Carenza, Pierluca and Marsh, M. C. David",
    title = "{Supernova limits on QCD axionlike particles}",
    eprint = "2405.00153",
    archivePrefix = "arXiv",
    primaryClass = "hep-ph",
    doi = "10.1103/PhysRevD.110.043019",
    journal = "Phys. Rev. D",
    volume = "110",
    number = "4",
    pages = "043019",
    year = "2024"
}

@article{Agrawal:2024ejr,
    author = "Agrawal, Prateek and Nee, Michael and Reig, Mario",
    title = "{Axion Couplings in Heterotic String Theory}",
    eprint = "2410.03820",
    archivePrefix = "arXiv",
    primaryClass = "hep-ph",
    month = "10",
    year = "2024"
}

@article{Foster:2021ngm,
    author = "Foster, Joshua W. and Kongsore, Marius and Dessert, Christopher and Park, Yujin and Rodd, Nicholas L. and Cranmer, Kyle and Safdi, Benjamin R.",
    title = "{Deep Search for Decaying Dark Matter with XMM-Newton Blank-Sky Observations}",
    eprint = "2102.02207",
    archivePrefix = "arXiv",
    primaryClass = "astro-ph.CO",
    reportNumber = "LCTP-21-05",
    doi = "10.1103/PhysRevLett.127.051101",
    journal = "Phys. Rev. Lett.",
    volume = "127",
    number = "5",
    pages = "051101",
    year = "2021"
}

@article{Cohen:2016uyg,
    author = "Cohen, Timothy and Murase, Kohta and Rodd, Nicholas L. and Safdi, Benjamin R. and Soreq, Yotam",
    title = "{\ensuremath{\gamma} -ray Constraints on Decaying Dark Matter and Implications for IceCube}",
    eprint = "1612.05638",
    archivePrefix = "arXiv",
    primaryClass = "hep-ph",
    reportNumber = "MIT-CTP-4863, MIT-CTP 4863",
    doi = "10.1103/PhysRevLett.119.021102",
    journal = "Phys. Rev. Lett.",
    volume = "119",
    number = "2",
    pages = "021102",
    year = "2017"
}

@article{Blanco:2018esa,
    author = "Blanco, Carlos and Hooper, Dan",
    title = "{Constraints on Decaying Dark Matter from the Isotropic Gamma-Ray Background}",
    eprint = "1811.05988",
    archivePrefix = "arXiv",
    primaryClass = "astro-ph.HE",
    reportNumber = "FERMILAB-PUB-18-627-A",
    doi = "10.1088/1475-7516/2019/03/019",
    journal = "JCAP",
    volume = "03",
    pages = "019",
    year = "2019"
}

@article{Liu:2023nct,
    author = "Liu, Hongwan and Qin, Wenzer and Ridgway, Gregory W. and Slatyer, Tracy R.",
    title = "{Exotic energy injection in the early Universe. II. CMB spectral distortions and constraints on light dark matter}",
    eprint = "2303.07370",
    archivePrefix = "arXiv",
    primaryClass = "astro-ph.CO",
    reportNumber = "MIT-CTP/5524",
    doi = "10.1103/PhysRevD.108.043531",
    journal = "Phys. Rev. D",
    volume = "108",
    number = "4",
    pages = "043531",
    year = "2023"
}

@article{Calore:2022pks,
    author = "Calore, Francesca and Dekker, Ariane and Serpico, Pasquale Dario and Siegert, Thomas",
    title = "{Constraints on light decaying dark matter candidates from 16~yr of INTEGRAL/SPI observations}",
    eprint = "2209.06299",
    archivePrefix = "arXiv",
    primaryClass = "hep-ph",
    doi = "10.1093/mnras/stad457",
    journal = "Mon. Not. Roy. Astron. Soc.",
    volume = "520",
    number = "3",
    pages = "4167--4172",
    year = "2023"
}

@article{Perez:2016tcq,
    author = "Perez, Kerstin and Ng, Kenny C. Y. and Beacom, John F. and Hersh, Cora and Horiuchi, Shunsaku and Krivonos, Roman",
    title = "{Almost closing the \ensuremath{\nu}MSM sterile neutrino dark matter window with NuSTAR}",
    eprint = "1609.00667",
    archivePrefix = "arXiv",
    primaryClass = "astro-ph.HE",
    doi = "10.1103/PhysRevD.95.123002",
    journal = "Phys. Rev. D",
    volume = "95",
    number = "12",
    pages = "123002",
    year = "2017"
}

@article{Roach:2022lgo,
    author = "Roach, Brandon M. and Rossland, Steven and Ng, Kenny C. Y. and Perez, Kerstin and Beacom, John F. and Grefenstette, Brian W. and Horiuchi, Shunsaku and Krivonos, Roman and Wik, Daniel R.",
    title = "{Long-exposure NuSTAR constraints on decaying dark matter in the Galactic halo}",
    eprint = "2207.04572",
    archivePrefix = "arXiv",
    primaryClass = "astro-ph.HE",
    doi = "10.1103/PhysRevD.107.023009",
    journal = "Phys. Rev. D",
    volume = "107",
    number = "2",
    pages = "023009",
    year = "2023"
}

@article{Ng:2019gch,
    author = "Ng, Kenny C. Y. and Roach, Brandon M. and Perez, Kerstin and Beacom, John F. and Horiuchi, Shunsaku and Krivonos, Roman and Wik, Daniel R.",
    title = "{New Constraints on Sterile Neutrino Dark Matter from $NuSTAR$ M31 Observations}",
    eprint = "1901.01262",
    archivePrefix = "arXiv",
    primaryClass = "astro-ph.HE",
    doi = "10.1103/PhysRevD.99.083005",
    journal = "Phys. Rev. D",
    volume = "99",
    pages = "083005",
    year = "2019"
}

@article{Reece:2024wrn,
    author = "Reece, Matthew",
    title = "{Extra-Dimensional Axion Expectations}",
    eprint = "2406.08543",
    archivePrefix = "arXiv",
    primaryClass = "hep-ph",
    month = "6",
    year = "2024"
}

@article{Foster:2022nva,
    author = "Foster, Joshua W. and Park, Yujin and Safdi, Benjamin R. and Soreq, Yotam and Xu, Weishuang Linda",
    title = "{Search for dark matter lines at the Galactic Center with 14~years of Fermi data}",
    eprint = "2212.07435",
    archivePrefix = "arXiv",
    primaryClass = "hep-ph",
    reportNumber = "MIT-CTP/5505",
    doi = "10.1103/PhysRevD.107.103047",
    journal = "Phys. Rev. D",
    volume = "107",
    number = "10",
    pages = "103047",
    year = "2023"
}

@article{Kalashev:2020hqc,
    author = "Kalashev, Oleg and Kuznetsov, Mikhail and Zhezher, Yana",
    title = "{Constraining superheavy decaying dark matter with directional ultra-high energy gamma-ray limits}",
    eprint = "2005.04085",
    archivePrefix = "arXiv",
    primaryClass = "astro-ph.HE",
    reportNumber = "INR-TH-2020-026",
    doi = "10.1088/1475-7516/2021/11/016",
    journal = "JCAP",
    volume = "11",
    pages = "016",
    year = "2021"
}

@article{Das:2023wtk,
    author = "Das, Saikat and Murase, Kohta and Fujii, Toshihiro",
    title = "{Revisiting ultrahigh-energy constraints on decaying superheavy dark matter}",
    eprint = "2302.02993",
    archivePrefix = "arXiv",
    primaryClass = "astro-ph.HE",
    reportNumber = "YITP-23-08",
    doi = "10.1103/PhysRevD.107.103013",
    journal = "Phys. Rev. D",
    volume = "107",
    number = "10",
    pages = "103013",
    year = "2023"
}

@article{Boyarsky:2006ag,
    author = "Boyarsky, Alexey and Nevalainen, Jukka and Ruchayskiy, Oleg",
    title = "{Constraints on the parameters of radiatively decaying dark matter from the dark matter halo of the Milky Way and Ursa Minor}",
    eprint = "astro-ph/0610961",
    archivePrefix = "arXiv",
    reportNumber = "CERN-PH-TH-2006-225",
    doi = "10.1051/0004-6361:20066774",
    journal = "Astron. Astrophys.",
    volume = "471",
    pages = "51--57",
    year = "2007"
}

@article{Boyarsky:2007ay,
    author = "Boyarsky, Alexey and Iakubovskyi, Dmytro and Ruchayskiy, Oleg and Savchenko, Vladimir",
    title = "{Constraints on decaying Dark Matter from XMM-Newton observations of M31}",
    eprint = "0709.2301",
    archivePrefix = "arXiv",
    primaryClass = "astro-ph",
    doi = "10.1111/j.1365-2966.2008.13266.x",
    journal = "Mon. Not. Roy. Astron. Soc.",
    volume = "387",
    pages = "1361",
    year = "2008"
}

@article{Boyarsky:2006fg,
    author = "Boyarsky, Alexey and Neronov, A. and Ruchayskiy, O. and Shaposhnikov, M. and Tkachev, I.",
    title = "{Where to find a dark matter sterile neutrino?}",
    eprint = "astro-ph/0603660",
    archivePrefix = "arXiv",
    doi = "10.1103/PhysRevLett.97.261302",
    journal = "Phys. Rev. Lett.",
    volume = "97",
    pages = "261302",
    year = "2006"
}

@article{Manzari:2024jns,
    author = "Manzari, Claudio Andrea and Park, Yujin and Safdi, Benjamin R. and Savoray, Inbar",
    title = "{Supernova Axions Convert to Gamma Rays in Magnetic Fields of Progenitor Stars}",
    eprint = "2405.19393",
    archivePrefix = "arXiv",
    primaryClass = "hep-ph",
    doi = "10.1103/PhysRevLett.133.211002",
    journal = "Phys. Rev. Lett.",
    volume = "133",
    number = "21",
    pages = "211002",
    year = "2024"
}

@article{Poulin:2016anj,
    author = "Poulin, Vivian and Lesgourgues, Julien and Serpico, Pasquale D.",
    title = "{Cosmological constraints on exotic injection of electromagnetic energy}",
    eprint = "1610.10051",
    archivePrefix = "arXiv",
    primaryClass = "astro-ph.CO",
    doi = "10.1088/1475-7516/2017/03/043",
    journal = "JCAP",
    volume = "03",
    pages = "043",
    year = "2017"
}

@article{Borsanyi:2016ksw,
    author = "Borsanyi, Sz. and others",
    title = "{Calculation of the axion mass based on high-temperature lattice quantum chromodynamics}",
    eprint = "1606.07494",
    archivePrefix = "arXiv",
    primaryClass = "hep-lat",
    reportNumber = "DESY-16-105",
    doi = "10.1038/nature20115",
    journal = "Nature",
    volume = "539",
    number = "7627",
    pages = "69--71",
    year = "2016"
}

@article{Giudice:1998bp,
    author = "Giudice, G. F. and Rattazzi, R.",
    title = "{Theories with gauge mediated supersymmetry breaking}",
    eprint = "hep-ph/9801271",
    archivePrefix = "arXiv",
    doi = "10.1016/S0370-1573(99)00042-3",
    journal = "Phys. Rept.",
    volume = "322",
    pages = "419--499",
    year = "1999"
}

@article{Jain:2025vfh,
    author = "Jain, Mudit and Sheridan, Elijah and Marsh, David J. E. and Heyes, Elli and Rogers, Keir K. and Schachner, Andreas",
    title = "{Bayesian inference on Calabi--Yau moduli spaces and the axiverse: experimental data meets string theory}",
    eprint = "2512.00144",
    archivePrefix = "arXiv",
    primaryClass = "hep-th",
    month = "11",
    year = "2025"
}

\clearpage
\appendix

\section{Heterotic compactification scan}
We study compactifications of weakly coupled heterotic string theory on CY 3-fold hypersurfaces of toric varieties. We construct these manifolds 
by triangulating reflexive polytopes of dimension 4, which have been enumerated in the KS database \cite{Kreuzer:2000xy}, using \texttt{CYTools} \cite{Demirtas:2022hqf}.
Triangulations of KS polytopes have been  used extensively to study axiverses arising in Type IIB string theory from dimensional reduction of the Ramond-Ramond field $C_4$ on O3/O7 orientifolds   \cite{Gendler:2023kjt,Kreuzer:2000xy, Demirtas:2018akl, Demirtas:2021gsq,Gendler:2023hwg, Mehta:2021pwf,Benabou:2025kgx}. In this case $C_4$ is reduced on a basis of prime toric divisors of the integral homology group $H_4$, and the manifolds obtained from the KS ensemble contain at most $h^{1,1}=491$ axions. 
We do not identify any manifolds in the KS ensemble with more than $11$ axions satisfying the weakly coupled heterotic constraints for our benchmark coupling $\alpha_{\rm GUT}^{-1} = 27$.

For each $h^{1,1} \le 16$, we scan over \textit{favorable} polytopes and for each we sample FRSTs using the \texttt{CYTools} function \texttt{ntfe\_frsts} with \texttt{triang\_method="grow2d"}. This ensures that no two triangulations returned have the same restriction to 2-faces \cite{MacFadden:2023cyf} (identical 2-face restrictions give the same manifold). 
The number of favorable KS polytopes and corresponding FRST triangulations used in our scan is listed for each $h^{1,1}$ in Table \ref{tab:number_polytops_triangulations}. Our scan is topologically exhaustive  for $h^{1,1} \le 8$.

For each CY 3-fold, we compute the SKC \cite{Demirtas:2018akl}, \textit{i.e.} the cone in Kähler moduli space for which all of the effective curve volumes are at least equal to a fixed constant $c$ (see the SM for details). In our fiducial analysis we take $c=1$, though we also consider smaller values as a systematic test.
In previous works extracting parameters of the 4D axion EFT from KS compactifications, calculations are typically performed at the tip of the SKC.  In this work, we calculate the 4D axion EFT scanning over generic points within the SKC, as we detail below. (Note that we do not impose any lower bound on divisor volumes, as these do not appear in the 4D axion EFT.) 
For a  FRST to generate an acceptable heterotic compactification manifold, {\it i.e.} for which the $\alpha'$ expansion is under control, we must impose that there is at least one point in the SKC with $\mathcal{V}_6=\mathcal{V}^\star_6 \equiv \frac{g_s^2}{\alpha_{\rm GUT}}\sim 27$ in string-length units. 
(We set $g_s = 1$ and $\alpha_{\rm GUT}^{-1} = 27$ and discuss the $g_s$ and $\alpha_{\rm GUT}^{-1}$ dependence of our results later in this work.)  
We check this condition numerically by minimizing $|\mathcal{V}_6-\mathcal{V}_6^\star|$ within the SKC using the differential evolution global optimization algorithm~\cite{Storn:1997uea}. We also perform a less computationally expensive scan (in some cases over a larger set of FRSTs; note that Table \ref{tab:number_polytops_triangulations} reports the global scan coverage $N_\mathrm{FRST}$)  for $\mathcal{V}_6=\mathcal{V}_6^\star$ attained along the generators of the SKC, though this underestimates the maximal curve volume if it is attained in the interior of the cone.
SUSY imposes an additional restriction on the moduli of the CY 3-fold. The D-term equations require the gauge bundle to be polystable with vanishing slope (the Donaldson-Uhlenbeck-Yau condition) \cite{Leedom:2025mlr}. In general, for a given manifold this condition restricts the set of allowed vector bundles, but it is automatically satisfied for a standard embedding, which we assume as our fiducial choice. Furthermore, anomaly cancellation requires, according to \eqref{eq:full_bianchi}, that $c_2(TX_6)-c_2(V)$ is an effective class. Again, for a given CY this does not hold for an arbitrary vector bundle, but is satisfied for a standard embedding.

For each $h^{1,1}$, the number of FRST classes which satisfy these conditions is listed in Table \ref{tab:number_polytops_triangulations}. In total, we find 2027 heterotic-compatible  FRST classes, and do not identify any such compactifications with $h^{1,1} > 10$. Of course, we cannot exclude the possibility that there are acceptable manifolds in the cases for which we have not scanned over all possible FRST classes  (i.e., for $h^{1,1} > 8$). However, we expect that such examples become increasingly rare at large $h^{1,1}$ because the SKC
becomes increasingly narrow. This is because the number of inequalities defining the SKC grows with $h^{1,1}$, which generically pushes the tip of the SKC  further from the origin of the Kähler cone, {\it i.e.} towards larger $\mathcal{V}_6$.

\begin{table}[t]
\centering
\begin{tabular}{|l|l|l|l|l|l|l|}
\hline
$h^{1,1}$ & $N_\mathrm{poly}$ &  $N_\mathrm{FRST}$ & \multicolumn{2}{c|}{$N_\mathrm{FRST}$ (heterotic)} & $N_\star$ & $\max(\mathrm{Vol}(C_2))$ \\
\cline{4-5}
 & & & Global & Gen. & & \\
\hline
1 &  5   &  5 & 5 & 5 & 0 & 5.45 \\
2  & 36      & 39        & 39 & 39 & 4 & 27 \\
3  & 243     & 309       & 196 & 192 & 3 & 27 \\
4  & 1{,}185 & 2{,}106   & 535 & 491 & 0 & 12.5 \\
5  & 4{,}897 & 15{,}266  & 567 & 432 & 0 & 8.72 \\
6  & 16{,}608 & 102{,}693 & 371 & 272 & 0 & 8.78 \\
7  & 48{,}221 & 738{,}841 & 196 & 134 & 0 & 8.83 \\
8  & 120{,}759 & 5{,}225{,}315 & $112$ & 86 & 0 & 3.94 \\
9  & 264{,}558 & 100{,}000 & 4 & 62 & 0 & 1.23 \\
10 & 515{,}319 & 90{,}193 & 2 & 2 & 0 & 4 \\
11 & 261{,}541 & 54{,}733 & 0 & 0 & 0 & --- \\
12 & 86{,}860  & 33{,}448 & 0 & 0 & 0 & --- \\
13 & 84{,}923  & 16{,}211 & 0 & 0 & 0 & --- \\
14 & 82{,}939  & 5{,}153 & 0 & 0 & 0 & --- \\
15 & 80{,}415  & 1{,}745 & 0 & 0 & 0 & --- \\
16 & 78{,}756  & 744 & 0 & 0 & 0 & --- \\
\hline
\end{tabular}
\caption{ 
As a function of $h^{1,1}$, the number of favorable KS polytopes $N_\mathrm{poly}$, \textit{sampled} FRST classes $N_\mathrm{FRST}$ for the global stochastic scan,  FRST classes which give a heterotic compactification for at least one point in the SKC (via the global scan, and via the scan along SKC generators), the number $N_\star$ of those FRSTs for which the QCD axion mass deviates non-negligibly from the MI value, and the maximal curve volume across those FRSTs appearing in the leading $h^{1,1}$ instantons.
(Note that we count FRSTs as equivalent if they induce identical triangulations on every (labeled) 2-face of the polytope. We do not quotient by polytope automorphisms or check for equivalence of FRSTs across different polytopes; the number of physically distinct manifolds is therefore smaller than the number of FRSTs listed in some cases.) We fix $\alpha_\mathrm{GUT}=1/27$, $g_s=1$.  
For each value of $h^{1,1}$ we sample at least one FRST from all favorable polytopes. We explore all 
6,084,574
FRST classes which exist with $h^{1,1} \le 8$. For $9 \le h^{1,1} \le 16$ we sample a subset of all possible FRST classes. In total we scan over 6,386,801  
FRST classes and our heterotic ensemble contains 2027 classes.  }
\label{tab:number_polytops_triangulations}
\end{table}

For each CY 3-fold in our ensemble, we perform a numerical scan over the K\"ahler moduli space to search for deviation of the QCD axion mass from the MI value. To do so, we scan over the SKC to find the point which realizes the largest possible effective curve volume subject to $\mathcal{V}_6=\alpha_{\rm GUT}^{-1}$ and the requirement that all curve volumes be larger than unity in string units. 
In more detail, we minimize
$\alpha = -\max_i\{\mathbf{Q} \cdot \mathbf{t}\}$ using differential evolution, with the
K\"ahler parameters sampled from $t^i \in [-100,100]$. For configurations with $\alpha < -20$
we evaluate the QCD axion mass and record the compactification data. This procedure is applied to KS CY hypersurfaces with $h^{1,1}\leq 8$. 

Note that to compute axion masses we use the approximate diagonalization procedure described in Ref.~\cite{Gendler:2023kjt}. After going to the approximate mass eigenbasis obtained from truncating the potential for MD axions to the $h^{1,1}$ heaviest instantons,
we integrate out heavy MD axions with masses $m_\mathrm{MD}>10^4 m_\mathrm{MI}$ (as they mix negligibly with the MI axion). We then diagonalize the reduced mass matrix for the remaining axions (including the MI axion), accounting for the QCD instanton contribution. 

\vspace{-0.2in}
\section{Leptogenesis constraints}

Here, we provide further details on the leptogenesis constraints shown in Fig.~\ref{fig:gayy_heterotic}.
 Heavy axions may lead to a period of early matter domination (EMD), which causes entropy dilution. 
This entropy injection will dilute any primordial lepton asymmetry~\cite{Fukugita:1986hr,Davidson:2008bu,Buchmuller:2004nz}.  
Thermal leptogenesis
requires Majorana neutrinos with $M_\nu > 10^9$~GeV~\cite{Davidson:2002qv} that freeze out after
inflation. The QCD axion isocurvature constraint from
Planck~\cite{Planck:2018jri} limits $H_\mathrm{I} \lesssim 10^9$~GeV
for $f_a \sim 1.1\times 10^{16}$~GeV, 
assuming the axion is all of the DM,
which in turn implies $M_\nu
\lesssim 10^{14}$~GeV in order for the neutrinos to have acquired a thermal relic abundance given the maximal reheating temperature post-inflation.  The observed baryon-to-photon ratio is {\it overproduced}, by an amount linear in $M_\nu$, for $M_\nu \gtrsim 10^{9}$ GeV (conservatively assuming no washout); thus, the primordial lepton asymmetry could be overproduced by as much as five orders of magnitude but not more.  Requiring that the heavy-axion-induced EMD period give less than five orders of magnitude of entropy dilution sets the constraints in Figs.~\ref{fig:gayy_heterotic} and~\ref{fig:gayy_typeIIB}.

In more detail,
in the limit where the lightest of the heavy right-handed neutrinos has a mass $M_\nu$ significantly smaller than the rest, the baryon-to-photon ratio is approximately bounded from above by \cite{Davidson:2002qv}
\begin{equation}
\label{eq:eta_B}
\eta_B \lesssim 10^{-10}\left(\frac{M_\nu}{10^{10} \mathrm{GeV}}\right)\left(\frac{\kappa_w}{0.1}\right)\left(\frac{m_3-m_1}{0.05 \mathrm{eV}}\right) \,,
\end{equation}
where $\kappa_w$ is the washout efficiency and $m_1$ ($m_3$) the smallest (largest) of the light neutrino masses. The  present-day value is $\eta_B^{\mathrm{obs}} \equiv n_B/n_\gamma \simeq 6.1 \times 10^{-10}$ \cite{Planck2018Parameters}.

To compute the resulting constraint on heavy axions, we assume that in the absence of heavy axions the Universe would be radiation-dominated prior to BBN. A heavy axion will decay dominantly to photons and gluons (we give the tree-level rate $\Gamma$ in the SM).
We consider the axion abundance produced via misalignment with an $\mathcal{O}(1)$ initial angle (for freeze-in/out production see the SM), such that the energy density is
$\rho_a \approx \rho_{a,i}\left(\frac{T}{T_\mathrm{osc}}\right)^3 $ for $T < T_{\rm osc}$, 
with the initial energy density $\rho_{a,i}\approx\frac{1}{2}\theta_i^2f_a^2m_a^2$, and the oscillation temperature set by $3H(T_\mathrm{osc})=m_a$.
In the instantaneous decay approximation, the baryon asymmetry is diluted by an amount $r_B$ at the decay time $t_{\rm dec} = \Gamma^{-1}$ because the  decay of the
(dominant) axion fluid reheats the photon bath while the
comoving baryon number $n_B \propto R^{-3}$ is conserved. At
$t_\mathrm{dec}$,
\begin{align}
r_B \equiv \frac{n_\gamma^\mathrm{new}}{n_\gamma^\mathrm{old}}\bigg|_{t_\mathrm{dec}}
 = \left(\frac{\rho_\gamma^\mathrm{new}}{\rho_\gamma^\mathrm{old}}\right)^{\!3/4}\bigg|_{t_\mathrm{dec}},
\label{eq:rB}
\end{align}
where we use $n_\gamma \propto T^3 \propto \rho_\gamma^{3/4}$.
The pre-existing radiation simply redshifts through the EMD era,
$\rho_\gamma^\mathrm{old}(t_\mathrm{dec})
 = \rho_\gamma(t_\mathrm{EMD})(R_\mathrm{EMD}/R_\mathrm{dec})^4$, with $t_{\rm EMD}$ the time at which the heavy axion density dominates the Universe's energy budget and leads to EMD,
while the decay products dominate after decay,
$\rho_\gamma^\mathrm{new}(t_\mathrm{dec}) = \rho_a(t_\mathrm{dec})
 \approx 3\Gamma^2 \mpl^2$, with
$\rho_a(t_\mathrm{dec}) = \rho_a(t_\mathrm{EMD})(R_\mathrm{EMD}/R_\mathrm{dec})^3$.
Using $\rho_\gamma(t_\mathrm{EMD}) = \rho_a(t_\mathrm{EMD})$ and
$H \propto R^{-3/2}$ in matter domination, \eqref{eq:rB} collapses to
\begin{align}
r_B \approx \left(\frac{H_\mathrm{EMD}}{\Gamma}\right)^{\!1/2} \,.
\end{align}

We have $H_{I} \gtrsim T_\mathrm{RH}^2/\mpl$, which, given the QCD axion isocurvature constraint on $H_I$, implies $T_{\mathrm{fo}}\lesssim T_{\mathrm{RH}}\lesssim 10^{13}$ GeV. 
Approximating freeze-out as instantaneous, the washout parameter scales as $\kappa_w \propto e^{-M_\nu/T_\mathrm{fo}}$, such that
we may conservatively  bound the baryon asymmetry by setting $M_\nu = T_{\mathrm{fo}}$; \eqref{eq:eta_B} then gives 
\begin{equation}
    r_B < 10^4 \left(\frac{M_\nu}{10^{13} \,\mathrm{GeV} }\right)\,.
    \label{eq:Rb_bound}
\end{equation}
This gives the leptogenesis constraint (for $m_a \lesssim H_I$)
\begin{align}
g_{a\gamma\gamma} &>
1.4 \times 10^{-18}
\,\mathrm{GeV}^{-1}\,\left(\frac{m_a}{10^{10} \,\mathrm{GeV}}\right)^{-\frac{1}{3}} \left(\frac{\alpha_\mathrm{GUT}^{-1}}{25}\right)^{-\frac{2}{3}} \notag \\
&\times \theta_i^{\frac{2}{3}}n_\mathrm{EM}^{\frac{2}{3}}
\left(\frac{M_\nu}{10^{13} \,\mathrm{GeV}} \right)^{-\frac{1}{3}}\,,
\label{eq:gayy_min_leptogenesis}
\end{align}
with $n_{\rm EM}$ the heavy axion's anomaly coefficient (see the SM).

\clearpage
\unappendix
\onecolumngrid

\begin{center}
  \textbf{\large Supplementary Material for 
  ``Heterotic String Theory Suggests
  a QCD Axion Near 0.5 neV''
  }\\[.2cm]
  {Joshua N.\ Benabou, Giulio Alvise Dainelli, Mario Reig,
  and Benjamin R.\ Safdi}
\end{center}

This Supplementary Material (SM) is organized as follows.
Sec.~\ref{SM:theory} provides extended details of the  4D axion
EFT in weakly coupled heterotic string theory. We discuss there PQ quality, gaugino condensation, and NS5-brane effects.
Sec.~\ref{SM:anomaly_coeff} discusses large anomaly coefficients and perturbative unitarity bounds on the QCD axion mass.
Sec.~\ref{SM:KS_scans} gives extended methodology and results for the scan of heterotic compactifications in the KS
ensemble.
Sec.~\ref{SM:leptogenesis} gives details on leptogenesis in the heterotic axiverse.
Sec.~\ref{SM:non-toric} discusses the QCD axion mass in heterotic string theory compactified on non-toric CY 3-folds.
Sec.~\ref{SM:duality_typeI} discusses the QCD axion mass through the lens of the duality between heterotic string theory, Type I string theory, and M theory.
Sec.~\ref{SM:duality_Ftheory} discusses expectations for the QCD axion mass in F-theory.
Lastly, Sec.~\ref{SM:figs} collects supplementary figures that illustrate analysis variations.

\setcounter{secnumdepth}{3}
\section{Axions from heterotic string theory}
\label{SM:theory}

In heterotic string theory axions may arise from multiple sources, including dimensional reduction of the bulk $B$-field $B_2$, its 10D dual $B_6$, field-theory axions from complex scalars, and NS5-brane axions from a self-dual 2-form $\tilde{B}_2$ on the NS5 worldvolume (see \cite{Buchbinder:2014qca,Agrawal:2024ejr,Reig:2025dqb,Leedom:2025mlr} for recent studies).  In this work we focus on the bulk $B_2$ (MD) and $B_6$ (MI) axions, \textit{i.e.} closed string modes present in the perturbative theory. NS5-brane axions will be briefly discussed later in this section. Some aspects of the heterotic axion EFT reviewed in this section are also presented in Ref. \cite{Leedom:2025mlr}.

We assume a compactification on $X_6 \times M_4$ with $X_6$ a CY 3-fold. Let us focus on heterotic $E_8 \times E_8$ (we comment on the case of $SO(32)$ later on). The gauge invariant field strength associated to the $2$-form field $B_2$ is modified by the Chern-Simons (CS) 3-form \cite{Polchinski:1998rr}:
\begin{equation}
H=d B_2-\frac{\alpha^{\prime}}{4}\left(\omega_{3}(A)-\omega_{3}(\Omega)\right)\,,
\end{equation}
where $\omega_{3}(A)$, $\omega_{3}(\Omega)$ are the Yang-Mills and Lorentz CS 3-forms, associated to the gauge field $A$ and the spin connection $\Omega$, respectively \cite{Polchinski:1998rr}. 
In the absence of NS5-branes, $H$ satisfies the modified Bianchi identity
\begin{equation}\label{eq:Bianchi_identity_with_field_strength}
d H=\frac{\alpha^{\prime}}{4}(\operatorname{Tr} R \wedge R-\operatorname{Tr} F \wedge F)\,,
\end{equation}
with $R$ the  2-form curvature of the spin-connection and  $F=dA + A\wedge A$ the gauge field strength.  In terms of Chern classes the Bianchi identity reads
\begin{equation}
c_2(V)+[W]=c_2(T X_6)\,,
\label{eq:full_bianchi}
\end{equation}
with $V$ the vector bundle associated to the gauge group, $T X_6$ the tangent bundle of $X_6$, and $[W]$ the curve class wrapped by NS5 branes. In the following, for concreteness we assume spacetime-filling NS5 branes are not present such that $[W]=0$. We do not expect that our conclusions about the possible range of the QCD axion mass are modified by relaxing this assumption, as we discuss later.

The heterotic gauge fields live in a vector bundle $V$ over $X_6$; in general the specification of $V$ restricts the possible choices of $X_6$. The situation is  simplified by assuming a ``standard embedding" \cite{Green:1987mn}: {\it i.e.}, using that the structure group of $TX_6$ is $SU(3)$, for any CY 3-fold we may choose $V=TX_6$, such that the spin connection is embedded in the gauge connection of a single $E_8$ factor via $SU(3) \hookrightarrow SU(3) \times E_6 \hookrightarrow E_8$. The result is that in 4D the  $E_8$ gauge group is broken to $E_6$.  In general, including non-standard embeddings of the spin connection, if the structure group of the gauge bundle is $G_V$, the unbroken 4D gauge group is $G_{4 \mathrm{D}}=\operatorname{Comm}_{E_8}\left(G_V\right)$; this allows more freedom for phenomenologically viable GUTs.
Note that our main conclusions are independent of the choice of embedding; for computing the shift of the QCD axion mass from the MI value due to light MD axions we assume the standard embedding for concreteness.

\subsection{The model independent axion}
The couplings to gauge bosons of the MI axion $a$ follow from the Bianchi identity~\eqref{eq:Bianchi_identity_with_field_strength}. The MI axion couples universally to all the unbroken gauge groups in the 4D EFT and has a decay constant  given by \eqref{eq:MI_fa_main}.
In the case where the MI axion becomes the QCD axion this leads to a sharp prediction for the QCD axion mass around 
\begin{equation}
    m_a\in [5.2,8.3] \times 10^{-10}\,\mathrm{eV}\,,
    \label{eq:MI_mass}
\end{equation}
with the range resulting from model dependence in the value of $\alpha_{\rm GUT}^{-1} \in [25,40]$ as discussed in the main Letter; if instead $\alpha_{\rm GUT}^{-1} \in [25,30]$ as in SUSY GUTs, this narrows to $[5.2,6.3] \times 10^{-10}$ eV.

In this work we assume the QCD axion solves the Strong $CP$ problem.\footnote{See~\cite{Benabou:2025viy} for a recent discussion of other possible solutions.} 
It is natural to ask, however, whether relaxing this assumption allows one to lower the MI mass by taking  $\alpha_\mathrm{GUT}$
even larger than $\sim 1/25$. For example, adding full GUT multiplets increases $\alpha_\mathrm{GUT}$ while leaving the unification scale unchanged. In any scenario, the minimal MI mass allowed by increasing $\alpha_\mathrm{GUT}$ is  $4.5 \times 10^{-10}$ eV, which is only slightly below the lower bound in \eqref{eq:MI_mass}. This is because, if $\alpha_\mathrm{GUT}>1/21$, the Euclidean NS5-brane instanton contribution to the MI axion mass exceeds the QCD instanton contribution.\footnote{In principle, gravitational instantons can contribute to the MI axion potential \cite{Svrcek:2006yi}. 
It is presently unknown how to calculate this effect from first-principles, but estimates in Ref.~\cite{Alonso:2017avz} suggest that it is subdominant to the NS5-brane contribution.} In detail, we require
\begin{align}\label{eq:instanton_scale_NS5}
\kappa\frac{16\pi}{\alpha_\mathrm{GUT}}m_{3 / 2} M_{\rm GUT}^3 e^{-2 \pi / \alpha_{\mathrm{GUT}}} \approx \Lambda_\mathrm{NS5}^4 < \chi_\mathrm{top}\,.
\end{align}
This is not satisfied even if $m_{3/2}$ is as low as $10$ TeV unless $\alpha_\mathrm{GUT} < 1/21$. Note that if the dominant contribution to the MI axion mass is from the NS5-brane instanton, then the MI axion lies strictly below the QCD axion line in the $(m_a,g_{a\gamma\gamma})$ plane.

The prefactor $\kappa$ appearing in~\eqref{eq:instanton_scale_NS5} is related to possible chiral suppression of the instanton potential and deserves careful discussion. 
The prefactor $\kappa$ is related to the Pfaffian of non-perturbative superpotential contributions --- {\it i.e.}, the fermionic one-loop determinant. Pfaffians are generally  difficult to compute, as they depend on how the fermionic zero modes associated to the different moduli are lifted via mass insertions or interactions. In Type IIB, for example, the Pfaffian of the Dirac operator on a $D3$-brane instanton worldvolume, evaluated at its intersections with $D7$-branes, is a holomorphic function of complex structure moduli and charged matter field VEVs that is only tractable in special limits, and can even vanish at codimension-one loci in moduli space. 
Although NS5-branes are less understood than $D3$-brane instantons, the fact that they correspond to small gauge instantons of the 10D gauge group allows us to understand the induced axion potential at the qualitative level. In particular, one expects that once the Kähler and complex structure moduli are stabilized (as they should be given that they control the size of the CY and hence the GUT gauge coupling), the fact that higher dimensional operators like $qqql$ and 4-quark operators lift all the fermionic zero modes associated to charged matter suggests that $\kappa$ is $\mathcal{O}(1)$, with the gravitino mass (already explicit in~\eqref{eq:instanton_scale_NS5}) being the only suppression factor in the NS5-brane potential. For this reason, unless explicitly stated, we will take $\kappa \sim 1$ throughout.
A value $\kappa \ll 1$ would require either additional zero modes protected by an approximate symmetry, or a mechanism to suppress otherwise-allowed operators, neither of which is generic. Note that this is the most conservative assumption in terms of QCD axion quality: a smaller $\kappa$ would relax the upper bound on $\alpha_\mathrm{GUT}$ and would relax the lower bound on $m_a$ relative to ~\eqref{eq:MI_mass}.

Throughout this work we assume gravity-mediated SUSY, so that the soft/superpartner masses are of order $m_\mathrm{soft} \sim m_{3/2}$. Collider bounds on the superpartner masses then impose $m_{3/2}\gtrsim \mathcal{O}(1)$ TeV.
Gauge-mediated SUSY breaking in principle could allow for $m_{3/2}\ll m_{\rm soft}$ but  requires a messenger mass $M_\mathrm{mess} \ll M_\mathrm{Pl}$, which is difficult to realize in string theory, particularly in heterotic. 
While gravity mediation
is  the generic expectation in string constructions, in order to evaluate the robustness of \eqref{eq:MI_mass},
in Fig.~\ref{fig:minQCD} we consider both scenarios.

Finally, we note that even if $\kappa \ll 1$, one expects that non-perturbative corrections to the Kähler potential will induce an axion potential. These contributions are even harder to compute, but parametrically one expects that the amplitude of the axion potential in this case scales as\footnote{Note that for the Euclidean NS5-brane instanton contribution to the Kähler potential we do not expect suppression from fermionic zero-modes beyond that induced by a light gravitino mass. In general, non-perturbative corrections to the Kähler potential are less understood than in the case of corrections to the superpotential. In Type IIB, for example, a subset of these corrections are computed in the $\mathcal{N}=2$ theory before placing D-branes and before orientifolding. As such, they are insensitive to suppression from the D-brane position zero-modes and complex structure zero modes.} \cite{Conlon:2006tq,Robles-Llana:2007bbv}
\begin{align}
\Lambda_\mathrm{NS5}^4\sim m_{3 / 2}^2 M_s^2 e^{-2 \pi / \alpha_{\mathrm{GUT}}}\,.
\label{eq:Kähler_pot_ns5}
\end{align}
In this case, requiring the NS5-brane contribution to the QCD axion mass to be subdominant relative to the QCD instanton contribution bounds the mass to be above $3.6 \times 10^{-10}$ eV (taking the prefactor in \eqref{eq:Kähler_pot_ns5} to be unity).

\subsubsection*{Threshold corrections and the upper bound on $m_a$}
The upper bound on $m_a$ in \eqref{eq:MI_mass} corresponds to $\alpha_\mathrm{GUT}^{-1}=40$, which is where the SM gauge couplings approximately unify without SUSY. In more detail, running the 2-loop Standard Model gauge couplings (see Ref.~\cite{Arason:1991ic}) from $M_Z$ to high scales, 
the $U(1)_Y$ and $SU(3)_c$ couplings meet at\footnote{We use $M_Z = 91.1880 \pm 0.0020~{\rm GeV}$,  $\sin^2\theta_W(M_Z)=0.23122\pm0.00006$, $\alpha_s(M_Z)=0.1177\pm0.0009$, $\alpha^{-1}(M_Z)=127.930\pm0.008$~\cite{ParticleDataGroup:2024cfk}. The error is dominated by the uncertainty in $\alpha_s(M_Z)$.}  $\alpha_{13}^{-1}=40.48 \pm 0.02$
at the scale $\mu = (1.98 \pm 0.07) \times 10^{14}~{\rm GeV}$.

The other two combinations of Standard Model gauge couplings meet at even smaller values.  It is natural to ask whether $\alpha_\mathrm{GUT}$ could be made even smaller than $\alpha_{13}$ via, {\it e.g.}, threshold corrections from massive particles appearing at intermediate energy scales.  As we discuss below, this cannot occur. Furthermore, note that grand unification at the scale $\mu_{13}$ is  naively in conflict with proton decay (although in principle this can be avoided via localization of matter in extra dimensions). To respect proton decay constraints, unification should instead occur at $\mu \sim M_\mathrm{GUT}=2\times 10^{16}$ GeV
~\cite{Hisano:2022qll,Super-Kamiokande:2020wjk}.
In this case, the gauge coupling at the unification scale is bounded above by the Standard Model value of $\alpha_1(M_\mathrm{GUT})^{-1}= 37.46 \pm 0.06$.

If additional particles beyond the MSSM are introduced with masses below $M_{\rm GUT}$, they modify the running of the gauge couplings and generally produce negative threshold corrections,
\begin{equation}
\Delta \alpha_i^{-1}
=
-\sum_\psi \frac{b_i^\psi}{2\pi}
\ln\!\left(\frac{M_{\rm GUT}}{M_\psi}\right),
\end{equation}
where $b_i^\psi$ is the contribution of the particle $\psi$ to the one–loop beta function coefficient, and $M_\psi$ the mass of $\psi$. When fermions or scalars are added, their effect is to reduce $\alpha_i^{-1}(M_{\rm GUT})$ with respect to the unified gauge coupling $\alpha_{\rm GUT}^{-1}$. Note that if the new particles fill complete GUT multiplets, the corrections are universal and increase the unified gauge coupling without changing the GUT scale.

Threshold corrections can also arise if some particles in the MSSM spectrum are heavier than the SUSY scale. In this case one can have $\Delta\alpha_i^{-1}>0$, which generally increases the inferred value of $\alpha_{\rm GUT}^{-1}$. In this sense, scenarios such as Split SUSY can be understood as using low-scale threshold corrections --- {\it i.e.}, using the fact that not all superpartners sit at the same scale --- to enhance the value of $\alpha_{\rm GUT}^{-1}$ with respect to the MSSM while keeping $M_{\rm GUT}$ approximately fixed.

Intermediate Pati-Salam breaking patterns, in which $SO(10) \to G_{\rm PS} \equiv SU(4)_C \times SU(2)_L \times SU(2)_R$ at a scale $M_U$ and $G_{\rm PS} \to G_{\rm SM}$ at a lower scale $M_I$, can populate a slightly broader range of $\alpha_{\rm GUT}^{-1}$. The PS gauge bosons running between $M_I$ and $M_U$ — in particular the enhanced asymptotic freedom of $SU(4)_C$ relative to $SU(3)_C$ and the presence of $SU(2)_R$ — drive $\alpha_{\rm GUT}^{-1}$ at $M_U$ above its one-step value, with PS-scale matter content providing a smaller modulation. In non-SUSY scenarios $\alpha_{\rm GUT}^{-1}$ typically lands in $[37, 45]$ for $M_I \sim 10^{11}\text{–}10^{13}$ GeV~\cite{Bertolini:2009qj,Hartmann:2014fya}, while SUSY versions span $[25, 35]$~\cite{Aulakh:2004hm,Hartmann:2014fya}. Wilson-line breakings of $E_8 \to G_{\rm PS}$ followed by a lower-scale breaking to the Standard Model are realized in a number of heterotic constructions~\cite{Blumenhagen:2006ux,Anderson:2011ns,Anderson:2012yf}. Note that on the extreme end $\alpha_\mathrm{GUT}^{-1}=45$, the MI mass is $9.4 \times 10^{-10}$ eV.

\begin{figure}
    \centering
\includegraphics[width=0.5\linewidth]{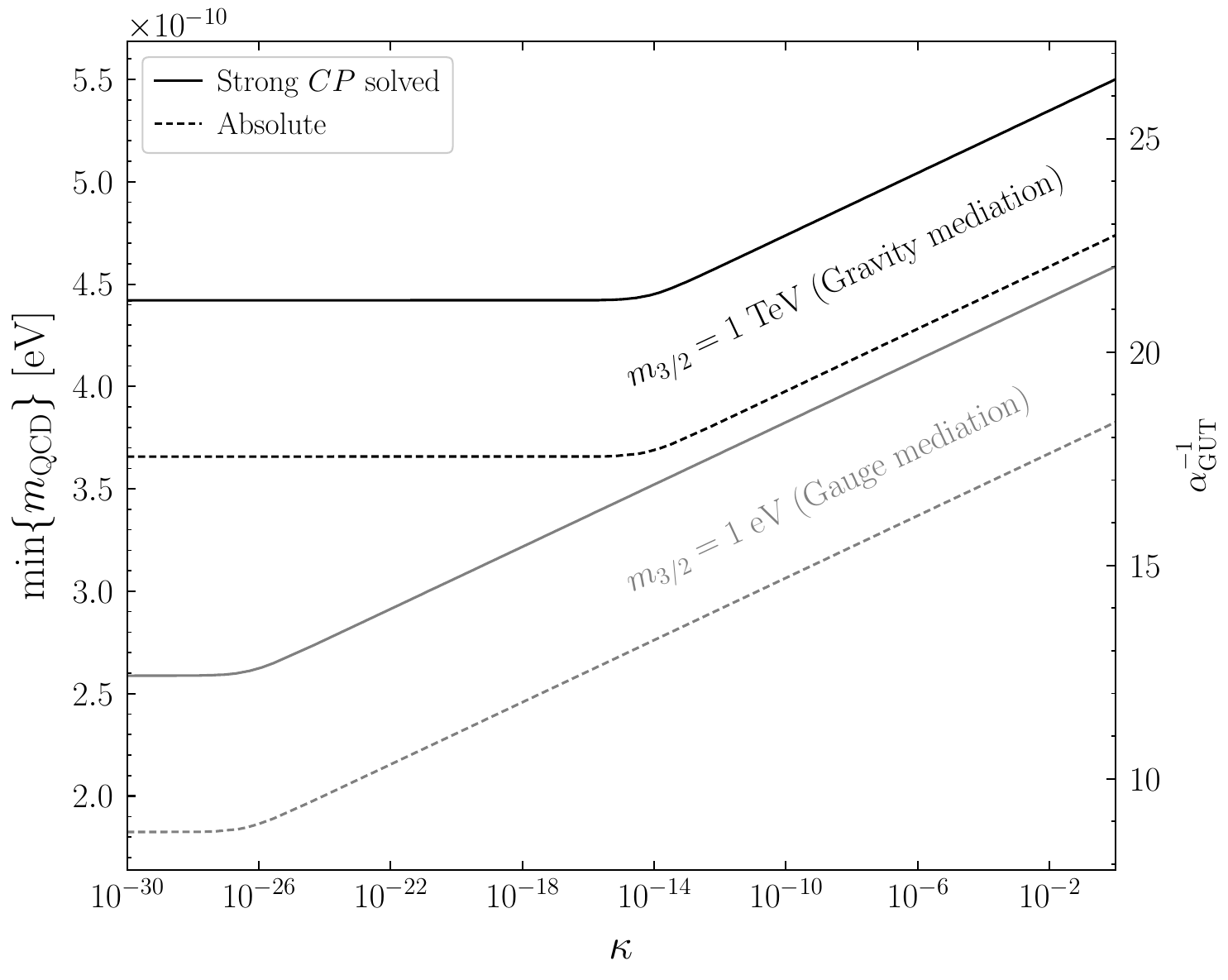}
    \caption{
    The smallest value of the QCD axion mass, varying the prefactor $\kappa$ of the superpotential contribution to the  NS5-brane instanton, assuming $m_{3/2}=$ 1 TeV  ($m_{3/2}=$ 1 eV ) as in gravity- (gauge-)mediated SUSY (black) (gray)\footnote{Note that in gauge mediation  $m_{3/2}$ is $\sim 1$ eV to $\sim 10^4$ TeV. To be maximally conservative we fix $m_{3/2} = 1$ eV, which
corresponds to the lowest messenger scale 
$\mathcal{O}(100)$ TeV allowed by direct collider 
searches while maintaining TeV-scale superpartners~\cite{Giudice:1998bp}.}. We compute the NS5-brane instanton potential as the sum of the superpotential \eqref{eq:instanton_scale_NS5} and the Kähler potential contribution \eqref{eq:Kähler_pot_ns5}. The smallest value is set by requiring $m_{3/2} \ge 1 $ TeV, and allowing $\alpha_\mathrm{GUT}$ to be the maximal value consistent with the QCD axion solving the Strong $CP$ problem (solid), or with the NS5 brane instanton being a subdominant contribution compared to QCD instantons to the QCD axion mass (dashed). This maximal value is shown on the right vertical axis. The minimum QCD axion mass tends towards a constant at small $\kappa$ where the Kähler potential contribution dominates over the superpotential contribution. Note that  in heterotic string theory, $\kappa = \mathcal{O}(1)$ is natural, though it may be smaller, while gauge mediation is difficult to achieve.
}\label{fig:minQCD}
\end{figure}
\subsection{Model-dependent axions}
Here we discuss the 4D axion EFT including MD axions. 
\\
\textbf{Coupling to gauge bosons.}
The couplings to gauge bosons of the MI and MD axions follow from the  Green-Schwarz counterterm~\cite{Choi:1985bz,Witten:1984Dg}.  Independently of $h^{1,1}$, only two linear combinations involving the MI and MD axions couple to gauge bosons as\footnote{In scenarios where there are non-trivial line bundles, there exists an additional axion linear combination that couples to gauge bosons via the anomaly. This combination is composed of MD axions, $\varphi = \sum_\alpha \tilde n_\alpha b_\alpha$, and its coupling to gauge bosons (\textit{i.e.} the coefficients $\tilde n_\alpha$) originates from holomorphic corrections to the gauge kinetic functions that only affect $U(1)$ gauge sectors \cite{Reig:2025dqb}. As $\varphi$ is only relevant for heterotic models where $U(1)_Y$ originates as a linear combination of $U(1)$ subgroups of both $E_8$ factors ~\cite{Blumenhagen:2006ux}, we do not consider this case further. At any rate, being composed of MD axions, $\varphi$ has a mass which is linked to the worldsheet instanton actions. For this reason we do not expect it to change our results.}~\cite{Agrawal:2024ejr,Reig:2025dqb}
\begin{equation}
    \mathcal{L} \supset \frac{\theta_1}{8\pi^2}\text{tr}_1F^2+\frac{\theta_2}{8\pi^2}\text{tr}_2F^2 \,.
    \label{eq:def_thetas_SM}
\end{equation}
Here $\text{tr}_{1,2} F^2$ contains the unbroken gauge groups from the first and second $E_8$ in the 4D EFT at the compactification scale after taking into account the embedding of the vector bundle and (possibly) discrete Wilson lines.

Since, for simplicity, we consider embedding of the Standard Model into the first $E_8$, only $\theta_1$ couples to Standard Model gauge bosons via the anomaly, and it does so in a GUT-symmetric way and becomes the QCD axion. Additional axions can couple to gauge bosons via mass mixing with this linear combination.
The linear combination coupled to gauge bosons from the first $E_8$ is then given by \eqref{eq:QCD_combo}.
The $n_i$ are integer MD axion anomaly coefficients that depend on the vector bundle, given by
\begin{align}
n_i & =\int_{X_6} \beta^{(i)}\left[\operatorname{tr}_1 \bar{F}^2-\frac{1}{2} \operatorname{tr} \bar{R}^2\right] 
=\int_{X_6} \beta^{(i)} \wedge\left(c_2\left(V_1\right)-\frac{1}{2} c_2(TX_6)\right)\,,
\label{eq:ni_def}
\end{align}
with $V_1$, $V_2$ the gauge bundles embedded in each $E_8$ factor.
For a standard embedding ($V_1=TX_6$), with trivial hidden bundle $V_2$, the Bianchi identity \eqref{eq:full_bianchi} then implies
\begin{align}
n_i = \frac{1}{2}\int_{X_6} \beta^{(i)} \wedge c_2(TX_6) \,.
\label{eq:anomaly_coeff}
\end{align} 

For a non-standard embedding, $n_i$ can be computed in explicit constructions ({\it e.g.}, monad bundles, spectral-cover models, or line-bundle sums). For certain manifolds, we can obtain a useful bound on $n_i$ which holds for any embedding, as follows. For SUSY-preserving bundles, we have $\int_{X_6} c_2(V_j) \wedge J \ge 0$ for $j=1,2$. If the divisor basis is \textit{nef} ({\it i.e.}, a basis in which each $\beta^{(i)}$
lies in the closure of the Kähler cone), then by continuity the previous inequality implies $\int_{X_6} \beta^{(i)} \wedge c_2\left(V_j\right) \geq 0$ for each $j=1,2$, so that also $\int_{X_6} \beta^{(i)} \wedge c_2(TX_6) \geq 0$. Consequently,
\begin{align}
|n_i| \le \frac{1}{2}\int_{X_6} \beta^{(i)} \wedge c_2(TX_6) \,,
\label{eq:anomaly_coeff_ineq}
\end{align} 
with the upper bound saturated by a standard embedding. In particular, as we discuss later, the divisor basis of  all manifolds in the ensemble of 375 ``favorable''  CICY manifolds used in this work is nef, as the Kähler cone is simply $\{t_i>0\}$. For these manifolds \eqref{eq:anomaly_coeff_ineq} holds. On the other hand, the divisor basis for a generic KS manifold is not nef and the anomaly coefficients are thus  not generally bounded by  \eqref{eq:anomaly_coeff_ineq}.

\textbf{Kähler potential.} The (leading order) Kähler potential for the dilaton and Kähler moduli is given by 
\begin{equation}
    K = - \mpl^2\ln(S+\bar S) - \mpl^2\ln\kappa\,,
\end{equation}
with $\kappa=\kappa_{ijk}t^it^jt^k=6\mathcal{V}_6$. To compute the MI axion decay constant in~\eqref{eq:MI_fa_main} we must compute $\frac{\partial^2K}{\partial S\partial \bar S}$, where $S=g_{\rm GUT}^{-2}+i\frac{a}{4\pi^2}$ is the $\mathcal{N}=1$ chiral superfield. For the MD axions, the field metric on Kähler moduli is given by
\begin{align}
    G_{ij} = - \frac{3}{4\pi^2}\mpl^2\left ( \frac{\kappa_{ij}}{\kappa}-\frac{\kappa_{i}\kappa_j}{\kappa^2} \right )\,,\,\,\, \text{with: }\kappa_{ij} = \kappa_{ijk}t^k \,\,\,\,\text{and: }\kappa_i = \kappa_{ijk}t^j t^k\,.
\label{eq:Kähler_metric}
\end{align}
This acts as the kinetic mixing matrix for MD axions.
Equivalently, one can define the Kähler metric as $\gamma_{ij}=\int_{X_6} \beta_i \wedge\star\beta_j$. Restoring the dimensionful parameters, the MD axion decay constants are given by
\begin{equation}
    f^{(i)}_{\rm MD}=\frac{\sqrt{\gamma_i}}{\sqrt{2\pi}}\frac{M_s}{g_s^2}\,,
\end{equation}
where $\gamma_i$ is an eigenvalue of $\gamma_{ij}$. Note that MD decay constants are largely insensitive to the details of the compactification and are generically close to the GUT scale.

\textbf{Worldsheet instantons.}
MD axions $b_i$ obtain masses from worldsheet instantons~\cite{Wen:1985jz}. These arise from Euclidean strings wrapping holomorphic 2-cycles in the CY and have no direct field-theory analogue in terms of small gauge instantons. In the absence of additional fermionic zero modes beyond the two universal ones (lifted by SUSY-breaking insertions), worldsheet instantons generate non-perturbative contributions to the superpotential $W$, which  depend on the complexified K\"ahler moduli.

We expand the K\"ahler form and the Kalb--Ramond field in a basis of harmonic $(1,1)$-forms $\beta_i$,
\begin{equation}
J = \sum_i t_i\,\beta_i,
\qquad
B_2 = \frac{1}{2\pi}\sum_ib_i\,\beta_i,
\end{equation}
where $t_i$ are the K\"ahler parameters and $b_i$ the MD axions. For an effective 2-cycle $C_\alpha$, we define the Mori charge matrix
\begin{equation}
Q_{i\alpha} \equiv \int_{C_\alpha} \beta_i ,
\end{equation}
so that the curve volume and the associated axion linear combination are
\begin{align}
\text{Vol}(C_\alpha)
&= \int_{C_\alpha} J
= t_i\, Q_{i\alpha}\,,\,\,\,
\int_{C_\alpha} B_2
= \frac{b_i}{2\pi}\, Q_{i\alpha}.
\end{align}
The complexified K\"ahler moduli are $T_i = t_i + i b_i$
and the instanton associated with $C_\alpha$ depends on the linear combination $T_\alpha = Q_{i\alpha} T_i$.

Worldsheet instantons are labeled by classes in the effective cone (Mori cone)
$\overline{\mathrm{NE}}(X_6)\ \subset\ H_2(X_6,\mathbb{R})$,
defined as the closed convex cone generated by effective curve classes. Equivalently, if $\{C_a\}$ denote the (extremal) generators of $\overline{\mathrm{NE}}(X_6)$, then any effective curve class can be written as
\begin{equation}
[C] \;=\; \sum_a n_a\, [C_a],
\qquad n_a \in \mathbb{R}_{\ge 0},
\end{equation}
(with $n_a\in \mathbb{Z}_{\ge 0}$ for integral curve classes). The K\"ahler cone is the dual cone in $H^{1,1}(X_6,\mathbb{R})$, {\it i.e.} the cone defined by positivity of effective curve volumes.
In terms of the Mori generators, the positivity conditions are the linear inequalities
\begin{equation}
\int_{C_a} J = t_i\, Q_{i a} > 0\qquad
\label{eq:def_Kählercone}
\end{equation}
for all effective curves $C_a$.

The worldsheet  instanton action is $S_{\rm ws}^{(\alpha)} = 2\pi\text{Vol}(C_\alpha) = 2\pi t_i Q_{i\alpha}$,
and summing over effective curve classes gives the  axion potential \eqref{eq:axion_pot_main}.
Assuming that the axion potential above comes from non-perturbative corrections to the superpotential, the UV scale is approximately given by 
\begin{equation}
(\Lambda^{(\alpha)}_\mathrm{UV})^4\approx A_\alpha m_{3/2}M_s^3 \,.
\end{equation}
Assuming gravity mediated SUSY breaking, $m_{3/2}\gtrsim \mathcal{O}(1)$ TeV. The prefactor $A_\alpha$ is determined by one-loop determinants and the zero-mode structure. In this work, we simply set $A_\alpha=1$.\footnote{For certain non-standard embeddings, it is possible that $A_\alpha=0$ due to Beasley-Witten cancellations \cite{Beasley:2003fx}. We do not study this possibility in this work.}
In the absence of non-perturbative corrections to the superpotential, an axion potential is still generated from non-perturbative corrections to the Kähler potential. While these are harder to compute, they are easier to generate as they do not require the saturation of all the zero modes.  In that case, however,  $\Lambda_\mathrm{UV}^4\approx m_{3/2}^2M_s^2$. 
\\
\\
\textbf{PQ quality.} 
Let us turn to the PQ quality of the QCD axion in heterotic constructions. First, if we ignore contributions to the potentials of the MI and MD axions other than from worldsheet instantons, then the quality of the QCD axion is perfect, as worldsheet instantons only contribute to the potentials of MD axions. In more detail, when the instanton expansion is truncated to the leading $h^{1,1}+1$ instantons, there generically appears a phase $\delta$ which cannot be removed via axion field redefinitions 
\begin{align}
V_\mathrm{MD}&=\sum_\alpha\left(\Lambda_{\mathrm{UV}}^{(\alpha)}\right)^4 \exp \left(-2 \pi t_i Q_{i \alpha}\right) \cos \left(Q_{i \alpha} b_i\right)
+ \left(\Lambda^{\cancel{\mathrm{PQ}}}_\mathrm{UV}\right)^4\exp \left(-2 \pi t_i Q_{i}'\right) \cos \left(Q_{i}' b_i +\delta \right)\,.
\end{align}
Minimizing this potential fixes the MD axion field values $b_i$.
The QCD axion potential is the sum of the above and the contribution from QCD instantons, $V_\mathrm{QCD}(\theta_1) + V_\mathrm{MD}(b_i)$, such that the MI axion field adjusts to minimize the full potential at $\theta_1=0$. 

However, as discussed previously, in reality Euclidean NS5-brane instantons wrapping the entire CY 3-fold generate non-perturbative contributions that explicitly break the continuous shift symmetry of the MI axion. 
These objects generate a non-perturbative superpotential which results in an axion potential of the form \eqref{eq:NS5_pot}.
This is shown in Fig.~\ref{fig:PQ_quality}. Note that for SUSY unification, $\alpha_\mathrm{GUT}^{-1}\lesssim 30$, which predicts a neutron EDM $d_N \gtrsim 10^{-40} \, e\cdot \mathrm{cm}$. The SNS nEDM experiment \cite{nEDM:2019qgk} projects sensitivity to $d_N=3\times10^{-28} \,e\cdot \mathrm{cm}$. The storage ring proton EDM experiment is projected to improve this further to $d_N \sim 10^{-29} \,e\cdot \mathrm{cm}$ \cite{pEDM:2022ytu}. Optimistically, measurements of the nuclear Schiff moment of radium-bearing molecules \cite{Arrowsmith-Kron:2023hcr,Wilkins:2023hua} are estimated to improve the current bound on  $|\bar{\theta}|$ by as much as 6 orders of magnitude, which translates to  $d_N\sim  10^{-32} \,e\cdot \mathrm{cm}$.\footnote{See \cite{Demirtas:2021gsq} for a discussion of the PQ quality of the QCD axion in compactifications on CY 3-folds in Type IIB string theory.}

In Fig.~\ref{fig:PQ_quality} we indicate the $(\alpha_{\rm GUT}^{-1}, m_{3/2})$ parameter space for three benchmark scenarios --- the TeV MSSM, Mini-Split SUSY, and Split SUSY --- obtained by
running of the gauge couplings through the relevant EFT thresholds, assuming gravity-mediated SUSY breaking. In the TeV MSSM, all superpartners sit at $\tilde{m}\sim m_{3/2}\in[1,10]$~TeV, bounded from below by LHC searches and from above by e.g., naturalness and the Higgs mass constraint~\cite{Aad:2012tfa,Chatrchyan:2012xdj}. In Mini-Split SUSY~\cite{Arvanitaki:2012ps}, scalars remain at $\tilde{m}\sim m_{3/2}\in[10^2,10^5]$~TeV while gauginos are lighter by a loop factor via anomaly mediation, $m_{\rm gaugino}\sim m_{3/2}/100$. In Split SUSY~\cite{Arkani-Hamed:2004ymt,Giudice:2004tc,Giudice:2011cg}, scalars decouple at $\tilde{m}\sim m_{3/2}$ up to $\sim 10^{10}$~GeV while gauginos and Higgsinos stay at the TeV scale.\footnote{For each scenario, the RGE fixes $\alpha_{\rm GUT}^{-1}$ as a function of $m_{3/2}$, which is related to the superpartner masses in the scenarios we consider. For the MSSM, we consider superpartner masses (hence $m_{3/2}$) at around few TeV, implying that $\alpha_{\rm GUT}^{-1}$ is nearly constant. For Mini-Split and Split SUSY, $\alpha_{\rm GUT}^{-1}$ drifts upward due larger superpartner masses. For simplicity we indicate in Fig.~\ref{fig:PQ_quality} by a rectangle the range of allowed values for $\alpha_{\rm GUT}^{-1}$ and $m_{3/2}$ independently, rather than making quantitative the underlying correlation. Furthermore, we assume $\mathcal{O}(1)$ $CP$-violating phases.}

\textbf{Gaugino condensation.}
We must also consider the effect of possible gaugino condensation from the second $E_8$. Upon dimensional reduction of the 10D action, a confining hidden sector plays two important roles. Firstly, the coupling between fluxes and gauginos induces a potential for the dilaton~\cite{DINE198555}. More importantly for us, a confining hidden sector induces a non-perturbative superpotential of the form
\begin{equation}
    W_\mathrm{NP}= -M_\mathrm{s}^3e^{-\frac{8\pi^2f_{\rm hidden}}{C_H}}\,,
\end{equation}
where $f_{\rm hidden}=S-n_iT_i$  
is the one-loop-corrected  gauge kinetic function of the second $E_8$, which contains the complex dilaton and the holomorphic corrections, 
and $C_H$ is the dual Coxeter number of the confining gauge group. The real part of $f_\mathrm{hidden}$ gives the UV gauge coupling of the second $E_8$, while the imaginary part is the axion linear combination $\theta_2$ that couples to the hidden sector. 

Consequently, gaugino condensation typically removes the entire $\theta_2$ linear combination from the spectrum of light axions (together with the scalar part, which also gains a mass). This axion gains a mass $m_{\theta_2}^2\sim \Lambda_H^4/f_{\theta_2}^2$, with $f_{\theta_2} \sim 10^{16} \, \,{\rm GeV}$ the decay constant and with $\Lambda_H$ roughly given by the confinement scale of the hidden sector; see Ref.~\cite{Reig:2025dqb} for the expected values of this scale for different gauge groups. 

In cases where all the MD axions obtain a large mass relative to the MI value (as in the ensembles of compactifications studied in this work, with a handful of exceptions), gaugino condensation makes the MI axion heavy and, consequently, the QCD axion does not solve the Strong $CP$ problem unless the minimum of the hidden sector axion potential coincides with the $CP$-conserving vacuum of QCD. Such scenarios do not  have any light axion (see Ref.~\cite{Leedom:2025mlr} for a detailed discussion). 
A simple way to avoid this obstruction is to break the second $E_8$ with a nontrivial vector bundle, or equivalently to Higgs it down to a non-confining subgroup, so that no hidden-sector gaugino condensate forms and the MI axion remains light.

By contrast, when at least one MD axion is light compared to the MI value, gaugino condensation can in principle be compatible with the QCD axion solving the Strong $CP$ problem. In detail, this requires the MD axion mass to satisfy $ m_a^{\mathrm{MD}}\lesssim10^{-5} m_{\mathrm{MI}}$. 
Let us consider the case where there is only one sufficiently light MD axion (note that we do not find cases with more than one such MD axion in our ensembles). After integrating out all other MD axions, the QCD axion is the linear combination of the MI and MD axions orthogonal to the one coupled to the gauge instanton of the confining sector,
\begin{align}
a_{\mathrm{QCD}}=\frac{n f_{\mathrm{MI}} a_{\mathrm{MI}}+f_{\mathrm{MD}} a_{\mathrm{MD}}}{\sqrt{n^2 f_{\mathrm{MI}}^2+f_{\mathrm{MD}}^2}} \,,
\end{align}
where $a_\mathrm{MI}$ ($a_\mathrm{MD}$) denotes the MI (MD) axion, and $n$ is the effective anomaly coefficient of the MD axion after transforming to the approximate mass eigenbasis. The decay constant associated to the QCD axion is then given by
\begin{equation}
    \frac{1}{f_{\mathrm{QCD}}^2}=\frac{4}{f_{\mathrm{MI}}^2+\frac{f_{\mathrm{MD}}^2}{n^2}} \,.
\label{eq:gaugino_cond_fqcd}
\end{equation}
In any scenario with $m_\mathrm{MD} \lesssim 10^{-5}m_\mathrm{MI}$, \eqref{eq:gaugino_cond_fqcd} gives $m_\mathrm{QCD}= 2 m_\mathrm{MI}/\sqrt{1 + f_\mathrm{MD}^2/(n^2 f_\mathrm{MI}^2)}$, 
which ranges from $2\,m_\mathrm{MI}$ as $f_\mathrm{MD}/n \to 0$ down to values below $m_\mathrm{MI}$ when $f_\mathrm{MD}>\sqrt{3}\,n \, f_\mathrm{MI}$. In the ensembles studied in this work, the latter condition is not satisfied (see Fig.~\ref{fig:gaugino_cond_mqcd}), so the bound $m_\mathrm{QCD} \geq m_\mathrm{MI}$ is preserved. In fact, we expect on general grounds that this condition is not satisfied in any heterotic axiverse. In particular, the electric axion weak gravity conjecture (see subsequent section) implies that $f_\mathrm{MD}/n\lesssim \sqrt{3/2}\mpl/S_\mathrm{inst}$ \cite{Benabou:2025kgx}, with $S_\mathrm{inst}$ the worldsheet instanton action associated to the light MD axion. For the MD axion to be light enough to not spoil PQ quality, we require $S_\mathrm{inst}\gtrsim 2\pi/\alpha_\mathrm{GUT}$, and thus $f_\mathrm{MD} \lesssim \sqrt{3}nf_\mathrm{MI}$.
\\\\
\textbf{Field theoretic axions.} In heterotic models with line bundles, the MI and MD axions may mix with axions from complex scalars~\cite{Choi:2011xt,Buchbinder:2014qca,Loladze:2025uvf} (see \cite{Petrossian-Byrne:2025mto} in the context of theories with open strings). Supersymmetric compactifications impose constraints on the VEV of these complex scalars, $|\Phi|$. In more detail, there exists a moduli-dependent D-term whose cancellation typically requires the VEV to take a non-zero value near $M_s$.
While near special regions in moduli space, $|\Phi|\ll M_s$, requiring that moduli fields are stabilized near this locus requires a large amount of tuning. Note that in any case, our lower bound on the QCD axion mass remains unchanged, as mixing with complex scalar axions can only increase the mass above the MI value (see main text).
\\\\
\textbf{Small string coupling limit $g_s^2\ll 1$.}
For a fixed GUT gauge coupling, in the limit $g_s \ll 1$, the total volume $\mathcal{V}_6$ becomes small, according to \eqref{eq:vol_GUT}. In the case $\mathcal{V}_6 <1$, barring special cancellations in the volume form, 
the 2-cycles become smaller than 1 in units of string length, which implies that the $\alpha^\prime$ expansion breaks down. In this limit, the SKC becomes very narrow and the QCD axion is increasingly aligned with the MI axion.

The MI axion is not affected by these effects because its mass comes from gauge instantons and from NS5-branes. The action of the latter, $S_{\rm NS5}\sim \frac{2\pi}{\alpha_{\rm GUT}}$, is not sensitive to  decreasing $g_s$. MD axions, on the other hand, have a shift-symmetry broken by worldsheet  instantons. In the limit of small $g_s$, the action of worldsheet  instantons is reduced as $g_s^2$ with respect to $S_{\rm NS5}$. Hence, the MD axions $b_i$ obtain a heavy mass  and can be integrated out. 
\\
\\
\textbf{Spacetime-filling NS5 branes.}
We have so far assumed that spacetime filling NS5-branes are absent. Here we justify that our conclusions concerning the QCD axion mass are unaffected by this assumption. NS5 branes modify the 4D axion EFT in two ways. Firstly, the anomaly coefficients for MD axions are shifted according to the modified Bianchi identity \eqref{eq:full_bianchi}. 

Secondly, the NS5 worldvolume contains its own self-dual 2-form $\tilde{B_2}$ which gives additional axions under dimensional reduction:
\begin{equation}
    \tilde{b}_r= \int_{C_2} \tilde{B}^{(r)}_2\,,
\end{equation}
with $C_2$ the effective 2-cycle wrapped by  the stack of NS5-branes. 
The non-perturbative axions $\tilde{b}_r$ also couple to gauge bosons. This can be deduced from the new Green-Schwarz-like counterterms, required to cancel anomalies~\cite{Blumenhagen:2006ux},
\begin{equation}
    S^{\rm GS}_{\tilde{B}}=\frac{1}{64\pi^3}N_r\int \tilde{B}^{(r)}_2\wedge (\text{tr}_1F^2- \text{tr}_2F^2)\,,
\end{equation}
indicating that, similar to the standard MD axions, the $\tilde b_r$ couple to gauge bosons of the different $E_8$ factors with a relative sign. 
This implies that the axion linear combinations coupled to gauge bosons in \eqref{eq:def_thetas_SM} are modified as
\begin{align}
     \theta_1 = a+\sum_i n_i^{(1)} b_i + \sum N_r\tilde{b}_r\,,\\
     \theta_2 = a+\sum_i n_i^{(2)} b_i - \sum N_r\tilde{b}_r\,.
\end{align}

The new non-perturbative 2-forms $\tilde B_2^{(r)}$ couple to non-critical strings. These are non-perturbative objects that break the shift-symmetry of $\tilde b_r$ and are better described in M-theory, where they correspond to M2-branes stretched between M5-branes or between an M5-brane and an $E_8$-brane. At large $g_s$, when the eleventh dimension is larger than the dimensions of the CY, the associated action is $S_\mathrm{non-crit} = \epsilon S_\mathrm{ws}$,
with $S_\mathrm{ws}$ the action of a worldsheet  instanton wrapping the same 2-cycle.
At strong coupling, the fact that $\epsilon<1$ can be understood as follows: fundamental strings correspond to M2-branes stretched between the two $E_8$-branes, and hence their length is generally larger than the length of a M2-brane between $E_8$ and M5-branes. This implies that the action of the worldsheet  instanton is generically larger than that of the non-critical string instanton for large $g_s$, which is the regime in which  these objects are best understood. 

Thus, in general we expect that the lightest axion which mixes with the MI axion, which could be one of the MD axions or a non-perturbative axion living in the worldvolume of an NS5-brane, is in fact a MD axion.  The primary effect of NS5-branes is therefore to modify the anomaly coefficient of that MD axion.
For further details on the couplings and masses of NS5-brane axions see \cite{Reig:2025dqb}.

\section{Large anomaly coefficients and unitarity}
\label{SM:anomaly_coeff}

In cases where a light MD axion mixes with the MI axion, the QCD axion decay constant depends on the MD axion anomaly coefficient \eqref{eq:fa_bound}. Here we show that perturbative partial-wave unitarity imposes an upper bound on the anomaly coefficient, and thus on the QCD axion mass. 

Let us first consider an axion coupling to QCD with
\begin{equation}
\mathcal{L} \supset N\frac{\alpha_s}{8 \pi} \frac{a}{f_a} G_{\mu \nu}^a \tilde{G}^{a \mu \nu}\,.
\end{equation}
Here $N$ is the integer QCD anomaly coefficient and $f_a$ is the axion decay constant (not the periodicity of the axion field).
Perturbative partial-wave unitarity imposes \cite{Benabou:2025kgx}
\begin{equation}
A(s) \sim \frac{\alpha_s^2}{64 \pi^2} 4 \frac{\left(N_c^2-1\right)}{\pi} \frac{N^2}{f_a^2} s \lesssim 1 \,.
\label{eq:unitarity}
\end{equation}
This relation, for a given Mandelstam parameter $s$, bounds the axion decay constant from below. In some models, as we show below, it also constrains the anomaly coefficient.

To gain intuition, let us consider a simple field-theory UV completion for the QCD axion, the KSVZ model. The axion arises as the Goldstone boson of $\Phi$ with the $U(1)$ symmetry spontaneously broken by the potential $V(\Phi)=\lambda\left(|\Phi|^2-\frac{f_a^2}{2}\right)^2$, after which the radial mode VEV is $\langle \Phi \rangle=f_a/\sqrt{2}$.
In this case, unitarity up to the scale $s=m_{\Phi}^2 = 2\lambda f_a^2$ imposes 
an upper bound on the anomaly coefficient
\begin{equation}\label{eq:bound_to_anom_coeff}
N<\sqrt{\frac{\pi^3}{\lambda}} \frac{1} {\alpha_s\left(m_{\Phi}\right)}\,,
\end{equation}
such that the upper bound may be as large as $\mathcal{O}(1000)$ if $\lambda$ is not tuned to be small.
We may interpret this bound as follows. The radial mode is the dynamical degree of freedom that regulates the amplitude. For large $N$, the effective decay constant $f_a/N$ becomes small. Alternatively, we can consider large $N$ as a limit where the axion interacts strongly with the instanton (whose charge is given by the anomaly coefficient). On the other hand, the mass of the radial mode does not change with $N$, such that for $N$ violating the inequality \eqref{eq:bound_to_anom_coeff}, we lose perturbative partial wave unitarity before the radial mode becomes dynamical, {\it i.e.} for $s < m_\Phi^2$. This theory, if consistent at all, requires a description that goes beyond the weakly coupled axion EFT that we consider. 

In the KSVZ model, a large anomaly coefficient arises from a large number of KSVZ quarks with color charge. These fermions also change the QCD gauge coupling running above the mass of the KSVZ quarks and, if asymptotic freedom is lost, can give Landau poles. If the KSVZ quark mass is comparable to the PQ scalar VEV, $m_Q\sim f_a$, the Landau pole will appear above this energy scale and the partial-wave unitarity is violated first. If the mass of KSVZ quarks is instead much smaller than the VEV, $m_Q=yf_a\ll f_a$, then QCD may be driven to a strongly-coupled regime before partial wave unitarity of the axion-mediated gluon scattering is lost. 

The situation in heterotic models differs from the KSVZ toy model. In the heterotic case, the anomaly coefficient $N$ is given by an $n_i$ in \eqref{eq:anomaly_coeff}. While $n_i$ does not directly relate to the number of charged fermions, as in the KSVZ model, in some cases it can contribute to the chiral index that determines the number of chiral fermions in the 4D EFT. This is the case for line bundle models, where the number of chiral representations is $\chi (L^q)=\int \frac{q^3}{6}c_1^3(L^q)+\frac{q}{12} c_1(L^q)\wedge c_2(TX_6)$ \cite{Reig:2025dqb}. 
For $c_2(TX_6)$ to contribute to the chiral index, the line bundles must have a non-vanishing first Chern class.
In the heterotic construction, the coefficients $n_i$ may be $\mathcal{O}(100)$. For example, the largest value of  $n_i$ we find in our ensemble of heterotic KS compactifications for $h^{1,1}\le 8$ is $ n_i = 80$, realized for a CY-3 fold with $h^{1,1}=3$, as we show in Fig.~\ref{fig:distribution_anomaly_coeffs}.

Imposing unitarity at the KK scale,  \eqref{eq:unitarity} reduces to (for the remainder of this section we write simply $n$ in place of $n_i$)
\begin{equation}
    f_a\geq n M_{\rm KK}\frac{\alpha_s}{\sqrt{2\pi^3}}\,,
    \label{eq:unitarity_fa}
\end{equation}
For  $f_a$ smaller than the bound of \eqref{eq:unitarity_fa},  the theory does not give a consistent weakly coupled axion EFT.
As $M_s\ge M_\mathrm{KK}$, we can obtain a conservative lower bound on $f_a$ by approximating $M_\mathrm{KK}\sim M_s$. Then, for example, an MD axion with decay constant comparable to the MI one, will violate \eqref{eq:unitarity_fa} unless $n\lesssim 90$. We do not find any examples in either of our ensembles of QCD axions aligned with light MD axions whose anomaly coefficient exceeds $26$  (see Table \ref{tab:manifolds}). On the other hand, it is possible that there exists a CY 3-fold heterotic compactification for which the QCD axion would be aligned with a light MD axion whose anomaly coefficients exceeds the bound \eqref{eq:bound_to_anom_coeff}.\footnote{In fact, as far as we are aware, it is an open question whether there are CY 3-folds realizing arbitrarily large values of the second Chern class components (this is open even for the ensemble of KS compactifications).} If such an example exists, it would give rise to a heterotic compactification with an axion EFT that fails to remain perturbative at high scales, and would be inconsistent with the Standard Model.  Importantly, \eqref{eq:unitarity_fa} gives an upper bound on $m_\mathrm{QCD}$ that is independent of $n$:
\begin{equation}
    m_\mathrm{QCD} \lesssim \left(3  \times 10^{-8}\, \mathrm{eV}\right)\left(\frac{\alpha_{\mathrm{UV}}^{-1}}{25}\right)^{\frac{3}{2}}  \,,
    \label{eq:unitarity_mQCD}
\end{equation}
as indicated in Fig. \ref{fig:meigens}. 

Ref.~\cite{Reece:2024wrn} conjectured the inequality \begin{align}\label{eq:string}
    f_a \gtrsim \frac{\sqrt{\alpha_{\mathrm{UV}}}}{2 \pi} M_s \,,
\end{align}
which follows from a conjectured upper bound on the tension of the associated axion string for any axion arising as the zero-mode of a higher-form gauge field.\footnote{The conjectured upper bound on the string tension is violated in some cases, {\it e.g.} by “co-scaling” axion strings; however, even in these examples the inequality \eqref{eq:string} continues to hold \cite{Reece:2025zva,ReeceRudeliusTudball:toappear}.
} 
As this bound concerns the tension of the axion string, it depends only on the fundamental period, $f_a$.  That is, we emphasize that the $f_a$ that appears in~\eqref{eq:string} is that defined through the periodicity of the axion field, while the scale that enters into the QCD axion mass determination is $f_a / n$.
Consequently, the conjecture in~\eqref{eq:string} translates to a conjectured bound on the QCD axion mass which is linear in $n$
\begin{equation}
m_a \lesssim 1.4 \,n \times 10^{-9} \mathrm{eV}\left(\frac{\alpha_{\mathrm{UV}}^{-1}}{25}\right)^{\frac{1}{2}} \,.
\label{eq:axion_string_tension_conjecture}
\end{equation}
Similarly, the magnetic weak gravity conjecture (WGC) imposes a bound on $m_\mathrm{QCD}$ which is linear in $n$. On the other hand, the electric WGC gives a lower bound on $m_a$, which is illustrated in Fig.~\ref{fig:meigens}. 
As shown in that figure, all of these inequalities are respected by the compactifications in our heterotic ensemble. See \cite{Benabou:2025kgx} for further discussion of these conjectured bounds.

\section{Heterotic compactifications from the Kreuzer-Skarke ensemble}
\label{SM:KS_scans}

\begin{table}[h]
\centering
\renewcommand{\arraystretch}{1.4}
\begin{tabular}{ccccccccc}
\hline\hline
FRST & Polytope & $\mathcal{V}_6$ & $K$ & $Q$ & SKC & Tip & $c_2(X_6)$ & $\max\{\mathrm{Vol}(C_2)\}$ \\
\hline
\\[-8pt]
1 &
$\left(\begin{smallmatrix}
 0 &  0 &  0 &  0 \\
-6 & -2 & -2 & -1 \\
 0 &  0 &  0 &  1 \\
 0 &  0 &  1 &  0 \\
 0 &  1 &  0 &  0 \\
 1 &  0 &  0 &  0 \\
-3 & -1 & -1 &  0 \\
-1 &  0 &  0 &  0
\end{smallmatrix}\right)$ &
$t_1 t_2^2 - \tfrac{4}{3}t_2^3$ &
$\left(\begin{smallmatrix}2&1\\1&0\end{smallmatrix}\right)$ &
$\left(\begin{smallmatrix}0&1\\1&0\\1&-2\end{smallmatrix}\right)$ &
$t_2\ge1,\ t_1\ge1,\ t_1-2t_2\ge1$ &
$(3,1)$ &
$(24,4)$ &
$\alpha_\mathrm{GUT}^{-1}-\frac{2}{3}$ \\[20pt]
2 &
$\left(\begin{smallmatrix}
 0 &  0 &  0 &  0 \\
 0 &  1 &  0 &  0 \\
 1 &  0 &  0 &  0 \\
-3 & -1 & -1 &  0 \\
-3 & -1 &  0 & -1 \\
 0 &  0 &  0 &  1 \\
 0 &  0 &  1 &  0 \\
-1 &  0 &  0 &  0
\end{smallmatrix}\right)$ &
$\tfrac{2}{3}t_1^3+t_1^2 t_2-\tfrac{1}{3}t_2^3$ &
$\left(\begin{smallmatrix}1&0\\1&-1\end{smallmatrix}\right)$ &
$\left(\begin{smallmatrix}1&1\\0&-1\\1&0\end{smallmatrix}\right)$ &
$t_1+t_2\ge1,\ {-t_2}\ge1,\ t_1\ge1$ &
$(2,-1)$ &
$(52,28)$ &
$\alpha_\mathrm{GUT}^{-1}-\frac{2}{3}$ \\[20pt]
3 &
$\left(\begin{smallmatrix}
 0 &  0 &  0 &  0 \\
 0 &  1 &  0 &  0 \\
 1 &  0 &  0 &  0 \\
-3 & -1 & -1 &  0 \\
-3 & -1 &  0 & -1 \\  
 0 &  0 &  0 &  1 \\
 0 &  0 &  1 &  0 \\
-1 &  0 &  0 &  0
\end{smallmatrix}\right)$ &
$\tfrac{2}{3}t_1^3+t_1^2 t_2$ &
$\left(\begin{smallmatrix}0&1\\1&0\end{smallmatrix}\right)$ &
$\left(\begin{smallmatrix}1&1\\0&1\\1&0\end{smallmatrix}\right)$ &
$t_1\ge1,\ t_2\ge1,\ t_1+t_2\ge1$ &
$(1,1)$ &
$(52,24)$ &
$\alpha_\mathrm{GUT}^{-1}-\frac{2}{3}$ \\[20pt]
4 &
$\left(\begin{smallmatrix}
 0 &  0 &  0 &  0 \\
 1 &  0 &  0 &  0 \\
-3 &  0 & -1 & -1 \\
 0 &  0 &  0 &  1 \\
 0 &  0 &  1 &  0 \\
-2 & -1 &  0 &  0 \\
 0 &  1 &  0 &  0 \\
-1 &  0 &  0 &  0
\end{smallmatrix}\right)$ &
$t_1^2 t_2$ &
$\left(\begin{smallmatrix}3&2\\0&1\end{smallmatrix}\right)$ &
$\left(\begin{smallmatrix}1&0\\0&1\\-2&3\end{smallmatrix}\right)$ &
$t_1\ge1,\ t_2\ge1,\ {-2t_1+3t_2}\ge1$ &
$(1,1)$ &
$(36,24)$ &
$\alpha_\mathrm{GUT}^{-1}$ \\[6pt]
\hline\hline
\end{tabular}
\caption{Geometric data for the four FRSTs with $h^{1,1}=2$ of KS polytopes for which the QCD axion mass deviates from the MI value. We list the polytope vertices (columns of the indicated matrix), the volume form, the K\"ahler cone generators $K$, the Mori charge matrix $Q$, the inequalities defining the SKC, the coordinates $(t_1,t_2)$ of the SKC tip, the second Chern class $c_2(X_6)$, and the maximal effective curve volume $\mathrm{max}\{\mathrm{Vol}(C_2)\}$ which appears in the $h^{1,1}=2$ leading worldsheet instantons. The SKC of manifold 1 is shown in Fig.~\ref{fig:skc_h11_2_optimal}. Note that manifolds 2 and 3 are equivalent triangulations of the same polytope and thus physically identical. Manifold 2 gives the same set of axion masses as manifold 1, only differing by a subleading instanton scale. All of the manifolds are K3-fibered over a $\mathbb{P}^1$ base.}
\label{tab:manifolds}
\end{table}

\begin{table}[h]                                                                                                                                
  \centering                                   
  \renewcommand{\arraystretch}{1.4}                                                                                                               
  \resizebox{\textwidth}{!}{
  \begin{tabular}{ccccccccc}                                                                                                                      
  \hline\hline                                                                                                                                    
  FRST & Polytope & $\mathcal{V}_6$ & $K$ & $Q$ & SKC & Tip & $c_2(X_6)$ & $\max\{\mathrm{Vol}(C_2)\}$ \\
  \hline                                                                                                                                          
  \\[-8pt]                                                        
  1 &                                                                                                                                             
  $\left(\begin{smallmatrix}                                      
   1 &  0 &  0 &  0 \\                                                                                                                            
  -3 & -1 & -1 & -2 \\
  -2 & -1 &  0 &  0 \\                                                                                                                            
  -2 &  0 & -1 &  0 \\                                            
   0 &  0 &  1 &  0 \\                                                                                                                            
   0 &  1 &  0 &  0 \\
   1 &  1 &  1 &  2                                                                                                                               
  \end{smallmatrix}\right)$ &                                                                                                                     
  $t_1 t_2 t_3$ &
  $\left(\begin{smallmatrix}0&1&0\\0&0&1\\1&1&1\end{smallmatrix}\right)$ &                                                                        
  $\left(\begin{smallmatrix}0&1&0\\1&0&0\\0&0&1\end{smallmatrix}\right)$ &                                                                        
  $t_1\ge1,\ t_2\ge1,\ t_3\ge1$ &                                                                                                                 
  $(1,1,1)$ &                                                                                                                                     
  $(12,12,12)$ &                                                                                                                                  
  $\alpha_\mathrm{GUT}^{-1}$ \\[20pt]                                                                                                             
  2 &
  $\left(\begin{smallmatrix}                                                                                                                      
   1 &  0 &  0 &  0 \\                                            
  -5 & -3 & -1 & -2 \\                                                                                                                            
   0 &  1 &  0 &  0 \\
   1 &  1 &  1 &  2 \\                                                                                                                            
  -2 &  0 & -1 &  0 \\                                            
   0 &  0 &  1 &  0                                                                                                                               
  \end{smallmatrix}\right)$ &
  $t_1 t_2 t_3 - t_2 t_3^{\,2}$ &                                                                                                                 
  $\left(\begin{smallmatrix}0&1&0\\2&1&1\\1&0&0\end{smallmatrix}\right)$ &                                                                        
  $\left(\begin{smallmatrix}0&0&1\\1&0&0\\0&1&0\\1&0&-2\end{smallmatrix}\right)$ &                                                                
  $t_3\ge1,\ t_1\ge1,\ t_2\ge1,\ t_1-2t_3\ge1$ &                                                                                                  
  $(3,1,1)$ &                                                                                                                                     
  $(12,12,0)$ &                                                   
  $\alpha_\mathrm{GUT}^{-1}-1 \xrightarrow{\mathrm{GV}} \alpha_\mathrm{GUT}^{-1}$ \\[20pt]                                                                                                           
  3 &                                                                                                                                             
  $\left(\begin{smallmatrix}                                      
   1 &  0 &  0 &  0 \\                                                                                                                            
  -2 & -1 &  0 &  0 \\                                            
  -2 &  0 & -1 &  0 \\                                                                                                                            
  -1 &  1 &  1 & -2 \\
   0 &  0 &  0 &  1 \\                                                                                                                            
   0 &  0 &  1 &  0 \\                                            
   0 &  1 &  0 &  0                                                                                                                               
  \end{smallmatrix}\right)$ &
  $t_1 t_2 t_3 - \tfrac{1}{2}\, t_2^{\,2}\, t_3 - \tfrac{1}{2}\, t_2\, t_3^{\,2}$ &                                                               
  $\left(\begin{smallmatrix}1&0&1\\1&1&0\\1&0&0\end{smallmatrix}\right)$ &                                                                        
  $\left(\begin{smallmatrix}1&0&-1\\1&-1&-1\\0&0&1\\1&0&0\\1&-1&0\\0&1&0\end{smallmatrix}\right)$ &
  $\begin{array}{@{}l@{}}                                                                                                                         
    t_1-t_3\ge1,\ t_1-t_2-t_3\ge1,\ t_3\ge1,\\                    
    t_1\ge1,\ t_1-t_2\ge1,\ t_2\ge1                                                                                                               
  \end{array}$ &                                                  
  $(3,1,1)$ &                                                                                                                                     
  $(12,12,12)$ &                                                  
  $\alpha_\mathrm{GUT}^{-1}-1$ \\[20pt]                                                                                                           
  \hline\hline                                                                                                                           
  \end{tabular}}                                                  
  \caption{As in Table \ref{tab:manifolds} but for the three $h^{1,1}=3$ FRSTs of KS    
  polytopes for which the QCD axion mass deviates non-negligibly from the MI value. For FRST 2, the maximal curve volume appearing in the         
  $h^{1,1}=3$ leading worldsheet instantons is modified if candidate curve classes with vanishing GV invariant are removed from the instanton     
  expansion (see text for details).}                                                                                                              
  \label{tab:manifolds_h11_3}                                     
  \end{table}

In this work we construct axiverses from compactifications of heterotic string theory on CY-3-fold hypersurfaces of toric varieties, sampled from the KS ensemble. The construction of compactifications using the \texttt{CYTools} software and the calculation of the QCD axion mass for a given compactification is summarized in the End Matter; here we give additional methodological details.

\subsection{QCD axion mass computation}

To compute worldsheet instanton actions, we first identify the Mori cone of effective curve classes
$\overline{NE}(X_6)$, computed in our chosen curve-class basis using the \texttt{CYTools} routine  \texttt{mori\_cone\_cap} (see Ref.~\cite{Demirtas:2018akl} for details).\footnote{This is an improved approximation of the Mori cone relative to previous works, which use \texttt{toric\_mori\_cone} to approximate the cone from the ambient toric variety, see Ref.~\cite{Demirtas:2018akl}. In particular, using the latter algorithm underestimates the number of heterotic-compatible FRSTs listed in Table \ref{tab:number_polytops_triangulations} by $\mathcal{O}(1)$ factors. However, for ease of computation, in Figs. \ref{fig:Nhet_alphaGUT_dependence},  \ref{fig:Nhet_gs_dependence}, and \ref{fig:Nhet_maxC2vol_dependence} only we use \texttt{toric\_mori\_cone} to construct the heterotic ensemble.} In our fiducial analysis, we generate the candidate curves supporting worldsheet instantons from positive linear combinations of extremal rays of $\overline{NE}(X_6)$. Note that while the extremal rays of $\overline{NE}(X_6)$ generate the cone over
$\mathbb{R}_{\ge 0}$, worldsheet instantons are labeled by \emph{integral} effective classes
$\beta\in \overline{NE}(X_6)\cap H_2(X_6,\mathbb{Z})$.
For a rational polyhedral cone, the semigroup of lattice points $\overline{NE}(X_6)\cap \mathbb{Z}^k$
need not be generated by the primitive ray generators alone: there can exist additional
indecomposable lattice points in the interior of the cone.
To account for all primitive effective degrees, we therefore compute the \emph{Hilbert basis} of $\overline{NE}(X_6)$
, {\it i.e.}\ the minimal set of lattice points in the cone such that every
integral effective class is a nonnegative integer combination of Hilbert basis elements, using
\texttt{hilbert\_basis}. Using the Hilbert basis elements as candidate curve degrees supporting
worldsheet instanton contributions -- which is a conservative choice\footnote{But note that not every element spanned by the Hilbert basis is necessarily effective \cite{Gendler:2026uux}.} -- we find no qualitative changes to our main conclusions (in particular, to the data in  Table \ref{tab:number_polytops_triangulations} and the axion mass spectra for the manifolds listed in Tables
\ref{tab:manifolds} and \ref{tab:manifolds_h11_3}).\footnote{On the other hand, sub-leading instantons are generically different when using Hilbert basis elements as candidate curve degrees. For example, for FRST 1 in Table \ref{tab:manifolds}, the Mori charge matrix is given by $\begin{pmatrix} 0 & 1 \\ 1 & -2 \end{pmatrix}$ using the Hilbert basis.} 

Furthermore, if the genus-0 Gopakumar-Vafa (GV) invariant $n^0_\beta$ \cite{Gopakumar:1998ii,Gopakumar:1998jq}
of a given curve class $\beta$ vanishes then it does not contain a holomorphic representative  and consequently does not support a worldsheet instanton contributing to the superpotential at leading order (note that the converse does not necessarily hold).\footnote{See Ref.~\cite{Alim:2021vhs} for a recent study of vanishing GVs on integer lattices within the Mori cone.} Accounting for this generically leads to larger worldsheet instanton actions (and thus lighter MD axions) appearing in the axion potential compared to the case where this condition is ignored. In principle, this could allow for MD axions sufficiently light to mix with the MI axion. In our fiducial analysis we ignore this effect.\footnote{A proper treatment would require accounting for the dependence of worldsheet instanton pre-potentials on GV invariants, which we leave to future work.} We verify however, that this effect is unlikely to qualitatively modify our results. To do so, we use the \texttt{CYTools} method \texttt{compute\_gvs}, based on algorithms introduced in \cite{Demirtas:2023als}. We verify for $h^{1,1} \le 3$ that, accounting for vanishing GV invariants, the set of FRSTs for which the QCD axion mass deviates from the MI value remains the same as that listed in Tables \ref{tab:manifolds} and \ref{tab:manifolds_h11_3}. The axion mass spectrum is  only affected for FRST 2 in Table \ref{tab:manifolds_h11_3}:  at $(t_1,t_2,t_3)=(\alpha_\mathrm{GUT}^{-1}+1,1,1)$ its leading Mori rays
are $\{(0,1,0),(0,0,1),(1,0,-2)\}$  with areas $\{1,1,\alpha_\mathrm{GUT}^{-1}-1\}$, the third ray has $n^0_\beta=0$; removing this candidate curve class from the instanton expansion substitutes a holomorphic subleading class and the maximal curve volume changes from $\alpha_\mathrm{GUT}^{-1}-1$ to
$\alpha_\mathrm{GUT}^{-1}$.

In our fiducial analysis, we assume a standard-embedding, such that anomaly coefficients are easily determined from \texttt{CYTools} as the integral of the second Chern class over the prime effective divisors (using \texttt{second\_chern\_class}). 
In heterotic compactifications, standard embedding ensures that the number $N_\mathrm{gen}$ of chiral $E_6$ generations  is fixed by the Euler characteristic of the manifold: $N_\mathrm{gen}=|\chi|/2$.  The number of Standard Model generations is given by the number of matter multiplets in the $\mathbf{27}$ representation of
$E_6$, up to smooth Wilson-line breaking on a freely acting quotient (note that Wilson line breaking requires a non-simply connected
manifold such that there are non-contractible
cycles). That is, if the CY 3-fold has a freely acting symmetry group $\Gamma$ of order $|\Gamma|$, the number of Standard Model generations is $N_\mathrm{SM}=N_\mathrm{gen}/|\Gamma|$~\cite{Green:1987mn}. We find that $17\%$ of the compactifications in our ensemble are consistent with obtaining three chiral generations via standard embedding (\textit{i.e.} with $\chi \in 6 \mathbb{Z}$), see Fig.~\ref{fig:euler_characteristic}.\footnote{This includes those listed in Table \ref{tab:manifolds}, which all have $\chi=-252$.} Note that for non-standard embeddings there is no such restriction on $\chi$, and the choice of embedding only enters in the computation of the QCD axion mass through the anomaly coefficients.

\begin{figure}
    \centering   \includegraphics[width=0.5\linewidth]{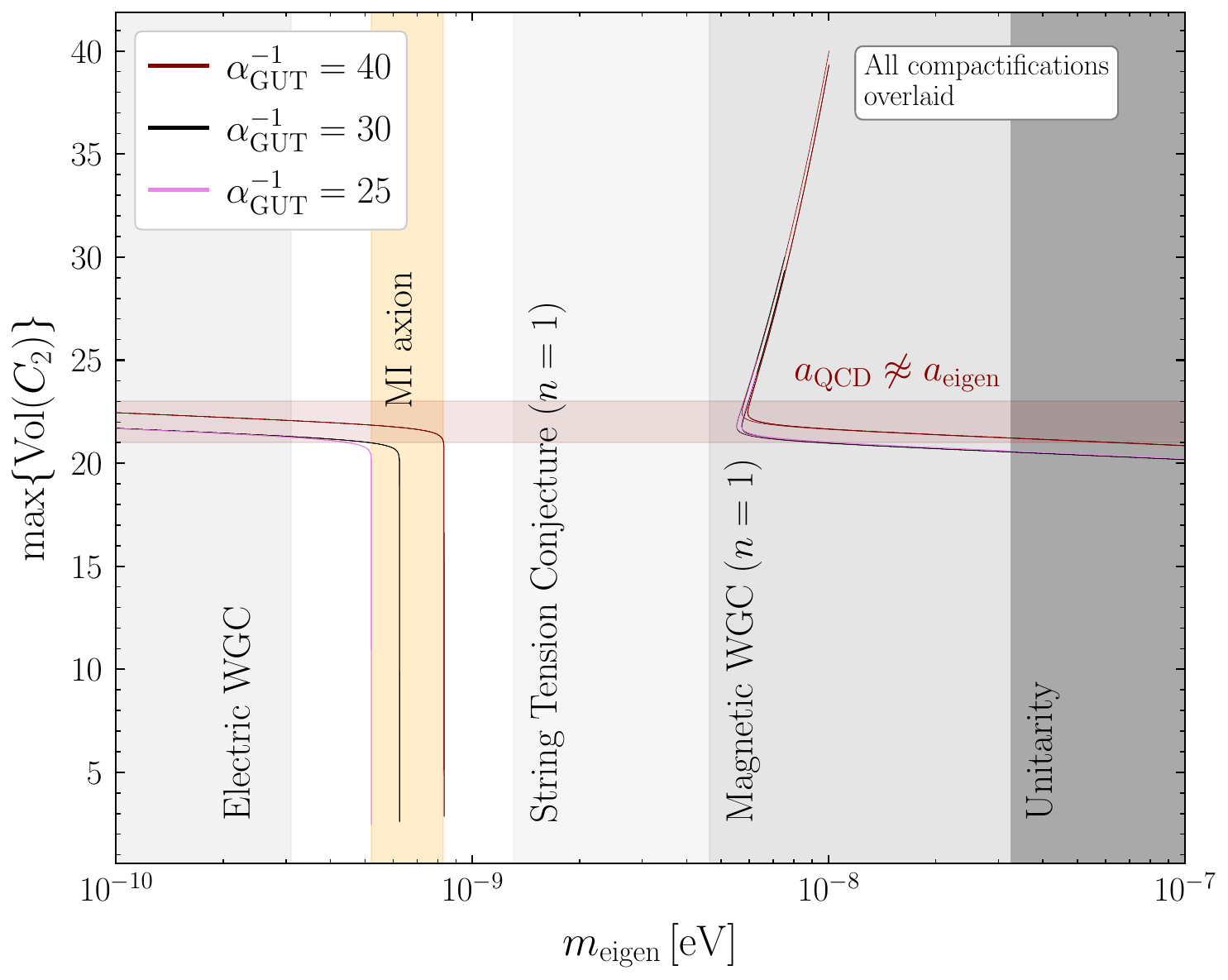}
    \caption{For the KS compactifications with $h^{1,1}=2$ for which the QCD axion mass deviates from the MI value  (see Table \ref{tab:manifolds}), the joint distribution of the masses of the two lightest axion mass-eigenstates and the maximal effective curve volume within the SKC (restricting to curve classes hosting the $h^{1,1}$ leading Euclidean worldsheet  instantons), assuming $g_s=1$. Note that the four FRSTs in Table \ref{tab:manifolds} are shown, yielding only two distinct trajectories. We indicate this assuming SUSY unification with $m_{3/2}=10$ TeV, for the extremal values of the allowed range of the UV gauge coupling ($\alpha_\mathrm{GUT}^{-1}=25,30$), as well as for high-scale SUSY with  $m_{3/2}=1$ PeV and $\alpha_{\mathrm{GUT}}^{-1}=40$.
    Following a given trajectory from smaller to larger curve volumes traverses an isosurface of $\mathcal{V}_6$ passing through the interior of the SKC, as shown in Fig. \ref{fig:skc_h11_2_optimal}.
     We indicate the regions excluded by the conjectured axion string tension bound \eqref{eq:axion_string_tension_conjecture}, the 0-form \textit{magnetic} WGC, and the \textit{electric} WGC (assuming $c=\sqrt{3/2}$, see text), assuming an anomaly coefficient $n=1$.\footnote{Note, however, that the anomaly coefficient for the axions in the compactifications shown is in all cases larger than unity, see Table ~\ref{tab:manifolds}.} We caution that the WGC bound applies for a single axion and is thus not directly applicable to this parameter space, while the string tension conjecture \eqref{eq:axion_string_tension_conjecture} is approximate. We also indicate the region excluded by unitarity of the QCD axion-mediate gluon scattering  \eqref{eq:unitarity}; note that this applies to the QCD axion, which is an approximate mass eigenstate only outside of the shaded horizontal band. For these bounds we take $\alpha_\mathrm{GUT}^{-1}=25$. 
}
    \label{fig:meigens}
\end{figure}

For each heterotic compactification, we compute the QCD axion mass as explained in the End Matter, with the result shown in Fig.~\ref{fig:meigens}. Specifically, we show the eigenvalues of the two lightest axion mass eigenstates for the compactifications with $h^{1,1}=2$ for which the QCD axion mass deviates from the MI value; the QCD axion is generically aligned with one of these eigenvalues, except for in a tuned region of moduli space, indicated by the shaded horizontal band.
In all cases we find that the points in moduli space which realize the largest effective curve volume lie on the boundary of the SKC (where at least one cycle size is equal to unity in string units). We show this for a compactification with $h^{1,1}=2$ in Fig.~\ref{fig:skc_h11_2_optimal}. In total, we find only 6 distinct KS manifolds for which the QCD axion mass deviates from the MI mass. More precisely, we identify 4 FRSTs with $h^{1,1}=2$, two of which are equivalent triangulations of the same polytope with identical axion mass spectra, and three physically distinct FRSTs with $h^{1,1}=3$. All of these manifolds are K3-fibered over a $\mathbb{P}^1$ base.\footnote{ To determine whether a CY 3-fold $X_6$ obtained from polytope $\Delta$ admits a K3 fibration, we
  check for the existence of a primitive lattice vector $m \in M$ satisfying two
  conditions~\cite{Avram:1996pj}: (i)~the projection of the reflexive dual polytope  $\Delta^*$ is
  $\pi_m(\Delta^*) = [-1,1]$, ensuring the base is $\mathbb{P}^1$, and (ii)~the
  fiber polytope $\Delta^* \cap \ker(m)$ is a three-dimensional reflexive polytope,
  ensuring the generic fiber is a K3 surface. For example, the three distinct polytopes in
  Table~\ref{tab:manifolds} are K3-fibered, with the unique fibration direction
  $m = (1,0,0,0)$ and a 30-point K3 fiber polytope.} Their properties are listed in Tables  \ref{tab:manifolds} and \ref{tab:manifolds_h11_3}, respectively.

\begin{figure}
    \centering    \includegraphics[width=0.5\linewidth]{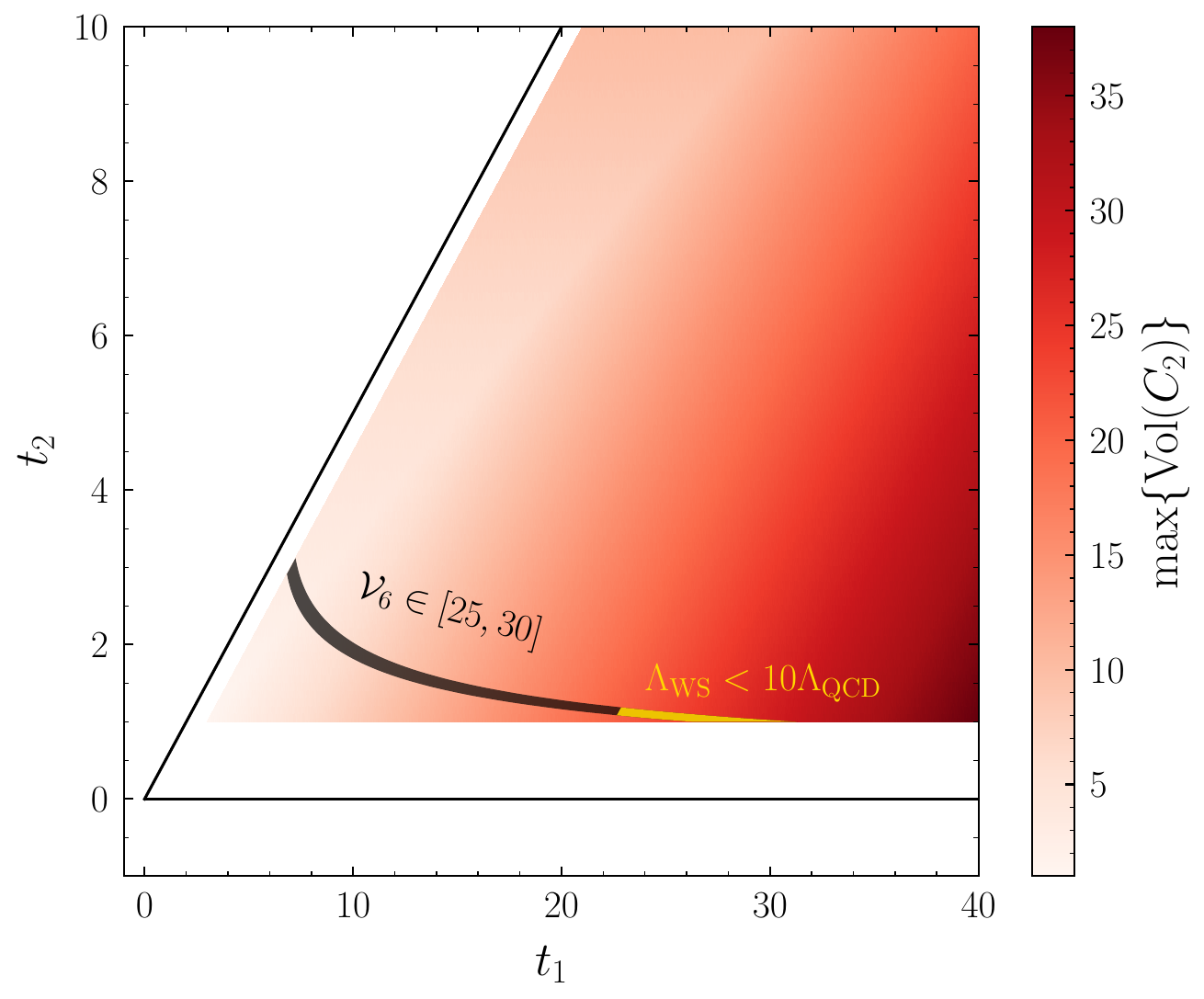}
    \caption{The Kähler cone, parametrized by its two generators  $t_1$ and $t_2$, and the SKC of the KS compactification with $h^{1,1}=2$ listed as FRST 1 in Table \ref{tab:manifolds} (the possible values of the QCD axion mass for the same compactification can be read off of Fig.~\ref{fig:meigens}). The region corresponding to the allowed values of $\alpha_\mathrm{GUT}$ is shaded black. We shade in gold the region of moduli space for which the QCD axion mass deviates non-negligibly from the MI value.
    }\label{fig:skc_h11_2_optimal}
\end{figure}

Our results suggest that there are likely points in moduli space which give deviations of the QCD axion mass from the MI value and which lie outside the SKC, where at least one curve volume is smaller than unity. Note that already for points which lie close to the boundary of the SKC, our calculations are possibly unreliable as corrections to the K\"ahler
potential (perturbative $\alpha'$, string loop, 
or non-perturbative corrections) which are suppressed
in the large volume limit, may become important \cite{Conlon:2006tq}. Note that, small-curve volume corrections to the decay constants of MD axions would not violate our lower bound on the QCD axion mass in weakly coupled heterotic string theory at the level of \eqref{eq:fa_bound}; on the other hand, one would need to account for corrections to the MI axion decay constant and  kinetic mixing between the MI and MD axions, which could arise beyond leading order.
As far as we are aware, it is not presently well-understood how to compute these corrections precisely. Nonetheless, for the purpose of illustration we consider relaxing the SKC to enforce only that the volume of effective curves exceed $c=0.5$, knowing that corrections are certainly important in this case. The distribution of QCD axion masses in this case is given in Fig.~\ref{fig:h11_2_light_masses_evolution_vary_SKC_c}. In this limit, the number of heterotic compactifications also grows to $\sim 10^5$ (see Fig.~\ref{fig:Nhet_maxC2vol_dependence}).

Let us now consider the $g_s$ dependence of our results. From \eqref{eq:vol_GUT}, smaller $g_s$ correspond to smaller volumes of the compactification manifold. The requirement of remaining within the SKC restricts the ensemble of compactifications to a smaller set for lower values of $g_s$. We show the number of KS compactifications and the maximal $h^{1,1}$ as a function of $g_s$ in Fig.~\ref{fig:Nhet_gs_dependence}. We consider values of $g_s$ larger than 1 ($g_s=1.2,1.5$). 
At these couplings tree-level SUGRA is no longer a reliable approximation, so the naive growth of both the number of heterotic compactifications and the range of QCD axion masses in this regime cannot be taken at face value. It is at best suggestive that the conclusions drawn from our weakly coupled search need not extend to strong coupling, a regime we address more properly via string dualities in Sec.~\ref{SM:duality_typeI}.

\begin{figure}
    \centering
\includegraphics[width=0.5\linewidth]{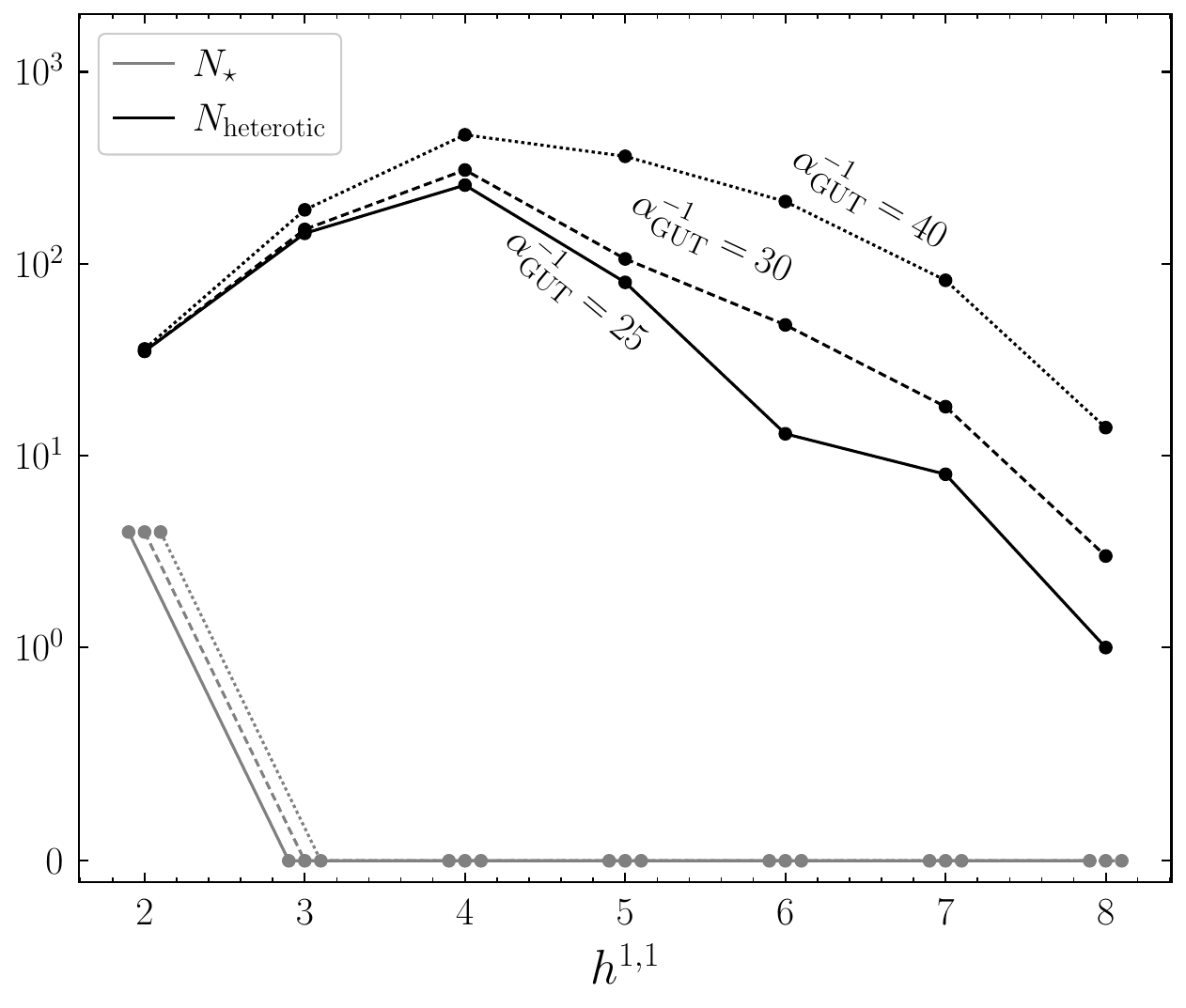}
    \caption{The number $N_\mathrm{heterotic}$ of KS CY 3-folds 
    compatible with heterotic compactifications (\textit{i.e.}, for which at least one point in Kähler moduli space satisfies \eqref{eq:vol_GUT}), varying $\alpha_\mathrm{GUT}$ (black), and the number $N_\mathrm{\star}$ of those compactifications for which the QCD axion mass deviates non-negligibly from the MI value (gray), for fixed $g_s=1$. We stagger the gray curves for clarity as they are identical. Note that for this figure, for computational ease, we approximate the Mori cone using \texttt{toric\_mori\_cone}, see text.}
\label{fig:Nhet_alphaGUT_dependence}
\end{figure}

Let us return to the compactifications under computational control. As discussed, all but $N_\star=4$ FRSTs in the KS heterotic ensemble have a QCD axion mass which is, to high accuracy, the MI value. In the tuned case where the instanton scale of the lightest MD axion coincides with the QCD confinement scale (see \eqref{eq:LambdaMD}), we have two axion mass eigenstates nearby in mass which couple sizably to QCD. The lighter eigenstate has a mass $m_{\rm ALP}$ below the MI axion mass; $m_{\rm ALP}$ is set by the dominant potential between that induced by NS5 branes and that of the MD potential. Furthermore, the photon coupling of this light ALP is mixing-suppressed by a factor $\sim m_{\rm ALP}^2/m_{\rm QCD}^2$. An example of such an ALP is represented in Fig.~\ref{fig:gayy_heterotic} by the black triangle below the QCD axion line.
(Note that this requires tuning in moduli space, see Fig.~\ref{fig:skc_h11_2_optimal}). 

This scenario gives a clear observational signature: the two axions could in principle be both detected in a lumped-element experiment; their misalignment abundances would also be modified due to mixing (see \cite{Gavela:2023tzu,Cyncynates:2021xzw,Dunsky:2025sgz,Ho:2018qur,Lee:2026umy} for a discussion). At large enough distances from the tip of the SKC, the QCD axion becomes aligned with the heavier mass eigenstate; for the four FRSTs for which this scenario can occur, the QCD axion mass in this case is always above $5\times 10^{-9}$ eV, see Fig.~\ref{fig:meigens}. The gap between this value and the MI mass is due to the anomaly coefficients having values $\mathcal{O}(10-30)$. Crucially, here we assume a standard embedding, such that anomaly coefficients are fixed by the second Chern classes of the manifold. For less trivial vector bundles, the anomaly coefficients are less constrained, such that in principle we may populate the entire range of masses between the MI mass and $\sim 10^{-8}$ eV, which is the upper bound if one assumes gauge coupling unification at the SUSY GUT scale \cite{Benabou:2025kgx}, as in Fig.~\ref{fig:vary_anomaly_coeff}.
\begin{figure}
    \centering   \includegraphics[width=0.5\linewidth]{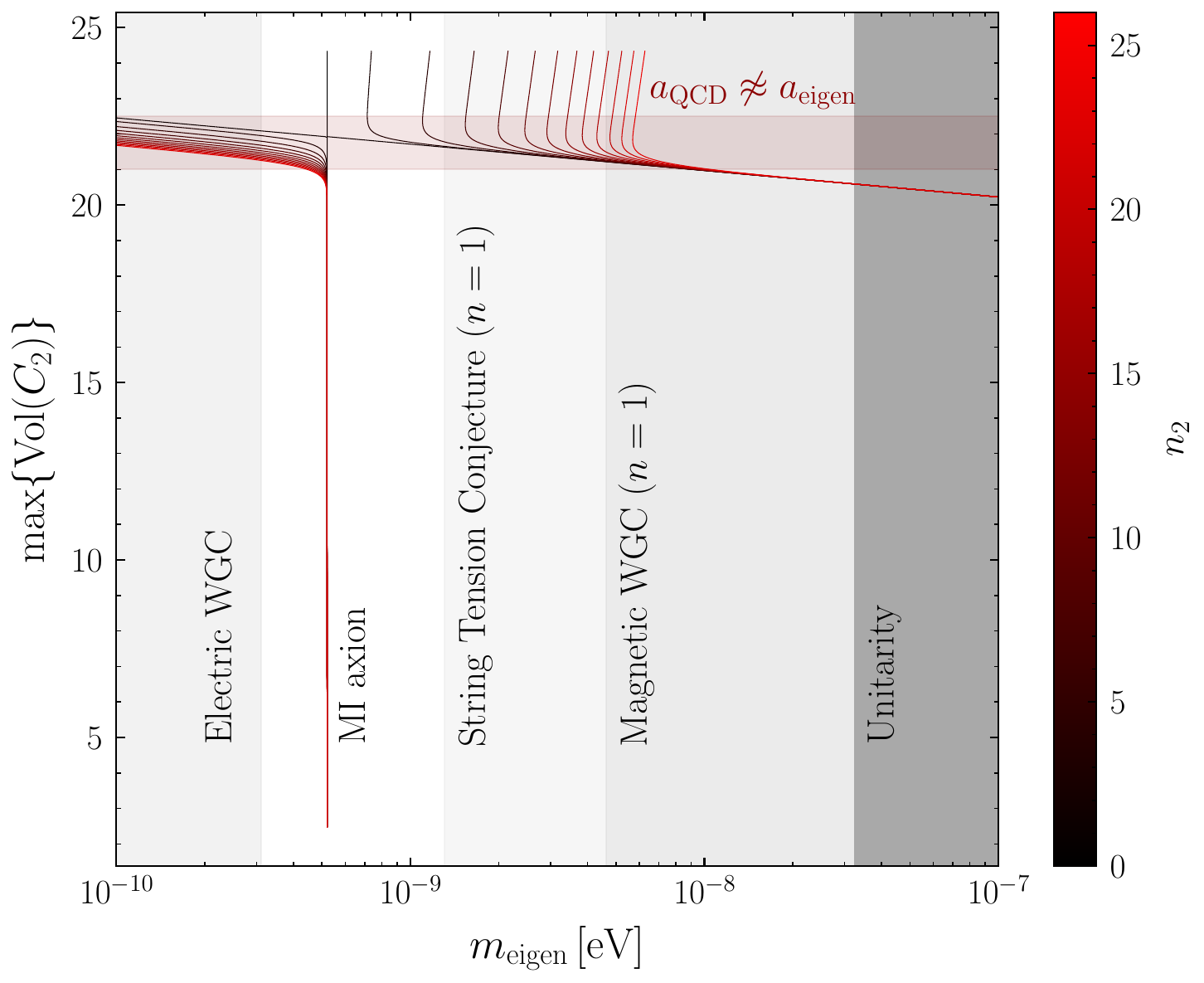}
    \caption{For  $g_s=1$, $\alpha_\mathrm{GUT}^{-1}=25$, and $m_{3/2}=10$ TeV, for the compactification with $h^{1,1}=2$ labeled FRST 3 in Table \ref{tab:manifolds}, the distribution of the two lightest axion mass-eigenstates masses, allowing for the anomaly coefficient $n_2$ (see \eqref{eq:anomaly_coeff}) to vary over all integers between 0 and its maximal value set by second Chern classes (note that the Kähler cone is the positive orthant for this manifold, such that the bound \eqref{eq:anomaly_coeff_ineq} applies). Plotting conventions are as in Fig.~\ref{fig:meigens}.
    }
\label{fig:vary_anomaly_coeff}
\end{figure}

\subsection{Heavy axion population}
\label{sec:heavy_axion}

In the ensembles studied in this work, the QCD axion is in almost all cases the lightest axion. The six distinct compactifications in Tables \ref{tab:manifolds} and \ref{tab:manifolds_h11_3} are an exception for which there exists an even lighter axion-like particle, lighter by a factor of at most $\sim 10^{13}$ in mass  (for $\alpha_\mathrm{GUT}^{-1}\lesssim 30$). This minimal axion mass is easily estimated from \eqref{eq:LambdaMD}, using that $f_a\sim M_s$, and that the maximal effective curve volume we find within the SKC across all compactifications is at most $\alpha_\mathrm{GUT}^{-1}$ (realized {\it e.g.} by FRST 4 in Table \ref{tab:manifolds}).
All of the other axions are heavy, with masses on the order of TeV or greater.  Note that the absence of axions much lighter than the MI value $\sim 5 \times 10^{-10}$ eV does not allow for signals from, {\it e.g.}, CMB birefringence \cite{Komatsu:2022nvu,Carralot:2026kps}
or fuzzy axion DM \cite{Sheridan:2024vtt} that may be expected in Type IIB compactifications under perturbative control.

We show axion-photon couplings in Fig.~\ref{fig:gayy_heterotic} for all KS heterotic compactifications, where we also indicate existing constraints from astrophysical and cosmological probes. We fix the point in moduli space to be along the ray connecting the Kähler cone origin to the tip of the SKC (except for the four FRSTs for which the QCD axion deviates from the MI value, for which we vary over the SKC). The support expands somewhat if we  vary over the full moduli space; we show this via Hamiltonian Monte Carlo of the SKC in Fig.~\ref{fig:gayy_heterotic_full_moduli_space}. In any GUT with standard embedding of the Standard Model gauge group, axion mass eigenstates with mass below that of the QCD axion must lie below the QCD axion line in $(m_a,g_{a\gamma\gamma})$ plane \cite{Agrawal:2022lsp,Agrawal:2024ejr}. This implies that existing constraints on low mass axion-like particles which are above the QCD axion line, such as from magnetic white dwarf polarization \cite{Dessert:2022yqq,Benabou:2025jcv} (see Ref.~\cite{Caputo:2024oqc} for a detailed review of astrophysical bounds), are irrelevant for heterotic axiverses. On the other hand, probes at masses above an eV which reach below this line are relevant and constrain the population of heavy axions. These include constraints which assume the ALP is all of the DM, such as from the CMB anisotropy \cite{Liu:2023nct}, as well as from decaying DM line searches using XMM Newton \cite{Foster:2021ngm,Boyarsky:2006ag, Boyarsky:2007ay, Boyarsky:2006fg}, INTEGRAL \cite{Calore:2022pks} and NuSTAR \cite{Perez:2016tcq, Roach:2022lgo, Ng:2019gch}. These constrain heavy axion down to the lower boundary of the region shaded in light gray in Fig.~\ref{fig:gayy_heterotic}.       

The heavy axion population in the ensembles studied in this work is qualitatively different from that in Type IIB constructions, such as those from \cite{Gendler:2023kjt, Sheridan:2024vtt,Benabou:2025kgx}. 
Most importantly, in Type IIB axiverse constructions using compactification on KS CY 3-folds, it is generic to populate a wide range of axion masses, including ultralight axions with masses below that of the QCD axion, as well as heavy axions, with an approximately log-uniform distribution of masses in between. Secondly, for the heavy axions in our heterotic ensemble, we have simply
\begin{align}
    g_{a\gamma\gamma}= \frac{8}{3}n_\mathrm{EM}\frac{\alpha_\mathrm{EM}}{2\pi f_a} \,,
    \label{eq:gayy}
\end{align} 
with $n_\mathrm{EM}\sim \mathcal{O}(1)$ the electromagnetic charge coefficient (obtained from the anomaly coefficients \eqref{eq:anomaly_coeff} after transforming to the approximate mass eigenbasis).\footnote{In Fig.~\ref{fig:gayy_heterotic} and similar figures we do not self-consistently account for the running of $\alpha_\mathrm{EM}$ (in reality, the QCD axion has $\alpha_\mathrm{EM}\sim 1/137$, while heavy axions have $\alpha_\mathrm{EM}\sim 1/27$). In addition, the QCD axion-photon coupling is corrected relative to \eqref{eq:gayy} via mixing with the neutral pion; however, for ease of visualization we also include this contribution for all the points along the blue dashed trajectories, though note that this is not physical.}
By contrast, in Type IIB axiverses constructed from KS ensemble, heavy axion photon couplings can be further suppressed by several orders of magnitude from weak kinetic mixing due to sparse intersection between a generic divisor and the divisor hosting QED \cite{Gendler:2023kjt}. This effect becomes more pronounced at larger $h^{1,1}$. Restricting to Type IIB axiverses compatible with unification at the SUSY GUT scale, for which $h^{1,1} \lesssim \mathcal{O}(50)$ \cite{Benabou:2025kgx}, this suppression is weak and photon couplings are comparable to the heterotic case, as in Fig.~\ref{fig:gayy_typeIIB}. 

Taken together, these facts mean that in the Type IIB KS axiverse, it is generic to have heavy, long-lived axions with masses in the $[\mathrm{MeV}, 10^{10} \,\mathrm{GeV}]$ range and photon couplings of order $10^{-25}\, \mathrm{GeV}^{-1}$. The present-day misalignment abundance of ALP DM for a stable axion is given by \cite{Blinov:2019rhb}
\begin{align}
\Omega_a h^2 & \simeq 0.12\left(\frac{f_a \theta_0}{1.9 \times 10^{13} \mathrm{GeV}}\right)^2\left(\frac{m_a}{1 \mu \mathrm{eV}}\right)^{1 / 2}
\left(\frac{90}{g_*\left(T_{\mathrm{osc}}\right)}\right)^{1 / 4} \,,
\end{align}
with $g_*$ the number of relativistic degrees of freedom. These axions therefore tend to be cosmologically problematic. For example, in Fig.~\ref{fig:gayy_typeIIB}, over $97\%$ of the sampled compactifications have at least one axion in conflict with decaying dark matter constraints (\textit{i.e.} lying in the light gray shaded region\footnote{Note that for these bounds we assume that the axion abundance is exactly that from misalignment, such that the constraints shut off for $m_a > H_I$.}). The heavy axion population in the heterotic axiverse generically decays before BBN and therefore does not overclose the Universe today (though it does generically lead to periods of early matter domination). On the other hand, these decays inject energy, which is constrained by the CMB and measurements of the light element abundances .

\section{Leptogenesis constraints}
\label{SM:leptogenesis}

In the End Matter we summarize our derivation of heavy axion constraints assuming thermal leptogenesis. For these constraints we use that heavy axions decay dominantly into photons and gluons (decays into lighter axions are suppressed \cite{Gendler:2023kjt}). We consider axions in our ensemble with $m_a \gtrsim 1.8\, \mathrm{GeV}$, such that the decay can be treated within perturbative QCD. With the photon coupling given by \eqref{eq:gayy}, the tree-level decay rate is given in the SM
\begin{align}
\Gamma=\frac{m_a^3 n_\mathrm{EM}^2} {64 \pi f_a^2}\bigg[
\left(\frac{\alpha_{\mathrm{EM}}(m_a) (E/N)}{2 \pi}\right)^2+ (N_c^2-1)\left(\frac{\alpha_\mathrm{s}(m_a) }{2 \pi}\right)^2 \bigg]
\,,
\end{align}
where $N_c=3$, $n_\mathrm{EM}$ is given  as in \eqref{eq:gayy}, and recall that $E/N=8/3$ for a GUT with standard embedding of the SM. 
We approximate $\alpha_{\mathrm{EM}}(m_a)\sim \alpha_\mathrm{s}(m_a) \sim \alpha_\mathrm{GUT}$.

In the End Matter we derive a bound on heavy axions assuming the abundance is produced via misalignment.\footnote{We do not consider the case of axion production from topological defects, as these are not generically expected to form for extra-dimensional axions \cite{Benabou:2023npn}} On the other hand, for $m_a>H_I$, the abundance is instead produced through freeze-in or freeze-out. Here we show that in this case, the heavy axion population in the heterotic axiverse is entirely compatible with thermal leptogenesis. Let us first consider freeze-in (our assumptions are as in the End Matter; in particular we fix the neutrino freeze-out temperature $M_\nu = T_\mathrm{fo} = 10^{13}$ GeV). At the maximal reheat temperature $10^{13}$ GeV, for all the axions in our ensembles we have $m_a \ll T_\mathrm{RH}$ such that the freeze-in is UV-dominated. For these temperatures, the dominant production channels are from scattering with gluons and quarks: $gg \to ga$, $q\bar{q} \to ga$, $qg\to qa$. The yield from these processes is (for $T_\mathrm{RH} \gg m_a$) \cite{Graf:2010tv,Salvio:2013iaa}
\begin{align}
    Y_a &\approx \frac{\zeta(3)  90 \sqrt{90}}{128 \pi^{10} g_{*S} g_*^{1 / 2}}  g_3^6  \ln \left(\frac{1.501}{g_3}\right)  \mathcal{F}(g_3) \frac{M_{\mathrm{Pl}} T_{\mathrm{RH}}}{f_a^2} \,,
\end{align}
with $g_{*S}$ the number of entropy degrees of freedom and $g_3=\sqrt{4\pi\alpha_s}$ the strong coupling constant. The factor $\mathcal{F}(g_3)$ encodes deviations from the Hard Thermal Loop approximation (which is retrieved by replacing $\mathcal{F}=1$), computed in Ref.~\cite{Salvio:2013iaa}.  
The inequality \eqref{eq:Rb_bound} then constrains the photon coupling as
\begin{align}
g_{a\gamma\gamma} &< \left(6.5 \times 10^{-11}\,\text{GeV}^{-1}\right) \left(\frac{\alpha_{\rm GUT}^{-1}}{25}\right) 
\left(\frac{1}{\ln \left(0.358\alpha_\mathrm{GUT}^{-1}\right)}\right)\left(\frac{\mathcal{F}(\alpha_\mathrm{GUT}^{-1})}{2}\right)^{-1}
\notag \\
&\times \left(\frac{T_{\rm RH}}{10^{13}\,\text{GeV}}\right)^{-1} \left(\frac{m_a}{10^{10}\,\text{GeV}}\right)^{1/2}\left(\frac{M_\nu}{10^{13} \,\mathrm{GeV}} \right) \,.
\label{eq:FI_bound}
\end{align}
Note that Ref. \cite{Salvio:2013iaa} computed $\mathcal{F}(\alpha_\mathrm{GUT}^{-1}=25)\sim 2$.
Opposite to the misalignment case, we now obtain an upper bound.\footnote{Note that we verify the axion decays after the EMD period begins.}

For sufficiently large photon couplings, the axion thermalizes and is instead produced via freeze-out. The freeze-out temperature is set by
\begin{align}
\frac{g_3^6 T_{\mathrm{th}}^3}{f_a^2} \sim \frac{T_{\mathrm{th}}^2}{M_{\mathrm{Pl}}} \Longrightarrow T_{\mathrm{th}} \sim \frac{f_a^2}{g_3^6 M_{\mathrm{Pl}}}\,,
\end{align}
Assuming relativistic freeze-out, which is justified for the heavy axions in our ensembles for which $T_\mathrm{th}>m_a$, the yield is $Y_a^{\mathrm{th}}
= 45 \zeta(3) / (2 \pi^4 g_{*S})$.
Finally, \eqref{eq:Rb_bound} leads to the lower  bound
\begin{align}
g_{a\gamma\gamma} &> \left(4.9 \times 10^{-20} \, \mathrm{GeV}^{-1} \right)  \left(\frac{m_a}{10^{10}\,\text{GeV}}\right)^{-\frac{1}{2}}\left(\frac{M_\nu}{10^{13} \,\mathrm{GeV}} \right)^{-1}\,,
\label{eq:FO_bound}
\end{align}
which, similar to the misalignment constraint, forbids the axion from being too long-lived and giving large entropy dilution. The combined constraints from freeze-in \eqref{eq:FI_bound} and freeze-out \eqref{eq:FO_bound}
lie in the region which is already excluded by decaying DM bounds (shaded in light gray in Fig.~\ref{fig:gayy_heterotic}).

\section{Non-toric Calabi-Yau compactifications}
\label{SM:non-toric}

Thus far we have considered CY 3-folds which are hypersurfaces of toric varieties. Here we compute the QCD axion mass in a selection of CY 3-folds which are not of this type.

First we consider the  product manifold $K3\times T^2$. 
Note that by the Beauville–Bogomolov decomposition theorem, the only CY 3-folds (in the loose sense of compact Kähler manifolds with vanishing first Chern class) that are product manifolds are $T^6$ and $K3\times T^2$.
However, these are not strictly CY 3-folds because they have reduced holonomy (which is a strict subgroup of $SU(3)$). Consequently, these manifolds are not viable for phenomenology as heterotic compactifications on these manifolds would give  $\mathcal{N}=4$ and $\mathcal{N}=2$
SUSY in 4D, respectively \cite{Gross:1984dd,Narain:1986am}.\footnote{On the other hand, orbifolding $K3\times T^2$ (or $T^6$) by an appropriate discrete subgroup $\Lambda \subset SU(3)$ yields manifolds that can have 4D $\mathcal{N}=1$ SUSY. Orbifolding modifies the volume form and $h^{1,1}$, though in many cases we still have $h^{1,1} \gg 1$ \cite{Fischer:2012qj, Hashimoto:2015zqm}. We leave the study of these orbifolds for future work.} Nonetheless, it is instructive to compute the QCD axion mass assuming a compactification on a product as the factorizability allows for light MD axions, and thus for the QCD axion mass to deviate from the MI value. Further, in this example we have $h^{1,1}=21$, whereas in our scan of the KS ensemble we only find manifolds with $h^{1,1}=2$ realizing this deviation.

We then consider complete intersection
Calabi–Yau (CICY) 3-folds. CICY manifolds have been used to construct Standard Models \cite{Anderson:2011ns,Anderson:2012yf,He:2010uj} (see Ref.~\cite{Buchbinder:2014qca} for a heterotic example). To our knowledge, axion properties have not previously been studied for compactification manifolds in this ensemble.
We do not identify any CICY compactifications for which the QCD axion mass differs non-negligibly from the MI value. The joint distribution of $(m_a,g_{a\gamma\gamma})$  is shown for this ensemble in Fig.~\ref{fig:gayy_heterotic}.

\subsection{Heterotic compactification on $K3 \times T^2$ }

$K3 \times T^2$ has $h^{1,1}=h^{1,1}(K3)+h^{1,1}(T^2)=20+1=21$ and the volume factorizes as
$\mathcal{V}_6=\mathcal{V}_{T^2}\mathcal{V}_{K3}$.
Let us focus on the MD axion associated to the $T^2$ cycle,
$b_{T}=\int_{T^2}B_2$. Its decay constant is~\cite{Svrcek:2006yi}
\begin{equation}
    f_{\rm MD}=\frac{l_s^2}{2\pi V_{T^2}}\frac{\mpl}{\sqrt 2}\,.
\end{equation}
As $\mathcal{V}_6$ factorizes, one can make $\mathcal{V}_{K3}\sim 1$, which implies that $\mathcal{V}_6 \sim \mathcal{V}_{T^2}$. In this case, the worldsheet instanton action breaking the shift symmetry of $b_T$ is $S_{\rm ws}\sim \frac{2\pi}{\alpha_{\rm GUT}}$ (for $g_s= 1$, which we assume below), similar to the NS5-brane action. Furthermore, the decay constant becomes 
\begin{equation}\label{eq:MD_axion_dec_const_K3xT2}
    f_{\rm MD}\approx \frac{1}{ g_s^2}\frac{\alpha_{\rm GUT} }{2\pi}\frac{\mpl}{\sqrt 2}\approx f_{\rm MI}\,.
\end{equation}

Note that in the limit $\mathcal{V}_{K3}\sim 1$, the $h^{1,1}(K3)=20$ MD axions that arise from integrating $B_2$ over 2-cycles in $K3$ obtain heavy masses from worldsheet instantons.
After integrating them out, the linear combination that couples to gauge bosons in the first $E_8$ is $\theta_1 = a+n_Tb_T$, where
the anomaly coefficient for the light MD axion is
\begin{equation}
    n_T=\frac{1}{16\pi^2}\int_{K3}(\text{tr}_1F^2-\frac{1}{2}\text{tr}R^2)=N_1-12\,.
\end{equation}
We define $N_1$ ($N_2$) as the instanton number on the first (second) $E_8$ and use the fact that the Euler characteristic of $K3$ is $\chi(K3)=24$ \cite{Svrcek:2006yi}.

Let us now study the breaking of shift-symmetry for $b_T$. As explained above, the worldsheet instanton action is large and the MD axion potential is
\begin{equation}
    V(b_T)=-m_{3/2}M_s^3e^{-2\pi/\alpha_{\rm GUT}}\cos (b_T)\equiv -\Lambda_{\rm ws}^4\cos (b_T)\,,
\end{equation}
where we used $S_{\rm ws}=2\pi/\alpha_{\rm GUT}$. For standard values of the unified gauge coupling (see above) we have $\Lambda_{\rm ws}^4\lesssim \Lambda_{\rm QCD}^4$, indicating that sizable mixing between MD and MI axions is possible, with $\theta_1 = a+n_Tb_T$ behaving as the QCD axion. The linear combination orthogonal to $\theta_1$ is a light ALP whose
coupling to gauge bosons is suppressed by a factor $m_{\rm ALP}^2/m_{\rm QCD}^2$ relative to that of the QCD axion and will lie below the QCD axion line (note that a light axion with this property is indicated for a $h^{1,1}=2$ KS compactification by the black triangle in Fig.~\ref{fig:gayy_heterotic}). 

As the QCD axion is given by $\theta_1$, the effective decay constant is given by
\begin{equation}
    \frac{1}{f_{\rm QCD}^2} = \frac{1}{f_{\rm MI}^2} + \frac{n_T^2}{f_{\rm MD}^2}\approx \frac{1+n_T^2}{f_{\rm MI}^2}\,,
\end{equation}
where we use \eqref{eq:MD_axion_dec_const_K3xT2}.
Altogether, the QCD axion mass is given by
\begin{equation}
    m_{\rm QCD}\approx m_\mathrm{MI}\sqrt{1+n_T^2}\,.
\end{equation}
Note that the Bianchi identity implies $N_1+N_2=24$, so that $|n_T|= |N_1 - 12|\leq 12$. Therefore, compactifying on $K3\times T^2$, the QCD axion mass is at most a factor of $\mathcal{O}(10)$ larger than the MI value. Interestingly, the fact that the mass departs from the MI value is correlated with the presence of a light ALP below the QCD axion line in the $(m_a,g_{a\gamma\gamma})$ plane. This is a generic requirement, and is exhibited in the examples from the KS ensemble listed in Tables \ref{tab:manifolds} and \ref{tab:manifolds_h11_3}. 

\subsection{Heterotic compactifications on complete intersection Calabi-Yau  3-folds}
\label{app:CICY}

A complete intersection Calabi--Yau (CICY) threefold is specified by a configuration matrix 
\begin{equation}
X_6\equiv \left[
\begin{array}{c|ccc}
\mathbb P^{n_1} & q^1_1 & \cdots & q^1_K\\
\vdots          & \vdots&        & \vdots\\
\mathbb P^{n_m} & q^m_1 & \cdots & q^m_K
\end{array}
\right],
\end{equation}
which describes the common zero locus of $K$  homogeneous polynomials
$P_a$ in the ambient space $\mathcal A=\prod_{r=1}^m\mathbb P^{n_r}$.
The entry $q^r_a$ is the degree of the polynomial $P_a$ with respect to the hyperplane class of the $r$-th projective factor.
Equivalently, $X_6=\{P_1=\cdots=P_K=0\}\subset \mathcal A$,
with $P_a$ a section of the line bundle $\mathcal O_{\mathcal A}(q^1_a,\dots,q^m_a)$.
The CY condition requires
\begin{equation}
\sum_{a=1}^{K} q^r_a = n_r+1\quad\forall r,
\end{equation}
and the 3-fold condition is
\begin{equation}
\sum_{r=1}^m n_r - K = 3.
\end{equation}

There are 7890 CICY manifolds, some of which are not toric \cite{Bull:2019cij}. In this work we use the augmented \href{http://www1.phys.vt.edu/cicydata/}{CICY dataset} constructed in Ref.~\cite{Anderson:2017aux}, which provides configuration matrices together with Hodge numbers, second Chern class data, and boolean flags indicating whether the Picard group and K\"ahler cone descend from the ambient space. We restrict to the 4874 manifolds satisfying \texttt{Favour=True} and \texttt{K\"ahlerPos=True}. In favorable configurations the Picard group of $X_6$ is generated by the restrictions of the ambient hyperplane classes, so that (in the favorable description) $h^{1,1}(X_6)=m$ and a divisor basis is $\beta_r = H_r|_{X_6}$,
where $H_r$ denotes the hyperplane class of the $r$-th factor $\mathbb P^{n_r}$.
For $\texttt{K\"ahlerPos=True}$, the K\"ahler cone of $X_6$ is the positive orthant in this basis, and hence the Mori cone is its dual $\mathbb{R}_{\ge 0}^{h^{1,1}}$; equivalently the Mori generators are the rows of the identity matrix in the basis dual to $\{\beta_r\}$ \cite{Anderson:2017aux}.
Imposing a SKC with minimal curve volume $c$ then reduces to
$t_r \ge c$
for the K\"ahler parameters $t_r$ in $J = \sum_r t_r \beta_r$.

The triple intersection numbers
\begin{equation}
\kappa_{rst}=\int_{X_6} \beta_r\wedge \beta_s\wedge \beta_t
\end{equation}
are determined combinatorially from the configuration matrix. Each defining equation corresponds to a divisor class in the ambient space,
\begin{equation}
[P_a]=\sum_{r} q^r_a H_r,
\end{equation}
so the cohomology class of the complete intersection is
\begin{equation}
[X_6]=\prod_{a=1}^{K}\left(\sum_r q^r_a H_r\right).
\end{equation}
The ambient cohomology ring obeys $H_r^{\,n_r+1}=0$ and is normalized by
\begin{equation}
\int_{\mathcal A}\prod_{r=1}^m H_r^{n_r}=1.
\end{equation}
Using the standard relation
\begin{equation}
\kappa_{rst}=\int_{\mathcal A} H_r H_s H_t\,[X_6],
\end{equation}
one expands $H_rH_sH_t[X_6]$ in the ambient cohomology ring and extracts the coefficient of the top monomial $\prod_u H_u^{n_u}$.
We show the distribution of intersection numbers in Fig.~\ref{fig:kappa_ijk_CICY}.
\begin{figure}
    \centering    \includegraphics[width=0.5\linewidth]{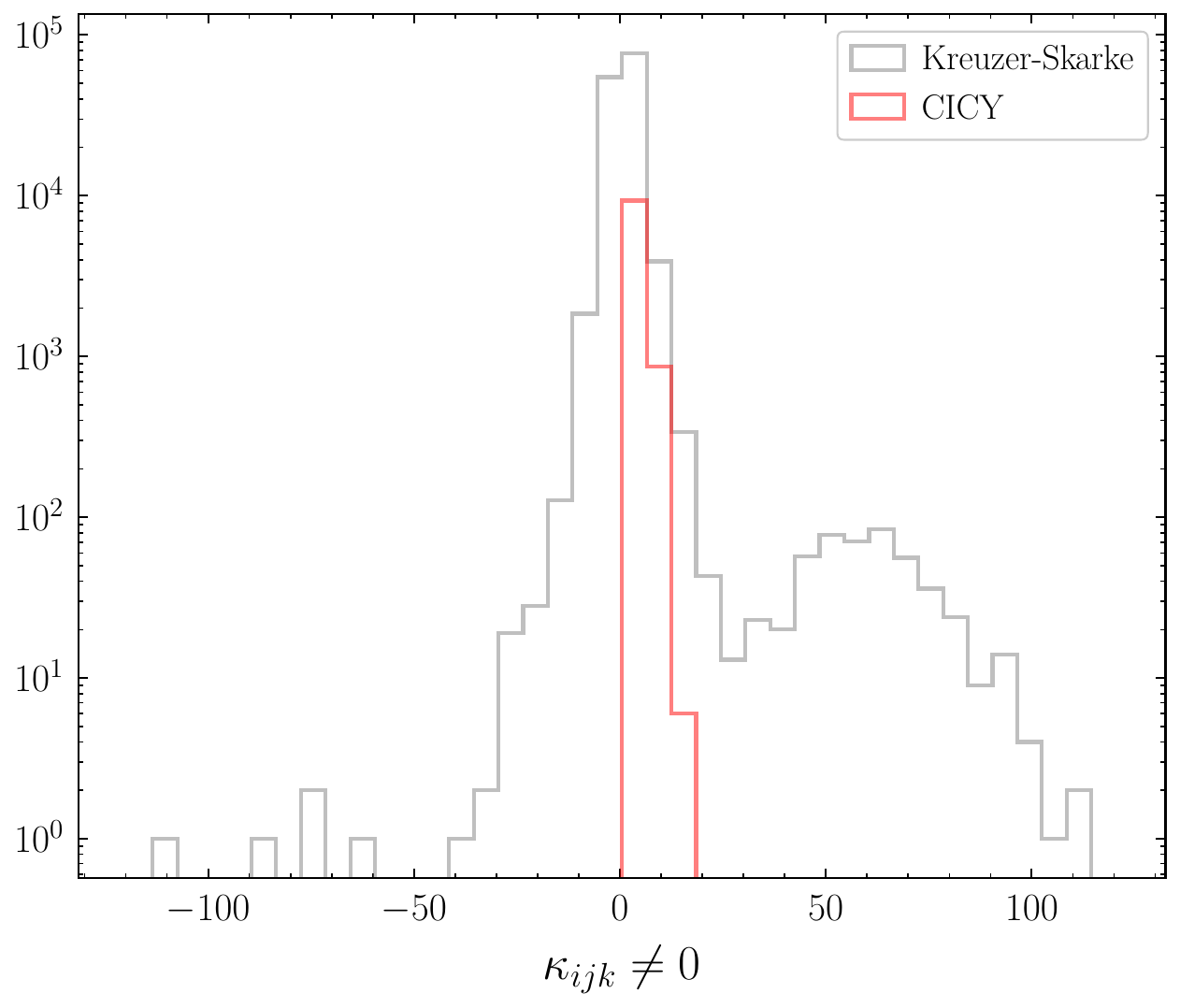}
    \caption{Distribution of triple intersection numbers in our heterotic ensemble constructed  from KS manifolds (gray) and from CICY  manifolds (red). As KS manifolds can have negative triple intersection numbers, cancellations in the volume form for $\mathcal{V}_6$ allow for larger effective curve volumes within the SKC compared to those within the simplicial cones for CICY manifolds.}
\label{fig:kappa_ijk_CICY}
\end{figure}

Since the Mori charge matrix is the identity in this basis, the curve volume appearing in the worldsheet instanton potentials giving the lightest MD axion is one of the Kähler parameters $t_i$. In the divisor basis $\{\beta_r\}$ the  intersection numbers are non-negative. Therefore, this volume is maximal within the SKC when all other parameters $t_j$ are at the boundary of the SKC, $t_j=c$. Fixing $\alpha_\mathrm{GUT}^{-1}=27$  
and $g_s=1$, we find 375 
acceptable heterotic compactification manifolds in this ensemble. Of these, the largest curve volume  we find is $13.33$. 
Consequently the QCD axion always has the MI mass for weakly coupled heterotic compactifications in this ensemble.

\section{The QCD axion mass in strongly coupled heterotic string theory}
\label{SM:duality_typeI}

Here we examine the validity of our results when we take the string coupling, $g_s$, to be large. While at weak coupling our analysis is essentially identical for both $SO(32)$ and $E_8\times E_8$ heterotic string theories -- our results depend on the geometry of the compact space, which determines decay constants and the worldsheet  instanton actions, none of which depend on the gauge group (with the only possible difference coming from the values of anomaly coefficients)  -- the situation changes slightly at strong coupling. 

For heterotic $SO(32)$, the large coupling limit has a description in terms of another weakly coupled string theory, the Type I superstring in 10 dimensions. The 10D action of these two theories is identical after the identification of various higher-form and moduli fields. We expect that our results apply to this theory when the visible sector is embedded into a subgroup of the $SO(32)$ gauge symmetry which is realized on $D9$-branes.\footnote{We do not consider cases where the Standard Model gauge groups are realized on $D5$-branes.} 

In particular, in Type I string theory the MI axion decay constant is 
given precisely by \eqref{eq:MI_fa_main}.
This can be found by dimensionally reducing the 10D Type I SUGRA action~\cite{Svrcek:2006yi}. Using the same conventions as above, the gauge-invariant field strength of $C_2$ is given by $F_3 = dC_2 - \frac{\alpha^\prime}{4}\omega_3$. The modified Bianchi identity for $F_3$ then ensures that the Type I MI axion couplings are universal to all the gauge bosons.
These conclusions can also be obtained from the fact that the $C_2$ in Type I maps to the $B_2$ 2-form field of heterotic $SO(32)$. From this mapping we also find that Type I contains MD axions that come from integrating the RR 2-form over curves, $\int_{\Sigma_2} C_2$. These MD axions  have couplings to gauge bosons that are analogous to the heterotic MD axion couplings from the Green-Schwarz mechanism, see \eqref{eq:def_thetas_SM}-\eqref{eq:ni_def}.

The exact duality between Type I and heterotic $SO(32)$, together with the mapping $C_2^{\rm Type\,I} \Leftrightarrow B_2^{\rm Het}$ has other important implications. First, the action of a Type I $D1$-brane wrapping $\Sigma_2$ -- and hence breaking the shift symmetry of the MD Type I axions -- coincides with the (heterotic) worldsheet instanton action, which breaks the shift symmetry of MD axions on the heterotic side. Additionally, the decay constant is also identical
in both theories. 

Altogether, this implies that, as in weakly coupled heterotic, in Type I
the QCD axion mass is bounded from below by the MI axion value. This lower bound on the mass is obtained when all the MD axions are heavy and mixing is negligible. Mixing is stronger when at least one of the MD axions is light, which in the Type I superstring can more readily occur at small string coupling, $g_s<1$ (corresponding to $g_s>1$ on the heterotic side).

Let us consider now the strong coupling limit of the  heterotic $E_8\times E_8$ string. We describe this limit by using M-theory, which has 11D SUGRA as the low-energy EFT \cite{Witten:1995ex,Horava:1995qa}. In this case, axions come from $C$, a 3-form gauge field present in the 11D supergravity multiplet. 
The MI axion arises as the 4D dual of the two-form coming from the $C$ field with one index tangent to the eleventh dimension and the other two along the 4D Minkowski space (as in the weakly coupled heterotic string). 
MD axions, on the other hand, come from integrating $C$ over the product of the eleventh dimension and a 2-cycle of the CY.

The axion-gauge boson coupling arises from the CS coupling of the 3-form field to its field strength \cite{Svrcek:2006yi}
\begin{equation}
  S_{11}\subset \int  C\wedge G\wedge G \,.
\end{equation}
Here $G$ is the 4-form field strength of $C$ that satisfies the modified Bianchi identity $dG \propto \delta(x^{11})(\text{tr}F^2-\frac{1}{2}\text{tr}R^2) $ \cite{Horava:1996ma}, which ensures that the quantized part of the axion couplings remains the same -- this is fixed by anomaly cancellation. On the other hand, axion decay constants depend on the details of the theory such as the size of the eleventh dimension. 

In M-theory, the compact space is a warped product $X_7=X_6\times S^1/\mathbb{Z}_2$, with $X_6$ a CY 3-fold and $\pi \rho$ the size of the $S^1/\mathbb{Z}_2$ interval. This implies that the volume of the CY is position-dependent $\mathcal{V}_6=\mathcal{V}_6(x^{11})$. This allows for different possibilities, depending on the $E_8$ boundary on which the Standard Model is placed. 
\\
\\
\textit{(i) Standard Model at the large boundary.} In this case the size of the eleventh dimension is bounded by the consistency of the theory (see \cite{Svrcek:2006yi} for a discussion). The size of the large CY is fixed by the GUT gauge coupling $\mathcal{V}_6(x^{11}=\rho)/l_{11}^6= \alpha_{\rm GUT}^{-1}$, such that requiring that the small CY has a volume $\mathcal{V}_6(x^{11}=0)/l_{11}^6\gtrsim 1$, implies that the interval is bounded as $\rho \lesssim \rho_{\rm max}$. This in turn bounds the axion decay constants as
\begin{equation}
    f_{\rm MI} \lesssim \frac{\alpha_{\rm GUT}}{2\pi\sqrt{q}}\mpl \,,\,\,\, f_{\rm MD}\gtrsim \frac{q\alpha_{\rm GUT}}{3\pi}\mpl\,.
\end{equation}
Here $q=\mathcal{O}(1)$ is an instanton number that depends on the model (e.g., on fluxes).
\\
\\
\textit{(ii) Standard Model at the small boundary.} In this case the size of the eleventh dimension is only bounded by the non-observation of proton decay~\cite{Reig:2025dpz}. Requiring that the cut-off of 11D SUGRA is above the GUT scale, $M_{11}\gtrsim M_{\rm GUT}$, implies that the MI and MD axion decay constants
\begin{equation}
    f_{\rm MI}\approx M_{11}\sqrt{\frac{3\alpha_{\rm GUT}^{1/3}}{4\pi q}}\sim M_{11} \,,\,\,\, 
    f_{\rm MD}\approx \mpl\frac{4}{3\pi q \alpha_{\rm GUT}^{1/3}}\left (\frac{l_{11}}{\pi \rho} \right )^2\,,
\end{equation}
are bounded from below. Imposing $\left (\frac{l_{11}}{\pi \rho } \right )\gtrsim 10^{-2}$ and $M_{11}\gtrsim M_{\rm GUT}$ to satisfy proton decay bounds, we obtain $f_{\rm MI}\gtrsim M_{\rm GUT}$ and $f_{\rm MD}\gtrsim 10^{14}$ GeV. This implies that the lower bound on $m_{\rm QCD}$ that we establish for weakly coupled heterotic also applies to this case. 
\\
\\
\textit{(iii) Flat eleventh dimension.} In the case where the interval is flat, we have $V_7=\mathcal{V}_6\pi \rho$. This implies that the MI axion has the same decay constant as in the weakly coupled heterotic case \eqref{eq:MI_fa_main}, and all our results obtained in the weakly coupled heterotic string case will hold.

We find that in the case of strongly coupled heterotic, a similar lower bound to the one obtained in the weakly coupled theory appears, but only after we impose the constraint (from proton decay searches) that $M_{11}\gtrsim M_{\rm GUT}$. We summarize  in Fig.~\ref{fig:string_dualities} the set of string theories for which the lower bound on the QCD axion mass discussed in this work holds.

\section{Duality with F-theory}
\label{SM:duality_Ftheory}

Exceptional gauge groups and $SO(10)$ with spinor representations are difficult to realize in weakly coupled Type IIB string theory but arise naturally in F-theory (see~\cite{Weigand:2018rez} for a review). It is therefore natural to ask whether the  results obtained for the heterotic string --- that special geometric conditions are required for $m_\mathrm{QCD}$ to deviate from the MI value --- extend to F-theory. Below we explain why it does not.

F-theory is defined by elliptically fibered CY 4-folds $Y_4 \to B_3$~\cite{Bershadsky:1996nh,Weigand:2018rez}, with non-abelian gauge symmetries arising from 7-branes wrapping divisors in $B_3$~\cite{Beasley:2008dc}. We work throughout in the Type IIB limit for simplicity, and consider closed-string axions from $C_4$ on divisors~\cite{Fallon:2025lvn,Nee:2026}. The key structural difference from the weakly coupled heterotic case, as we discuss in the main text, is that the visible gauge coupling is set by a local divisor volume rather than the bulk volume $\mathcal{V}_6$; GUT gauge group constraints then restrict the divisor topology~\cite{Beasley:2008dc,Beasley:2008kw,Donagi:2008ca} but do not generally fix $\mathcal{V}_6$, permitting a broad range for $f_\mathrm{QCD}$.

A geometric duality nonetheless exists~\cite{Bershadsky:1996nh,Donagi:2008ca}: heterotic on an elliptically fibered CY 3-fold over a surface $B_2$ is dual to F-theory on an elliptically fibered CY 4-fold whose base $B_3$ is a $\mathbb{P}^1$-bundle over $B_2$. The duality maps specific corners of the two moduli spaces, but does not imply that the heterotic lower bound on $m_\mathrm{QCD}$ holds throughout the F-theory landscape.

We illustrate these considerations with two examples. The first has $B_3 = \mathbb{P}^1 \times \mathbb{P}^2$, which is a (trivial) $\mathbb{P}^1$-bundle over $\mathbb{P}^2$ and therefore admits a heterotic dual. In this example the local divisor axion corresponds to the heterotic MI axion, and its decay constant is correspondingly constrained. The second example takes $B_3 = \mathrm{Bl}_p(\mathbb{P}^3)$, which has no heterotic dual and admits a Swiss-cheese volume form. Here the local axion decay constant does scale with the bulk volume, $f_a \propto M_s \propto \mathcal{V}_B^{-1/2}$, demonstrating that deviations from the MI value are generic in F-theory GUTs -- consistent with the results of Ref.~\cite{Fallon:2025lvn} for F-theory axiverses with thousands of axions. Our findings are summarized in Fig.~\ref{fig:F_theory}; both examples admit $SO(10)$ (as well as $E_6$ or $E_8$) gauge sectors~\cite{Chen:2010ts}.

\subsection{Factorizable base: \texorpdfstring{$\mathbb{P}^1\times \mathbb{P}^2$}{P1 x P2}}
\label{app:P1xP2}

Consider a F-theory compactification on an elliptically fibered Calabi--Yau 4-fold $  \pi:Y_4\to B_3$, $B_3=\mathbb{P}^1\times\mathbb{P}^2$.
Let \(H_1\) and \(H_2\) denote the pullbacks to \(B_3\) of the hyperplane classes of \(\mathbb P^1\) and \(\mathbb P^2\), respectively. The only non-vanishing triple intersection is
\begin{equation}
  \int_{B_3} H_1 H_2^2 = 1,
\end{equation}
while \(H_1^2=0\) and \(H_2^3=0\). Expanding the K\"ahler form as\footnote{In what follows we abuse notation by identifying divisor classes with their Poincaré dual (1,1)-forms.}
\begin{equation}
  J=t_1 H_1+t_2 H_2,
\end{equation}
the K\"ahler cone is simply $t_1>0$, $ t_2>0$.
The base volume is
\begin{equation}
  \mathcal{V}_B \equiv \frac16 \int_{B_3} J^3
  = \frac12\, t_1 t_2^2\,,
\end{equation}
and the divisor volumes are
\begin{equation}
  \tau_1 \equiv \frac12 \int_{B_3} J^2 \wedge H_1 = \frac12\, t_2^2,
  \qquad
  \tau_2 \equiv \frac12 \int_{B_3} J^2 \wedge H_2 = t_1 t_2 \,,
\end{equation}
so that
\begin{equation}\label{eq:vol_typeIIB_factorisable}
  \mathcal{V}_B=\frac{1}{\sqrt2}\,\tau_2\sqrt{\tau_1}.
\end{equation}

We take the visible seven-brane to wrap the divisor
$S\sim H_1 \simeq \mathbb P^2$, whose volume is $\tau_S=\tau_1=\frac12\,t_2^2$.
Fixing the visible-sector gauge coupling amounts to fixing \(\tau_S\). For example, imposing $\tau_S=25$
in string units fixes \(t_2=\sqrt{50}\), while \(t_1\) remains free. The base volume then becomes   $\mathcal{V}_B=25\, t_1$,
which can be made parametrically large by taking \(t_1\gg 1\) at fixed \(\tau_S\). The string scale therefore scales as
\begin{equation}
  M_s \sim \frac{M_P}{\sqrt{\mathcal{V}_B}}
  \propto t_1^{-1/2},
\end{equation}
up to the usual \(g_s\)- and \(2\pi\)-dependent prefactors.

Following the notation in \cite{Reece:2024wrn}, the dimensionful Kähler potential is given in terms of the volume by
\begin{equation}\label{eq:Kähler_dimensionful}
    K= \mpl^2 \,k(T,T^\dagger)\,,\,\,\,\text{ with: }\,\,k=-2\log \mathcal{V}_B\,.
\end{equation}
Here $T^i=\frac{\tau_i}{g_s}+i\frac{\theta_i}{2\pi}$ are chiral $\mathcal{N}=1$ supermultiplets. From \eqref{eq:Kähler_dimensionful}, one can derive the Kähler metric as $K_{ij}= \frac{\partial ^2K}{\partial T^i\partial T^{j\,\dagger}}$, which allows us to write the scalar kinetic term:
\begin{equation}
    -K_{ij}\partial_\mu T^i \partial^\mu T^j= -\frac{1}{4}\frac{\partial^2k}{\partial\tau_i\partial \tau_j}\left [\partial_\mu\tau_i \partial^\mu\tau_j+\frac{g_s^2}{4\pi^2}\partial_\mu\theta^i\partial^\mu\theta^j \right ]\,.
\end{equation}

The axion decay constants are obtained as the eigenvalues of the matrix:
\begin{align}
\label{eq:eigenvals}
    \frac{g_s^2}{8\pi^2}\mpl^2\frac{\partial^2k}{\partial\tau_i\partial\tau_j}\,.
\end{align}
Let $T_S=\frac{\tau_S}{g_s}+i \frac{a_S}{2\pi}$
denote the K\"ahler modulus associated with \(S\), where \(a_S\) is the RR four-form axion obtained by integrating \(C_4\) over the divisor \(S\). The K\"ahler potential using the volume in \eqref{eq:vol_typeIIB_factorisable} factorizes as
\begin{equation}
  k=-2\ln \mathcal{V}_B
   = \text{const}-2\ln\tau_2-\ln\tau_1.
\end{equation}
It follows that
\begin{equation}
  K_{11}\equiv \frac{\partial^2 K}{\partial \tau_1^2}=\frac{1}{\tau_1^2},
  \qquad
  K_{22}\equiv \frac{\partial^2 K}{\partial \tau_2^2}=\frac{2}{\tau_2^2},
  \qquad
  K_{12}=0.
\end{equation}
Thus the kinetic term of the axion \(a_S\) depends only on the local divisor volume \(\tau_S=\tau_1\), and is independent of the bulk modulus \(\tau_2\). In particular, once \(\tau_S\) is fixed by the gauge coupling, the canonically normalized decay constant of \(a_S\) is fixed in 4D Planck units up to order-one coefficients and does \emph{not} scale as \(M_s\) when the overall volume is varied.

This example therefore illustrates an important point: although the compactification volume and string scale can be scanned while keeping the visible-sector gauge coupling fixed, the local closed-string axion associated with \(S\) does not inherit the bulk scaling \(f_a\propto \mathcal{V}_B^{-1/2}\). The decay constant is plotted in Fig.~\ref{fig:F_theory}. In this sense the geometry behaves differently from a Swiss-cheese compactification that will be studied in the next section, and is analogous to the heterotic examples discussed in the main text.

An elliptically fibered CY fourfold over \(B_3\) can be written in Weierstrass form ~\cite{Bershadsky:1996nh,Weigand:2018rez}
\begin{equation}
  y^2=x^3+f x+g,
\end{equation}
with
\begin{equation}
  f\in H^0(B_3,\mathcal{O}(-4K_{B_3})),
  \qquad
  g\in H^0(B_3,\mathcal{O}(-6K_{B_3})).
\end{equation}
For \(B_3=\mathbb P^1\times \mathbb P^2\), $-K_{B_3}=2H_1+3H_2$,
and hence
\begin{equation}
  f\in H^0(B_3,\mathcal{O}(8,12)),
  \qquad
  g\in H^0(B_3,\mathcal{O}(12,18)).
\end{equation}
The discriminant
\begin{equation}
  \Delta=4f^3+27g^2
\end{equation}
determines the seven-brane locus.

To engineer a gauge algebra on the visible divisor \(S=\{w=0\}\sim H_1\), it is convenient to use Tate form~\cite{Katz:2011qp},
\begin{equation}
  y^2+a_1 xyz+a_3 y z^3
  =
  x^3+a_2 x^2 z^2+a_4 x z^4+a_6 z^6,
  \label{eq:Tate}
\end{equation}
with $  a_n\in H^0(B_3,\mathcal{O}(-nK_{B_3}))=H^0(B_3,\mathcal{O}(2n,3n))$.
A standard split \(D_5\) (\(\mathfrak{so}(10)\)) tuning is obtained by imposing~\cite{Katz:2011qp}
\begin{equation}
  (\mathrm{ord}_w a_1,\mathrm{ord}_w a_2,\mathrm{ord}_w a_3,\mathrm{ord}_w a_4,\mathrm{ord}_w a_6)=(1,1,2,3,5).
  \label{eq:D5_tuning}
\end{equation}
Explicitly, one may take
\begin{align}
  a_1&=b_1 w,\qquad
  a_2=b_2 w,\qquad
  a_3=b_3 w^2,\qquad \notag \\
  a_4&=b_4 w^3,\qquad
  a_6=b_6 w^5,
\end{align}
with
\begin{align}
  b_1 &\in H^0(B_3,\mathcal{O}(1,3)),\,
  b_2 \in H^0(B_3,\mathcal{O}(3,6)),  \notag \\
  b_3 &\in H^0(B_3,\mathcal{O}(4,9)),\,
  b_4 \in H^0(B_3,\mathcal{O}(5,12)), \notag \\
  b_6 &\in H^0(B_3,\mathcal{O}(7,18)).
  \label{eq:bi}
\end{align}
These bundles are effective, so the tuning is globally available on this base. The corresponding vanishing orders are
\begin{equation}
\mathrm{ord}_w(f,g,\Delta)=(2,3,7),
\label{eq:good_ord}
\end{equation}
corresponding to Kodaira type \(I_1^\ast\) and gauge algebra $\mathfrak{so}(10)$
along \(S\) ~\cite{Bershadsky:1996nh}.

\subsection{Swiss-cheese base: $\mathrm{Bl}_p(\mathbb P^3)$}
\label{app:blowupPthree}

Now let us study a F-theory base in the form of  a blow-up of \(\mathbb P^3\) at a point, $B_3=\mathrm{Bl}_p(\mathbb P^3)$.
We will see that the K\"ahler geometry has Swiss-cheese form, allowing the visible gauge coupling to be held fixed while the bulk volume is varied, and that the corresponding local closed-string axion scales as $f_a \propto M_s \propto \mathcal{V}_B^{-1/2}$ at (moderately) large volume. This example therefore makes explicit that deviations of the QCD axion mass from the MI value are generic in F-theory compactifications.

Let \(H\) denote the pullback of the hyperplane class of \(\mathbb P^3\), and let \(E\) denote the exceptional divisor. Geometrically, $E\simeq \mathbb P^2$,
so \(E\) is a rigid local divisor and is therefore a natural candidate to support the visible seven-brane sector.
The non-vanishing triple intersections are
\begin{equation}
  \int_{B_3} H^3=1,
  \qquad
  \int_{B_3} E^3=1,
\end{equation}
while the mixed intersections vanish,
\begin{equation}
  \int_{B_3} H^2E=\int_{B_3} HE^2=0.
\end{equation}
Expanding the K\"ahler form as
\begin{equation}
  J=t\,H-s\,E,
\end{equation}
the K\"ahler cone conditions are $s>0, t-s>0$. 
The base volume is then
\begin{equation}
  \mathcal{V}_B\equiv \frac16\int_{B_3} J^3
  =\frac16\left(t^3-s^3\right).
\end{equation}
The divisor volumes are
\begin{equation}
  \tau_b\equiv \frac12\int_{B_3}J^2\wedge H=\frac12\,t^2,
  \qquad
  \tau_s\equiv \frac12\int_{B_3}J^2\wedge E=\frac12\,s^2.
\end{equation}
and therefore
\begin{equation}
  \mathcal{V}_B=\frac{\sqrt2}{3}\left(\tau_b^{3/2}-\tau_s^{3/2}\right).
  \label{eq:VB}
\end{equation}
Thus \(B_3\) furnishes an explicit two-modulus Swiss-cheese geometry, with \(\tau_b\) controlling the bulk volume and \(\tau_s\) controlling the local blow-up divisor.

We place the visible seven-brane on the divisor $S=E\simeq \mathbb P^2$,
so that $\tau_S=\tau_s$.
Fixing the visible-sector gauge coupling amounts to fixing \(\tau_s\). One may then take $\tau_b\gg \tau_s$
so that the bulk volume becomes parametrically large while the visible divisor volume remains fixed:
\begin{equation}
  \mathcal{V}_B \simeq \frac{\sqrt2}{3}\,\tau_b^{3/2}
  \qquad
  (\tau_b\gg \tau_s).
\end{equation}
The string scale therefore scales as $M_s\sim \frac{M_P}{\sqrt{\mathcal{V}_B}}$.

Using the same notation as the previous subsection, from \eqref{eq:VB}
we find
\begin{equation}
  K_{ss}
  =\frac{\partial^2 K}{\partial \tau_s^2}
  =\frac{3\left(\tau_b^{3/2}+2\tau_s^{3/2}\right)}
         {2\sqrt{\tau_s}\left(\tau_b^{3/2}-\tau_s^{3/2}\right)^2}.
\end{equation}
In the large-volume regime \(\tau_b\gg \tau_s\), we have $  K_{ss}\sim 1/\mathcal{V}_B\sqrt{\tau_s}$. Hence the canonically normalized axion decay constant scales parametrically as \eqref{eq:eigenvals}
\begin{equation}
  f_{a_s}\sim  \frac{g_s}{\sqrt{8\pi^2}}M_P\sqrt{K_{ss}}
  \sim \frac{g_s}{\sqrt{8\pi^2}}\frac{M_P}{\sqrt{\mathcal{V}_B}\,\tau_s^{1/4}}
  \sim \frac{g_s}{\sqrt{8\pi^2}}\frac{M_s}{\tau_s^{1/4}}.
\end{equation}
Once \(\tau_s\) is fixed by the visible gauge coupling, this gives (see Fig.~\ref{fig:F_theory})
\begin{equation}
  f_{a_s}\propto M_s\propto \mathcal{V}_B^{-1/2}.
\end{equation}
This is precisely the scaling needed to lower the QCD axion decay constant by increasing the overall volume while keeping the visible-sector gauge coupling fixed.

The KK scale associated with the GUT divisor scales in the same way, $M_{\rm KK}^{\rm vis}\sim \frac{M_s}{\tau_s^{1/4}}$,
so fixing \(\tau_s=\mathcal{O}(10)\) keeps the local KK threshold at an \(\mathcal{O}(1)\) fraction of the string scale. Requiring that the KK modes of the GUT gauge bosons are sufficiently heavy to suppress fast proton decay ($M_{\rm KK}^{\rm vis}\gtrsim M_{\rm GUT}$),
therefore imposes only a lower bound on \(M_s\), leaving a broad range of allowed bulk volumes and corresponding axion decay constants.
See Fig.~\ref{fig:F_theory} for the range of allowed values of the axion decay constant and the QCD axion mass.

For the blow-up of a point in \(\mathbb P^3\),
\begin{equation}
  K_{B_3}=-4H+2E,
  \qquad
  -K_{B_3}=4H-2E,
\end{equation}
and therefore
\begin{equation}
  f\in H^0(B_3,\mathcal{O}(16H-8E)),
  \qquad
  g\in H^0(B_3,\mathcal{O}(24H-12E)).
\end{equation}
Let \(w=0\) be the equation of the exceptional divisor \(E\). In Tate form (see \eqref{eq:Tate}), the coefficients $a_n$ are 
\begin{equation}
  a_n\in H^0(B_3,\mathcal{O}(-nK_{B_3}))
      = H^0(B_3,\mathcal{O}(n(4H-2E))).
\end{equation}
Using the split \(D_5\) tuning \eqref{eq:D5_tuning} implemented as in \eqref{eq:bi}, the coefficients are
\begin{align}
  b_1 &\in H^0(B_3,\mathcal{O}(4H-3E)),\,
  b_2 \in H^0(B_3,\mathcal{O}(8H-5E)),\notag\\
  b_3 &\in H^0(B_3,\mathcal{O}(12H-8E))\,
  b_4 \in  H^0(B_3,\mathcal{O}(16H-11E)), \nonumber\\
  b_6 &\in H^0(B_3,\mathcal{O}(24H-17E)).
\end{align}
These line bundles are effective, so the tuning is globally available on this base. The resulting vanishing orders are as in \eqref{eq:good_ord},
corresponding to Kodaira type \(I_1^\ast\) and gauge algebra $\mathfrak{so}(10)$
on the seven-brane wrapping \(E\).

\section{Supplementary figures}
\label{SM:figs}

\begin{figure}[H]
    \centering
\includegraphics[width=0.5\linewidth]{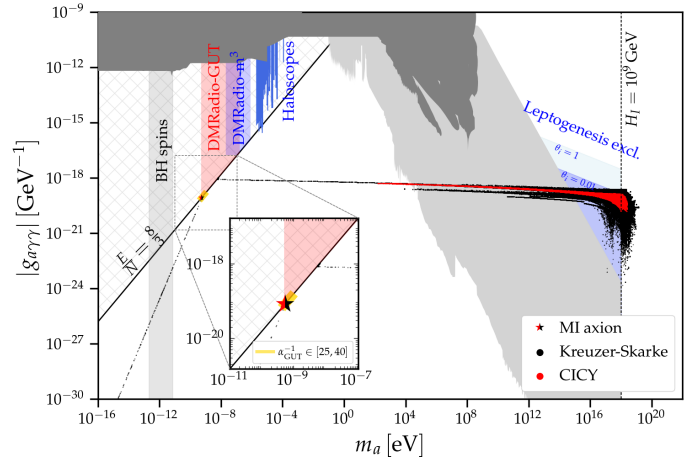}
    \caption{
    As in Fig.~\ref{fig:gayy_heterotic} but varying over the full hypersurface of the SKC for which $\alpha_\mathrm{GUT}^{-1}=25$ and $g_s=1$ via constrained Monte Carlo, with 1000 sampled points in the SKC per manifold. Of the 2,027,000  KS samples, 96\%  survive the BBN constraint and 9\% survive the leptogenesis constraint (with $\theta_i=1$ and $H_I=10^{9}$ GeV; for $H_I=10^{7}$ GeV, 56\% survive the leptogenesis constraint). For the 375,000 CICY manifolds, the corresponding survival fractions are 97\% and 2\%. 
}\label{fig:gayy_heterotic_full_moduli_space}
\end{figure}

\begin{figure}
    \centering
\includegraphics[width=0.9\linewidth]{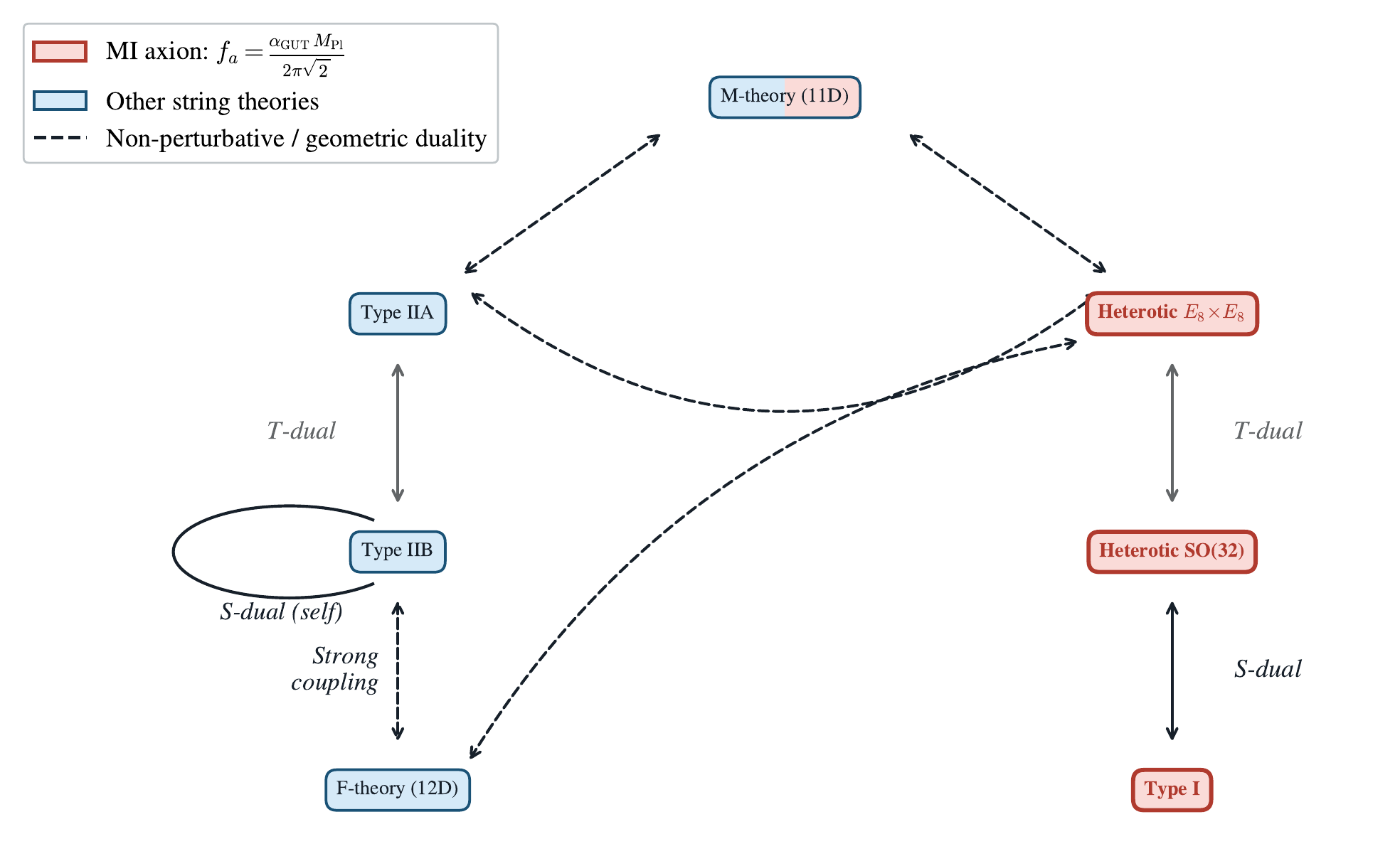}
    \caption{
    Map of dualities between string theories, with theories for which a MI axion is present (not present)  highlighted in red (blue). M-theory is highlighted in both colors as a MI axion is present or not depending on the limit of the theory. For theories with a MI axion, the QCD axion mass is at least the MI value, $m_a \gtrsim 5.2\times 10^{-10}$ eV. In all cases, assuming unification of the Standard Model gauge couplings at the SUSY GUT scale we expect $m_a \lesssim 10^{-8}$ eV \cite{Benabou:2025kgx}.  Dashed arrows indicate non-perturbative /geometric dualities: $\text{M-theory on } S^1 \leftrightarrow \text{Type IIA}$ (10D), $ \text{M-theory on } S^1/\mathbb{Z}_2 \leftrightarrow \text{Heterotic } E_8 \times E_8$ (10D), $\text{Heterotic on } T^2 \leftrightarrow \text{F-theory on K3} $ (in 8D), $ \text{Heterotic on } T^4\leftrightarrow$ $\text{Type IIA on K3}$ (6D), where in each case we indicate the spacetime dimension of the resulting effective theory after compactification in parentheses.
}\label{fig:string_dualities}
\end{figure}

\begin{figure}[ht]
    \centering
    \includegraphics[width=0.495\textwidth]{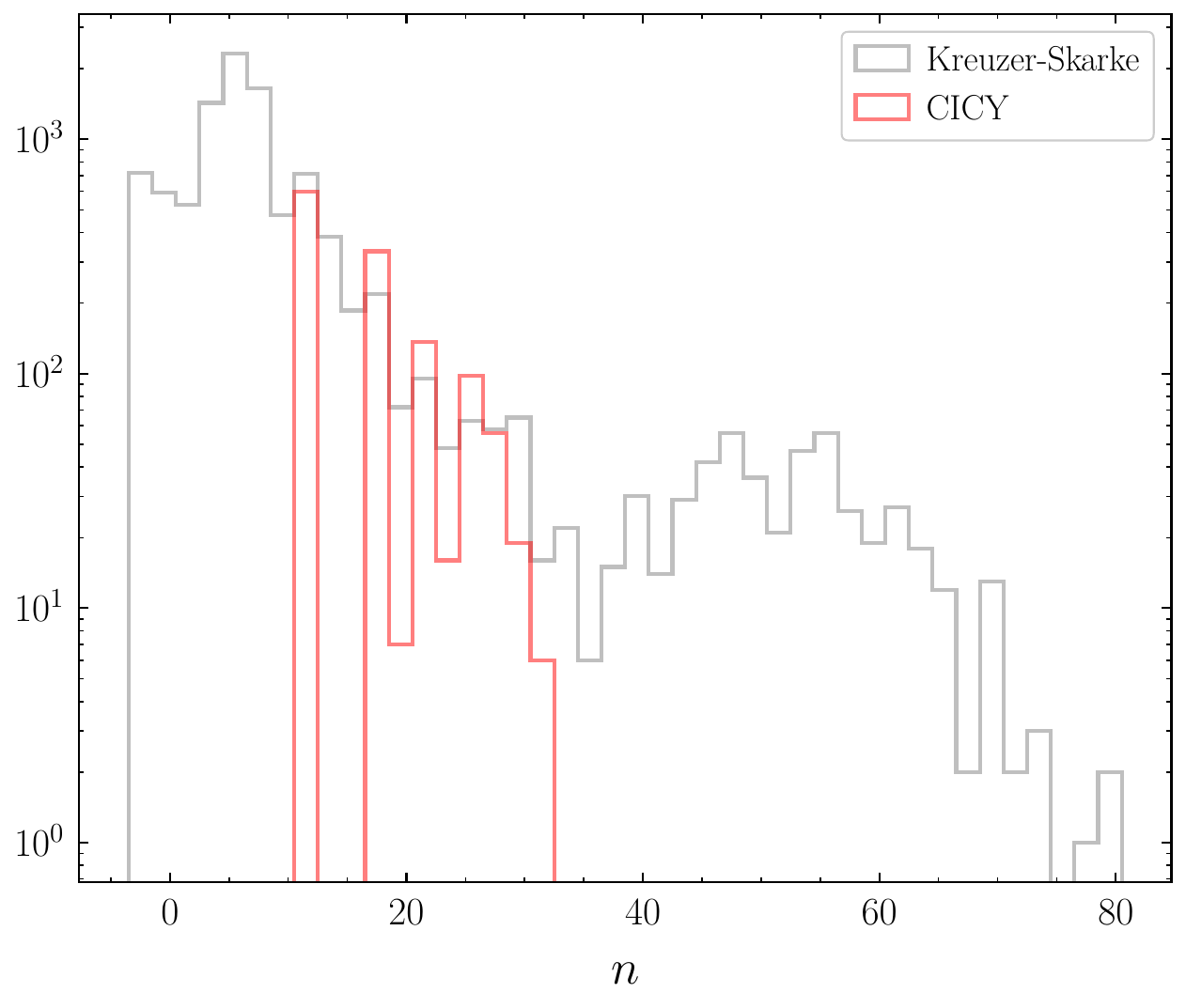}
    \caption{Distribution of all anomaly coefficients  (over all $h^{1,1}$), assuming a standard embedding of the GUT gauge group into $E_8 \times E_8$, in our ensemble of KS (CICY) heterotic compactifications (gray) (red) with $\alpha_{\mathrm{GUT}}^{-1}=25$ and $g_S=1$.} 
    \label{fig:distribution_anomaly_coeffs}
\end{figure}
\begin{figure}[ht]
    \centering
\includegraphics[width=0.495\textwidth]{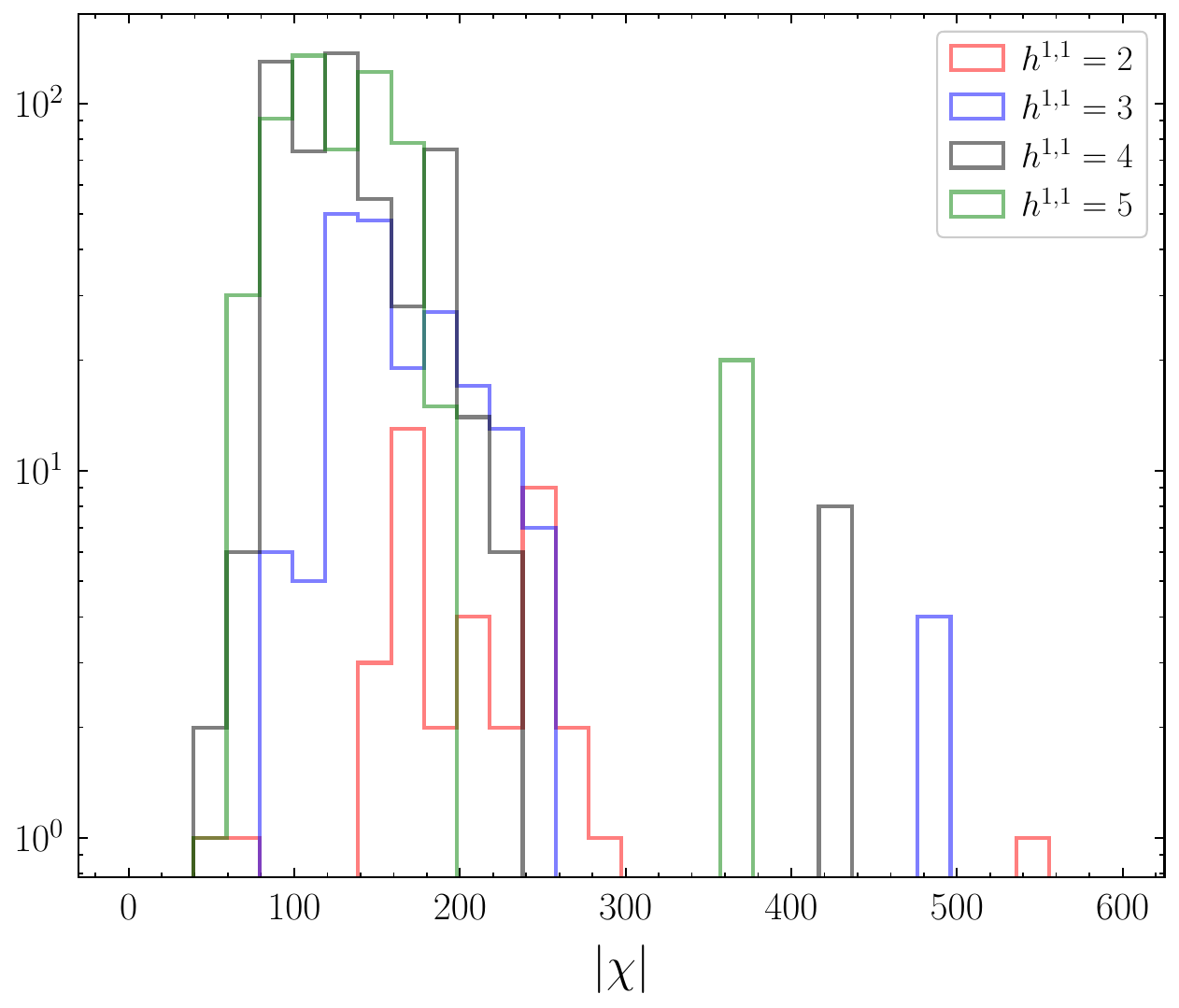}
    \caption{Euler characteristic (in absolute value) of manifolds in our ensemble of KS heterotic compactifications. Assuming a standard embedding, to obtain the three Standard Model matter generations we must impose $\chi \in 6\mathbb{Z}$, which is satisfied for ($17\%$) of the total 2027 in our ensemble.
    } \label{fig:euler_characteristic}
\end{figure}

\begin{figure}
    \centering   \includegraphics[width=0.5\linewidth]{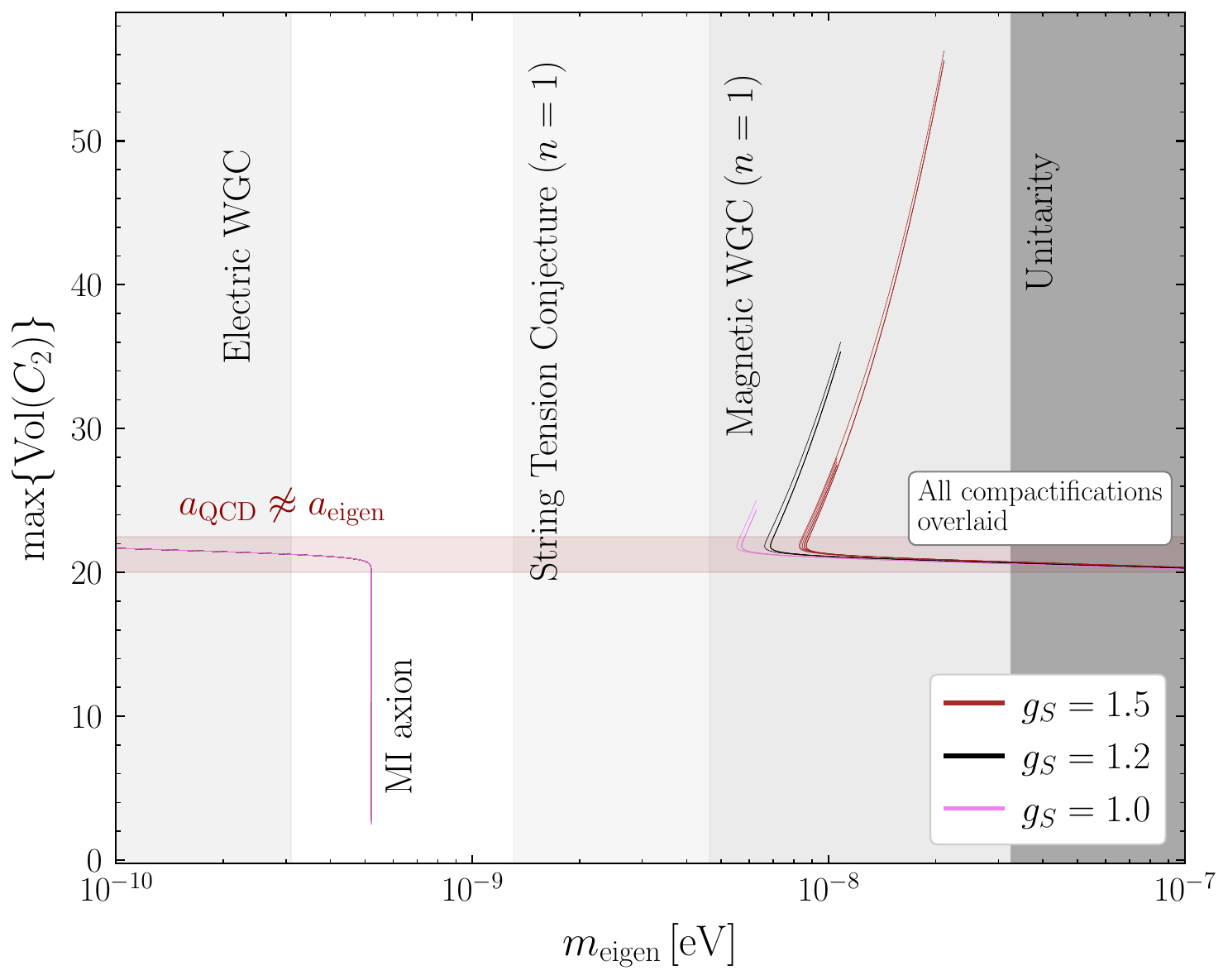}
    \caption{As in Fig.~\ref{fig:meigens}, but varying $g_s$ for fixed $\alpha_\mathrm{GUT}^{-1}=25$. For ease of visualization we only show manifolds with $h^{1,1}=2$.
    }
    \label{fig:meigens_gs}
\end{figure}

\begin{figure}
    \centering   \includegraphics[width=0.5\linewidth]{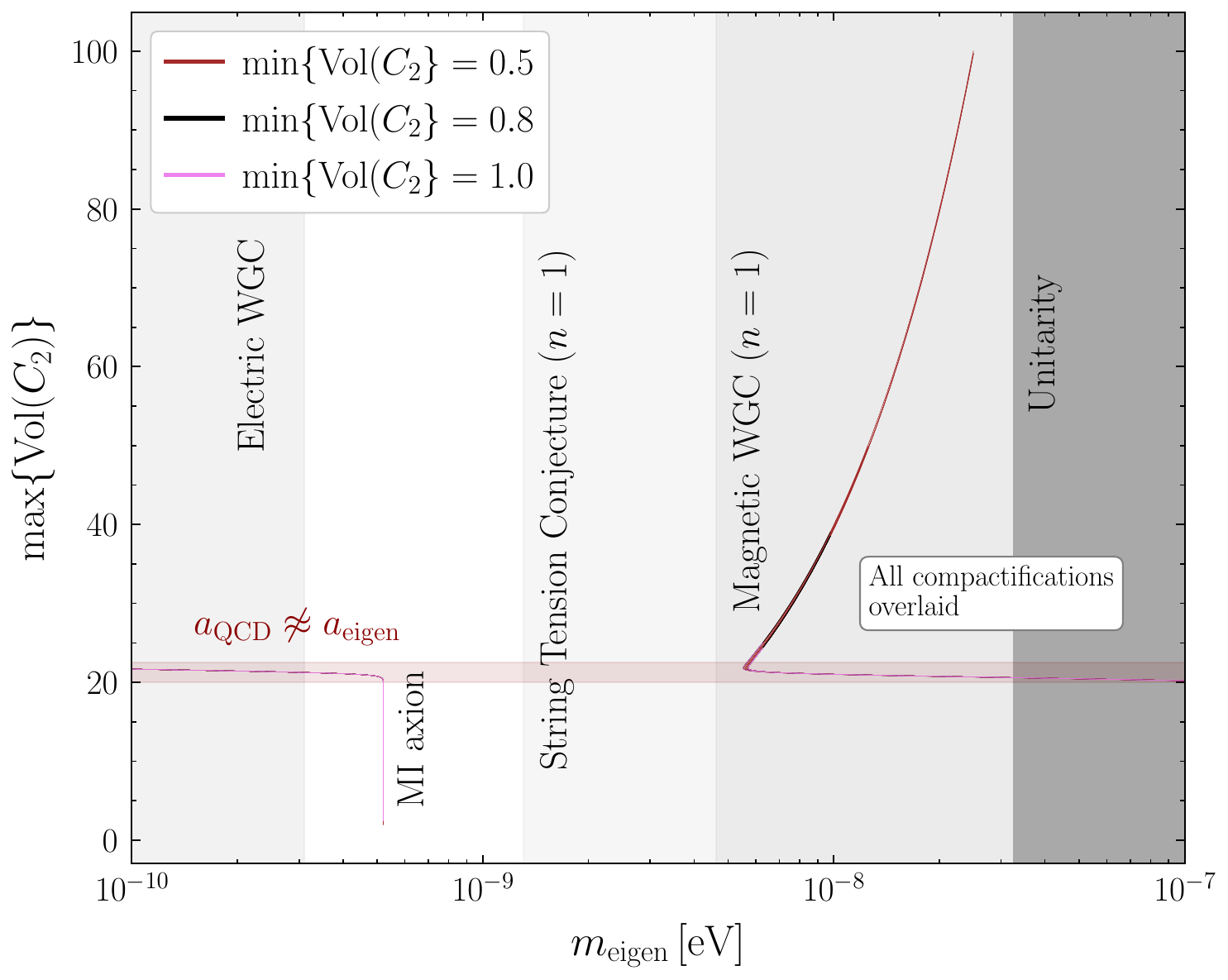}
    \caption{As in Fig.~\ref{fig:meigens}, but varying the minimum volume of effective curves $c$ which defines the SKC. We  fix $\alpha_\mathrm{GUT}^{-1}=25$ and $g_s=1$. For ease of visualization we only show manifolds with $h^{1,1}=2$.
    }
\label{fig:h11_2_light_masses_evolution_vary_SKC_c}
\end{figure}

\begin{figure}
    \centering
\includegraphics[width=0.5\linewidth]{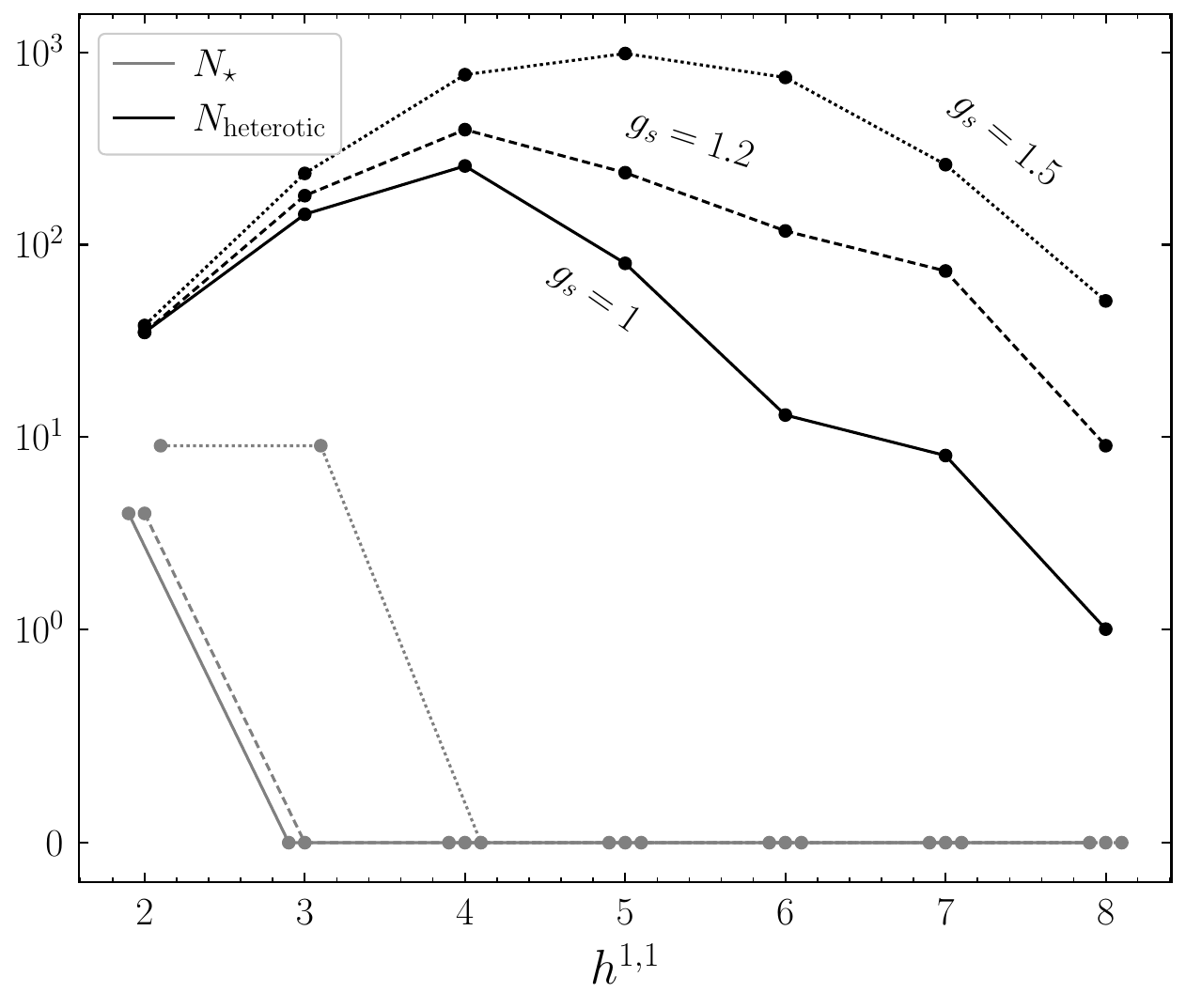}
    \caption{As in Fig.~\ref{fig:Nhet_alphaGUT_dependence}, but varying $g_s$, fixing $\alpha_\mathrm{GUT}^{-1}=25$.
    }
\label{fig:Nhet_gs_dependence}
\end{figure}

\begin{figure}
    \centering
\includegraphics[width=0.5\linewidth]{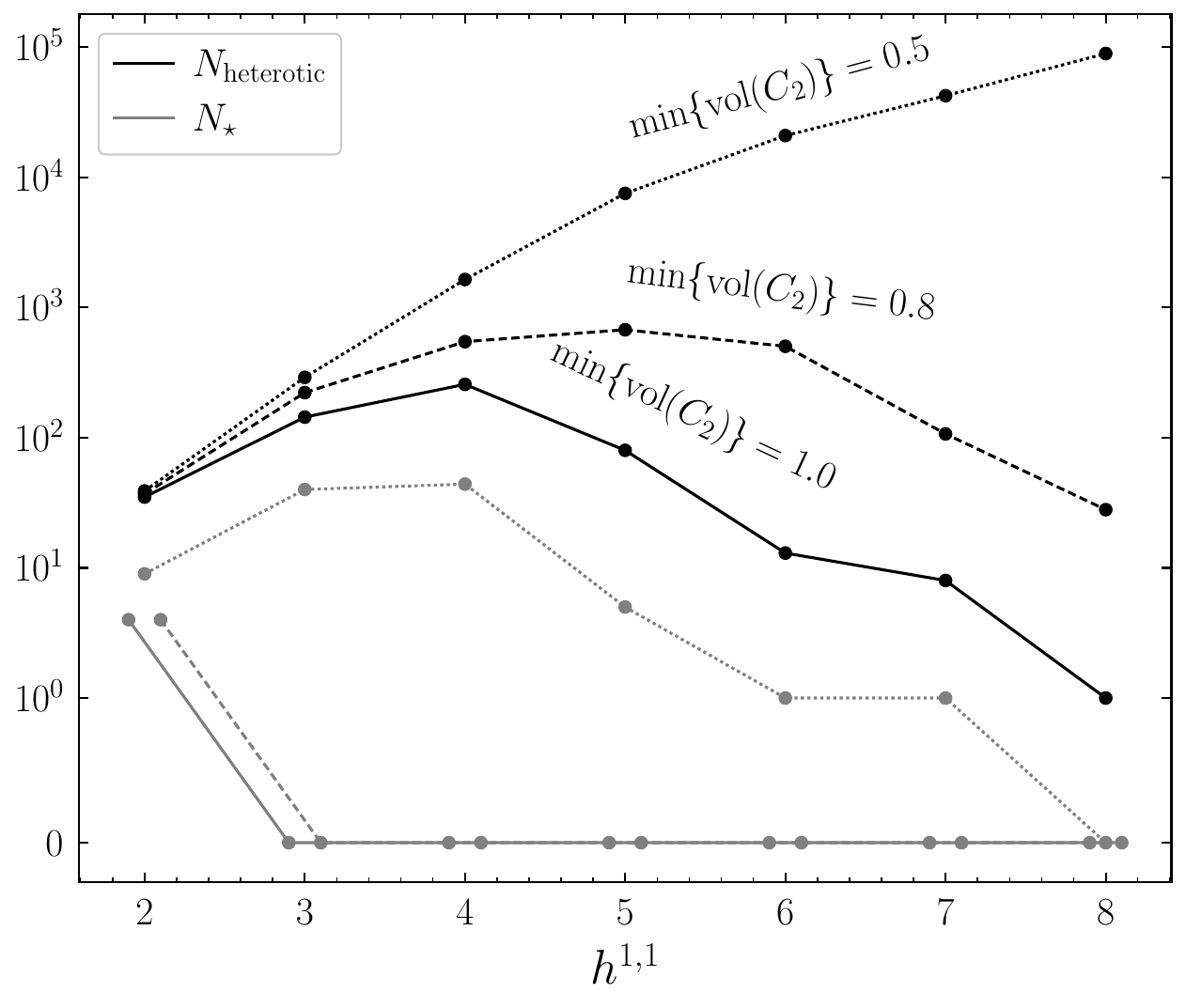}
    \caption{As in Fig.~\ref{fig:Nhet_alphaGUT_dependence}, but varying the minimum curve volume allowed within the SKC, fixing $\alpha_\mathrm{GUT}^{-1}=25$ and $g_s=1$.
    }
\label{fig:Nhet_maxC2vol_dependence}
\end{figure}

\begin{figure}
    \centering
\includegraphics[width=0.5\linewidth]{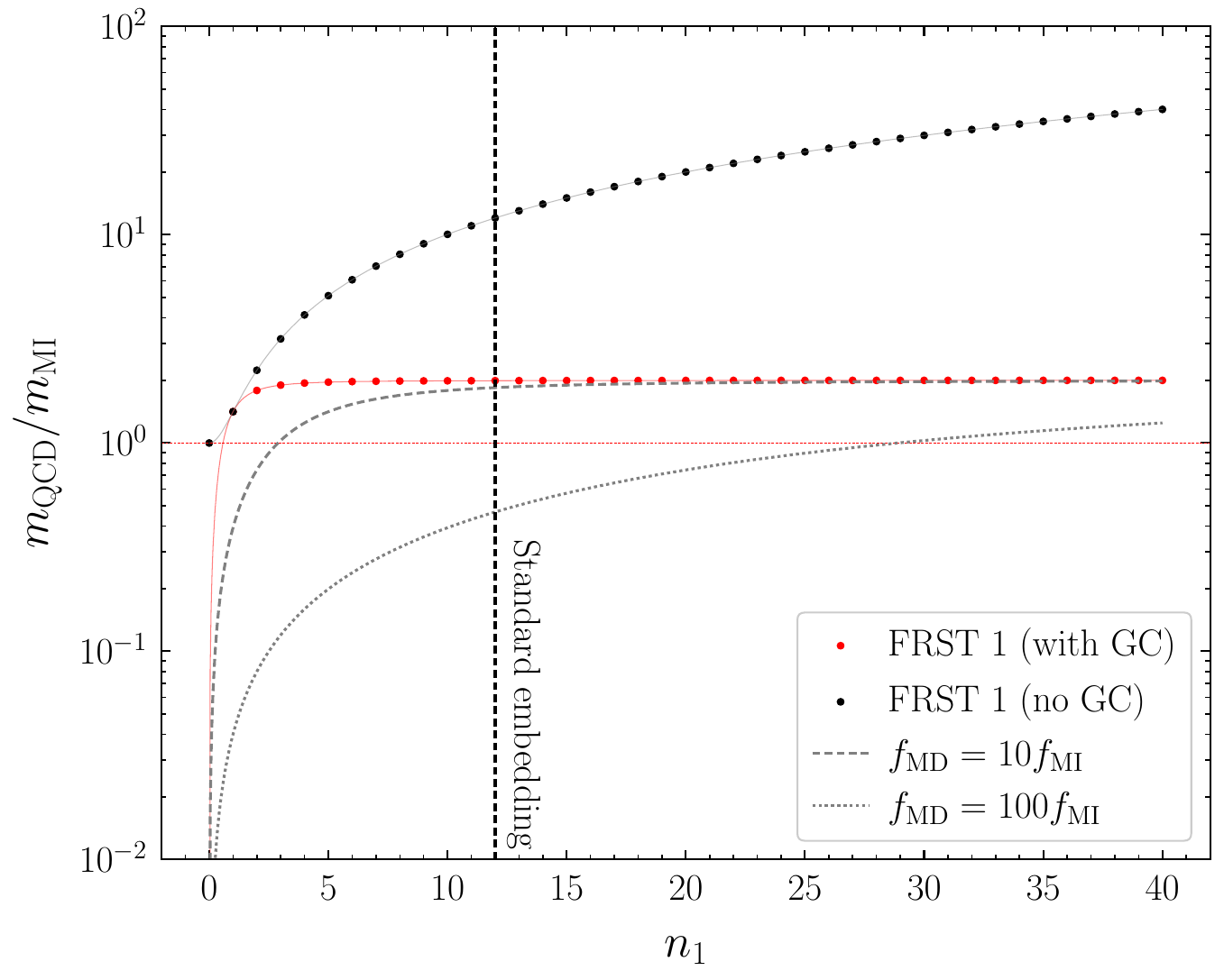}
\caption{The QCD axion mass, relative to the MI value, assuming an $E_8\times E_8$ heterotic compactification on the FRST 1 of Table \ref{tab:manifolds}. We indicate this assuming hidden-sector gaugino condensation occurs (does not occur) by the red (black) points. We fix $\alpha_\mathrm{GUT}^{-1}=30$, and vary one of the anomaly coefficients $n_1$ (red points)\footnote{The QCD axion mass is independent of the other anomaly coefficient $n_2$ (up to a suppressed contribution coming from mixing with the associated heavy MD axion).}, with $n_1=12$ corresponding to a standard embedding. For FRST 1, the lightest MD axion has a decay constant $f_\mathrm{MD}\simeq 1.01 \,f_\mathrm{MI}$, with $f_\mathrm{MI}$ the decay constant of the MI axion. We also indicate the locus of the QCD axion mass in this scenario for similar hypothetical compactifications which would have $f_\mathrm{MD}=10f_\mathrm{MI}$  and $f_\mathrm{MD}=100f_\mathrm{MI}$. As discussed in Sec. \ref{SM:theory}, we do not expect $f_\mathrm{MD}/f_\mathrm{MI}$ to exceed $\mathcal{O}(1)$ for compactifications on CY 3-folds, such that in practice we do not expect violations of our lower bound $m_\mathrm{QCD}>m_\mathrm{MI}$.}
\label{fig:gaugino_cond_mqcd}
\end{figure}

\begin{figure}
    \centering
\includegraphics[width=0.5\linewidth]{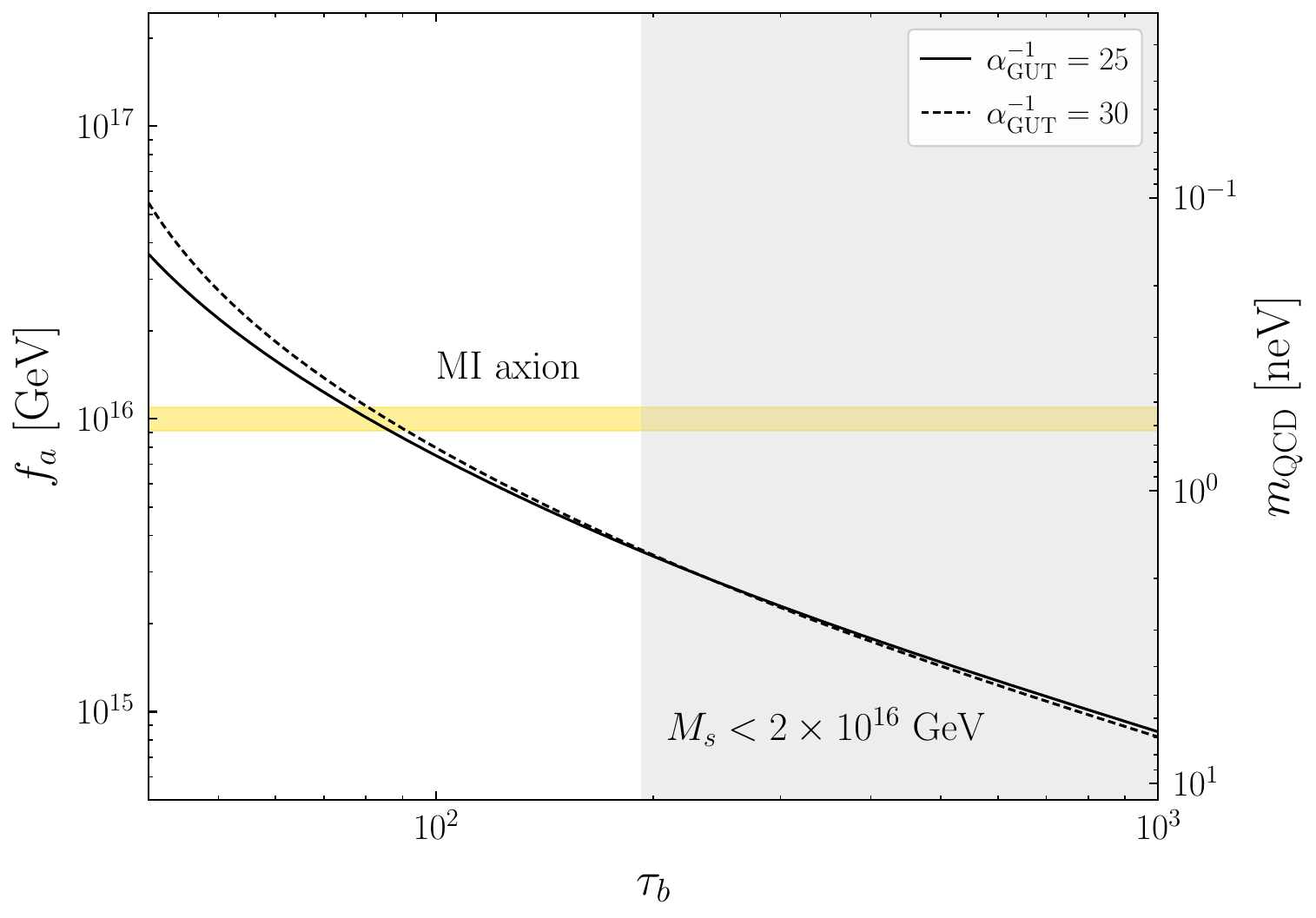}
\caption{The QCD axion decay constant and mass for a F-theory compactification with a Swiss-cheese base $Bl_p(\mathbb{P}^3)$ (solid). The compactification with base $\mathbb{P}^1 \times \mathbb{P}^2$ (see Sec. \ref{SM:duality_Ftheory} for details) has precisely the heterotic MI value (horizontal band). The region compatible with unification at the SUSY GUT scale, $M_s \gtrsim M_\mathrm{GUT}$, is shaded.} 
\label{fig:F_theory}
\end{figure}

\end{document}